\documentclass[12pt]{article}

\usepackage{fleqn}
\usepackage{epsfig}

\usepackage{graphicx}
\usepackage{dcolumn}
\usepackage{bm}
\usepackage{wrapfig,rotating}
\usepackage{amssymb,amsmath,array}
\usepackage{subfigure}

\def\beq{\begin{equation}}
\def\be{\begin{equation}}
\def\eeq{\end{equation}}
\def\ee{\end{equation}}
\def\bea{\begin{eqnarray}}
\def\eea{\end{eqnarray}}
\def\bq{\begin{quote}}
\def\eq{\end{quote}}

\def\gappeq{\mathrel{\rlap {\raise.5ex\hbox{$>$}}
{\lower.5ex\hbox{$\sim$}}}}
\def\lappeq{\mathrel{\rlap{\raise.5ex\hbox{$<$}}
{\lower.5ex\hbox{$\sim$}}}}
\def\Toprel#1\over#2{\mathrel{\mathop{#2}\limits^{#1}}}

\def\beq{\begin{equation}}
\def\eeq{\end{equation}}
\def\beqn{\begin{eqnarray}}
\def\eeqn{\end{eqnarray}}
\def\ppsi[#1,#2,#3,#4]{\psi^#1_{#2#3}(#4)}
\def\cchi[#1,#2]{\chi^{#1#2}}

\begin{document}

\begin{titlepage}
\vspace*{-2cm}
\begin{flushright}
DSF-NA/2008-15
\end{flushright}

{\Large
\begin{center}
\bf{Spectrum of positive and negative parity pentaquarks,
including $SU(3)_F$ breaking.}
\end{center}
}

\begin{center}
M. Abud, F. Buccella, D. Falcone, G. Ricciardi and F. Tramontano\\
{\textit{Dipartimento di Scienze Fisiche, Universit\`{a} di Napoli
``Federico II" \\ and INFN, Sezione
di Napoli, Via Cintia, I-80126 Napoli, Italy} } \\

\begin{abstract}
We present the spectrum of the lightest pentaquark states of both
parities and compare it with the present experimental evidence for
these states. We have assumed that the main role for their mass
splittings is played by the chromo-magnetic interaction. We have
also kept into account the $SU(3)_F$ breaking for their contribution
and for the spin orbit term. The resulting pattern is in good agreement
with experiment.
\end{abstract}

\vspace*{0.5cm}

\end{center}

\noindent Keywords: Pentaquarks, Chromo-magnetic interaction \\
\noindent PACS: 12.39.Ki, 12.40.Yx

\end{titlepage}

\noindent

\section{Introduction}

Exotic baryonic resonances in $KN$ scattering have been found by phase shift
analysis~\cite{AR1,AR2,AR3}. Evidence has also been claimed at CERN SPS for
the existence of a narrow $\Xi ^{-}\,\pi ^{-}$ baryon resonance with mass
$1.862\pm 0.002$ GeV at $4.0\,\sigma $~\cite{NA}. This state would be an
exotic baryon $\Xi ^{--}$ with isospin $I=3/2$, hypercharge $Y=-1$ and a
quark content $ddss\bar{u}$. The original observation of a narrow
exotic baryon resonance $\Theta ^{+}$ (with $I=0$ and $Y=2$ ) in two
independent experiments~\cite{nakano} was confirmed~\cite{N}.
Such discovery has motivated
several attempts to study it as a $uudd\bar{s}$
state~\cite{JW},\cite{SR},\cite{All}.
\newline
The circumstance that the previous evidence in photoproduction for the
$\Theta ^{+}$~\cite{N} has been recently disproved~\cite{CLAS} does not
seem to be the last word.
Recently, new results have became available, partly based on new data,
confirming seeing the $\Theta ^{+}$~\cite{news};
the results of a new run, which should increase the statistics
by 10, are expected at LEPS.\newline
The $N\pi $ $1/2^{+},I=1/2,Y=1$ resonance with mass $1358$ MeV discovered at 
BES~\cite{BES} in the decay $J/\psi \rightarrow p\bar{p}\pi ^{0}$ and the 
$P_{11}$(1860) and $P_{13}$(1900) states found in the photoproduction on proton
of $K\Lambda $ and $K\Sigma $~\cite{Nic} are natural candidates for a pentaquark 
interpretation.\newline
Exotic baryons, consisting of $4q$ and a $\bar{q}$ have been
studied~\cite{HS}, at the times of bubble chambers, the best
device to detect these particles. \\ 
In this paper, we evaluate their spectrum
with the assumption that the mass splittings between the different states
are due to the effect of the chromo-magnetic interaction; we also include
the effect of $SU(3)_{F}$ flavour symmetry breaking. Such simple model has
been proved remarkably successfully at describing the spectrum of the
standard baryons~\cite{DGG}, which transform as the $56$ representation of
flavour-spin $SU(6)_{FS}$~\cite{GR}. The same approach has already been
exploited to evaluate the spectrum of the positive and negative parity~$Y=2$ 
baryons~\cite{BFT} and the spectrum of the scalar mesons~\cite{BHRS}.\\
Here, we extend the analysis to the pentaquarks with one or more strange
constituents, that is to $Y<2$. As in~\cite{BFT}, we consider $4q$ in $S$ and 
$P$ wave, which give rise, together with the $\bar{q}$ in S-wave with
respect to them, to negative and positive parity states, in the last case
with an extension of the proposal of Jaffe and Wilckzek in~\cite{JW}. We
shall consider states exclusively of the type $(4q)\bar{q}$. We shall call
$p$ this state and $t$ the corresponding $(4q)$ subsystem.

\section{The chromo-magnetic interaction}

The hyperfine interaction arising from one gluon exchange between
constituents leads to a simple Hamiltonian involving the colour and spin
degrees of freedom: 
\begin{equation}
H_{CM}=\sum_{i}\,m_{i}-\frac{1}{4}\,\sum_{i<j}\,\frac{K_{ij}}{m_{i}\,m_{j}}
\,\sum_{a=1}^{8}\,\sum_{k=1}^{3}\,(\lambda _{a}\,\otimes \,\sigma
_{k})^{(i)}\,(\lambda _{a}\,\otimes \,\sigma _{k})^{(j)}
\end{equation}
where the index $i~(j)$ refers to the $i$th ($j$th) quark, $\lambda _{a}$ 
are the 8 Gell-Mann matrices, $\sigma _{k}$ the Pauli matrices, $m_{i}$ the
mass of the $i$th (anti)quark and $K_{ij}$ appropriate coupling constants;
the sum above depends on the spatial relative configuration of quarks $i$
and $j$, since one has to include only pairs which effectively interact with
each other via the short range QCD interaction.
It's natural to define the operators $O_{CM}^{(ij)}$ for the 2-body
chromo-magnetic operators by
\begin{equation}
O_{CM}^{(ij)}=\frac{1}{4}\,\sum_{a=1}^{8}\,\sum_{k=1}^{3}\,
(\lambda_{a}\,\otimes \,\sigma _{k})^{(i)}\,(\lambda _{a}\,\otimes
\,\sigma_{k})^{(j)}.
\end{equation}
Quarks belong to the fundamental representation of $SU(6)_{CS}$ ($6_{CS}$)
and transform as ($3_{C},\,2_{S}$) with respect to $SU(3)_C\,\times\,SU(2)_S$,
they are represented by a wave function
$\psi _{\alpha \,p}$, while the antiquark $\chi ^{\alpha \,p}$ trasforms in
the conjugate rapresentation ($\bar{3}_{C},\,2_{S}$), with $\alpha $ the
color index ($\alpha=1,2,3$) and $p$ the spin index ($p=1,2$). Since the
Hamiltonian involves only two body forces, we have just to consider the
action of the chromo-magnetic operator on a specific quark pair
$\psi_{\beta\,q}^{(1)}\,\psi _{\nu \,l}^{(2)}$ or a quark-antiquark pair
$\psi _{\beta\,q}\,\chi ^{\mu \,m}$ that is displayed below: 
\begin{eqnarray}  \label{qq}
\left(O^{(12)}_{CM}\,\psi^{(1)}\,\psi^{(2)} \right)_{\alpha\,p,\,\mu\,m}&=&
\psi _{\alpha \,p}^{\prime \, (1)}\,\psi _{\mu \,m}^{\prime \, (2)} \\
&=&\frac{1}{4}\,\sum_{\beta ,\nu =1,2,3}\,\sum_{q,l=1,2}\,
\sum_{\substack{ a=1,...,8 \\ k=1,2,3}}\,
\left[ (\lambda _{a})_{\alpha }^{\beta }\,(\sigma _{k})_{p}^{q}
\right] ^{(1)}\,\left[ (\lambda _{a})_{\mu }^{\nu }\,(\sigma _{k})_{m}^{l}
\right] ^{(2)}\,\psi _{\beta \,q}^{(1)}\,\psi _{\nu \,l}^{(2)} \notag \\
&=&\frac{1}{6}\,\psi _{\alpha \,p}^{(1)}\,\psi _{\mu \,m}^{(2)}-\frac{1}{3}
\,\psi _{\alpha \,m}^{(1)}\,\psi _{\mu \,p}^{(2)}-\frac{1}{2}\,\psi _{\mu
\,p}^{(1)}\,\psi _{\alpha \,m}^{(2)}+\psi _{\mu \,m}^{(1)}\,\psi _{\alpha
\,p}^{(2)}  \notag
\end{eqnarray}
\begin{eqnarray}  \label{qa}
\left(O^{(15)}_{CM}\psi^{(1)}\,\chi^{(5)} \right)_{\alpha \,p}^{\nu\,l}&=&
\psi _{\alpha \,p}^{\prime (1)}\,\chi ^{\prime \, \nu \,l (5)} \\
&=&\frac{1}{4}
\,\sum_{\beta ,\mu =1,2,3}\,\sum_{q,m=1,2}\,
\sum_{\substack{ a=1,...,8\\k=1,2,3}}\,
\left[ (\lambda _{a})_{\alpha }^{\beta }\,(\sigma _{k})_{p}^{q}
\right] ^{(1)}\,\left[ (\lambda _{a})_{\mu }^{\nu }\,(\sigma _{k})_{m}^{l}
\right] ^{(2)}\,\psi^{(1)}_{\beta \,q}\,\chi ^{\mu \,m (2)}  \notag \\
&=&\frac{1}{6}\,\psi^{(1)} _{\alpha \,p}\,\chi ^{\nu \,l (5)
}-\frac{1}{3}\,\delta_{p}^{l}\,\sum_{n=1}^{2}\,\psi^{(1)} _{\alpha \,n}
\,\chi ^{\nu \,n (2)}-\frac{1}{2} \,\delta _{\alpha }^{\nu }\,
\sum_{\rho =1}^{3}\,\psi^{(1)} _{\rho \,p}\,\chi^{\rho \,l (5)} \notag \\
&& +\delta _{\alpha }^{\nu }\,\delta _{p}^{l}\,\sum_{\rho=1}^{3}\,
\sum_{n=1}^{2}\,\psi^{(1)} _{\rho \,n}\,\chi ^{\rho \,n (5)}. \notag
\end{eqnarray}
The action of $H_{CM}$ on the pentaquark states $|\Phi_A\,\rangle$
(a complete set
of states for assigned flavour and spin of the pentaquark) is readily
obtained as follows. Since the $|\Phi_A\,\rangle$'s can be written as
\beq
|\Phi_A\,\rangle\,=\,\Theta_A^{(12)\,\alpha\,p,\,\mu\,m}\,
\psi^{(1)}_{\alpha\,p}\,\psi^{(2)}_{\mu\,m}
\eeq
where $\Theta^{\alpha\,p,\,\mu\,m}$ is trilinear in $\psi^{(3}$,
$\psi^{(4)}$ and the antiquark $\chi$ define
\beq \label{eqqq}
O_{CM}^{(12)}\,|\Phi_A\,\rangle\,=\,|\Phi_A\,\rangle'\,=
\,\Theta_A^{(12)\,\alpha\,p,\,\mu\,m}\,\psi'^{\,(1)}_{\alpha\,p}
\,\psi'^{\,(2)}_{\mu\,m}
\eeq
replacing $\psi'^{\,(1)}_{\alpha\,i}\,\psi'^{\,(2)}_{\mu\,m}$
according to Eq.\ref{qq}.
We get the new states $|\Phi'_A\,\rangle$ as
\beq
|\Phi'_A \rangle = \sum \, C_A^B\,|\Phi_B\,\rangle
\eeq
where $C_A^B=\langle\,\Phi_B\,|\Phi'_A\,\rangle\,=
\,\langle\,B|\,O^{(12)}\,|A\,\rangle$ are the
matrix elements of the operator $O^{(12)}$ between pentaquark
states. The same reasoning applies to the operator $O^{(i\,5)}$
for $q\bar{q}$ interaction.
\newline
In the flavour symmetry limit (i.e. $m_{i}=m$), the hyperfine interaction
reduces to a term proportional to 
\begin{equation}  \label{simlim}
\sum_{i<j}\,\sum_{a=1}^{8}\,\sum_{k=1}^{3}\,(\lambda _{a}\,\otimes \,\sigma
_{k})^{(i)}\,(\lambda _{a}\,\otimes \,\sigma _{k})^{(j)}
\end{equation}
which can be expressed in terms of the Casimir operators of $SU(6)_{CS}$,
$SU(3)_{C}$ and $SU(2)_{S}$~\cite{HS} denoted in the following by $C_{6}$,
$C_{3}$ and $C_{2}$, respectively. 

\subsection{Pauli Principle and Flavour Content of Pentaquarks}

The Pauli principle imposes the complete antisymmetry of the wave
function of the quarks in the tetraquark.
On the other hand, the requirement that the pentaquark
is a colour singlet enforces
the tetraquark wave function to transform as a $3_{C}$ . The only
representations occuring in the direct product of four $6_{CS}$'s, which
contain a $3_{C}$, are the $210_{CS}$, $105_{CS}^{\prime }$, $105_{CS}$ and
$\overline{15}_{CS}$ of $SU(6)_{CS}$. For all these
representations there is a $3_C$ with $S=1$ while a $3_C$ with $S=0$ is
present in $210_{CS}$ and $105$', which contains also a $3_C$ with $S=2$.
\newline
For a symmetric spatial wave function, as fo the tetraquark in S-wave,
the corresponding flavour wave functions must tranform congruentely 
in order to fulfill the Pauli principle and one gets straightforwardly the
correspondence between the colour-spin and $SU(3)_{F}$ flavour contents
\begin{equation}  \label{SU6SU3}
210_{CS}\leftrightarrow 3_{F}\qquad 105_{CS}\leftrightarrow \bar{6}
_{F}\qquad 105_{CS}^{\prime }\leftrightarrow 15_{F}\qquad \overline{15}
_{CS}\leftrightarrow 15_{F}^{\prime }
\end{equation}
In the case the tetraquark subsystem be in P wave, we need to take into
account, besides the flavour, also the spatial degrees of freedom. In that
case we get the following correspondances~\cite{BFT}: 
\begin{eqnarray}
210_{CS}^{(1)} \leftrightarrow \bar{6}_{F}\qquad 105_{CS}^{(1)}
\leftrightarrow 3_{F} &\qquad& 210_{CS}^{(2)} +105_{CS}^{\prime \, (1)}
\leftrightarrow 15_{F}+3_{F}\qquad  \label{SU6SU3} \\
105_{CS}^{\prime \, (2)} \leftrightarrow 15_{F}^{\prime }+\bar{6}_{F}
&\qquad& 105_{CS}^{(2)} +\overline{{15}}_{CS}\leftrightarrow 15_{F}
\end{eqnarray}
With the exception of the $\overline{15}_{CS}$, all the representations
appear twice, since there are two inequivalent ways of obtaining them,
namely:
\begin{eqnarray}
21_{CS} \otimes 21_{CS} &=& 126_{CS}+210_{CS}^{(1)} +105_{CS}^{(1)} \\
21_{CS} \otimes 15_{CS} &=& 210_{CS}^{(2)} +105_{CS}^{\prime \, (1)} \\
15_{CS} \otimes 15_{CS} &=& 105_{CS}^{(2)} +105_{CS}^{\prime \, (2)} 
+\overline{{15}}_{CS}
\end{eqnarray}

\subsection{Negative parity pentaquarks}

By composing the S-wave $t$'s with the $\bar{q}$, one gets the following
$1_C $ flavour spin multiplets: 

\begin{eqnarray}
8_F + 1_F, S &=& 1/2 + 1/2 + 3/2 \\
\overline{10}_F + 8_F , S &=& 1/2 + 3/2 \\
27_F + 10_F, + 8_F , S &=& 1/2 + 1/2 + 3/2 + 3/2 + 5/2 \\
35_F + 10_F, S &=& 1/2 + 3/2
\end{eqnarray}

Let us construct explicitely the pentaquark states relevant for the
calculation of the spectrum. Since we have at disposal only 3 flavours, at
least 2 quarks must have identical flavour, say $uu$. The more general state
would then correspond to the case the remaining pair differ in flavour from
each other and from $u$, so we can denote it by $ds$. The $uu$ pair must be
symmetric, so it is a $6_{F}$ to be combined with a $ds$ pair that can be a
$6_{F}$ (symmetric under the exchange $d\leftrightarrow s$)
or a
$\overline{3}_{F}$ (antisymmetric under the exchange $d\leftrightarrow s$).
As $6_{F}\otimes 6_{F}=\overline{6}_{F}+15_{F}^{\prime }+15_{SF}$ and
$6_{F}\otimes \overline{3}_{F}=3_{F}+15_{AF}$, we call the $15_{SF}$ the
representation appearing in the state $uu(ds)_{S}$ and $15_{AF}$ that
appearing in $uu(ds)_{A}$. The other representations $3_{F}$,
$\overline{6}_{F}$, $15_{F}^{\prime }$ appear unambigously.
So the states in this case
can be classified according to the spin and flavour of the tetraquark. The
more transparent way of getting the classification of the states is based on
a well known argument concerning the tranformation properties of the
tetraquark wave function under the group $SU(6)_{FS}\otimes SU(3)_{C}$.
Since the only occurring $3_{C}$ state is a $210_{FS}$, whose decomposition
under $SU(3)_{F}\otimes SU(2)_{S}$ is given by:

$(3_{F},1_{S})+(3_{F},3_{S})+(\overline{6}_{F},3_{S})
+(15_{F},1_{S})+(15_{F},3_{S})+(15_{F},5_{S})+(15_{F}^{\prime },3_{S})$,

we readily get the 17 states below\footnote{
When the two quarks in the second pair are equal, the $\bar{3}$ and the 
$15_{AF}$ are absent. In the case one of the two quarks, in the second pair, 
is equal to the ones in the first pair, no $3$ and $\overline{6}_{F}$ occur. 
If all quarks are the same, then only the $15_{F}^{\prime }$ is present.}:
\begin{center}
{\scriptsize 
\begin{tabular}{|c|c|c|}
\hline
Pentaquark Spin & Flavour-Spin Tetraquark & Number of states \\ \hline
&& \\
1/2 & ($3_{F},S=0,1) \,;\, (\overline{6}_{F},S=1) \,;\, (15_{F}^{\prime
},S=1) \,;\, (15_{SF},S=0,1) \,;\, (15_{AF},S=0,1$) & 8 \\
&& \\
\hline
&& \\
3/2 & ($3_{F},S=1) \,;\, (\overline{6}_{F},S=1) \,;\, (15_{F}^{\prime },S=1)
\,;\, (15_{SF},S=1,2) \,;\, (15_{AF},S=1,2$) & 7 \\
&& \\
\hline
&& \\
5/2 & ($15_{SF},S=2) \,;\, (15_{AF},S=2$) & 2 \\
&& \\
\hline
\end{tabular}
}
\end{center}
As a matter of fact, in order to operate with the chromo-magnetic 
Hamiltonian Eq.'s.(\ref{qq},\ref{qa}) we need the explicit expression of the 
wave functions, which for the sake of completeness are given in Appendix~A.
Let's write explicitely the expression of $m^{(S)}$ for our conventional
state $uuds\overline{q}$ 
\beq  \label{ondaS}
m^{(S)} =2\,m_{u}+m_{d}+m_{s}+m_{\overline{q}} + K^{S} S 
\eeq
where $S$ is a $17 \times 17$ matrix, which splits into $8 \times 8$, $7 \times 
7$ and $2 \times 2$ matrices corresponding to spin $1/2$, $3/2$ and $5/2$,
respectively. The matrix elements of $S$ between the states
in Appendix A may be computed through the use of Eq.'s.(\ref{qq},\ref{qa}) 
and their values are reported in Appendix B.
By an appropriate choice of $\bar{q}$ and also with suitable change of the
set $uuds$, one can apply Eq.(\ref{ondaS}) to any negative parity
pentaquark. For instance, the $I_{3}=1/2$, $Y=-3$, $J=1/2$ and $3/2$ 
states have the quark content $ssss$ and $\bar{q}$.

\subsection{The Flavour Symmetry Limit}

As mentioned before, in the case we can disregard the breaking of $SU(3)_{F}$,
the hyperfine interaction Eq.(\ref{simlim}) can be expressed in terms of a
purely gruppal expression involving the quadratic Casimir operators.

A weaker, and more useful, limit is when all the quarks have the same
constituent mass (we assume isospin invariance $m_u=m_d$), while the
antiquark may be a light or strange one, corresponding to the $Y=+2$ baryons
(and some cases with $Y\leq 1$). In that limit the mass of a negative parity
pentaquark state is~\cite{BFT}:\footnote{
Our normalization for the Casimir operators, at difference with~\cite{HS},
is the one, which takes the value $n$ for the adjoint representation of
$SU(n)$.} 
\begin{eqnarray}
m^{(S)} &=&{\large 4}m_{q}+m_{\overline{q}}\,+\, \frac{K^{S}}{
m_{q}m_{\overline{q}}}\, \left[ C_{6}(p)-C_{6}(t)-\frac{1}{3}C_{2}(p)+ \frac{
1}{3}C_{2}(t)-\frac{4}{3}\right] \notag \\
&& -\, \frac{K^{S}}{m_{q}^{2}}\, \left[ C_{6}(t)-\frac{1}{3}C_{2}(t)-\frac{
26}{3}\right]  \label{eq1}
\end{eqnarray}
$K^{S}$ being the chromo-magnetic coupling constant
for $qq$ and $q\overline{q}$ (all in S-wave),
$m_{q} $ the common quark mass and $m_{\overline{q}}$ the antiquark mass. The
above expression in (\ref{eq1}) shows that the lightest states have large 
$SU(6)_{CS}$ Casimir for the $4q$ and as small as possible for the pentaquark.

Hypercharge $Y=+2$ baryon resonances (notice that in this case $\overline{q}=
\overline{s}$), called $Z^{\star }$, have been revealed in $K\,N$
interactions. The $Z^{\star }$ resonances $D_{03}$ and $D_{15}$ (the two
lower indexes stand for the isospin and twice the spin, respectively) have
negative parity and have possibly been revealed within mass ranges
$m_{D_{03}}=1788-1865$~\cite{AR1,AR3} and
$m_{D_{15}}=2074-2160$~\cite{AR2,AR3}.
From the spin content of the $3_C$ tetraquarks given before and from the
tensor products:
\bea
105_{CS}\,\otimes \,\bar{6}_{CS}\,&=&\,560_{CS}\,\oplus 70_{CS}\, \\
105_{CS}^{\prime }\,\otimes \,\bar{6}_{CS}\,&=&\,540_{CS}\,\oplus
70_{CS}\,\oplus \,20_{CS}.
\eea
the spin $S=5/2$ and the isospin $I=1$ of the $D_{15}$ state imply that the
pentaquark is in the $540_{CS}$ with the respective tetraquark being in the
$105_{CS}^{\prime }\,\ $of \ $SU(6)_{CS}$. By inserting in Eq.~(\ref{eq1})
the Casimir values : $C_{2}(t)=C_{2}(5)=6$, $C_{6}(t)=C_{6}(105^{\prime
})=26/3$, $C_{6}(p)=C_{6}(540)=49/4$ and $C_{2}(p)=C_{2}(6)=35/4$, the
chromomagnetic contribution to the $D_{15}$ mass turns out to be
$K^{S}\,\left( \frac{4}{3\,m_{q}\,m_{s}}+\frac{2}{m_{q}^{2}}\right)$.
\newline
Similar reasonings hold for $D_{03}$, the pentaquark state being in the
$560_{CS}$ and the tetraquark in the $105_{CS}$. The Casimir values involved
in the calculation are $C_{2}(t)=2$, $C_{6}(t)=C_{6}(105)=32/3$,
$C_{6}(p)=C_{6}(560)=57/4$ and $C_{2}(p)=C_{2}(4)=15/4$.
The observed mass difference
\beq
m_{D_{15}}-m_{D_{03}}=-\frac{1}{3}\,\frac{K^{S}}{m_{q}\,m_{s}}
\,+\,\frac{10}{3}\,\frac{K^{S}}{m_{q}^{2}}
\eeq
implies a larger value for the mass of the $D_{15}$ in agreement
with experiment.\\

\section{``Open door'' channels for pentaquarks}

It has been observed for the first time by Jaffe~\cite{J} that some
$qq \bar{q} \bar{q}$ mesons may decay into two ordinary mesons (PP, PV, VV)
\footnote{By P we mean a pseudoscalar meson, by V a vector meson.} by simple
 separation of the constituents: he called these channels ``open door''. 
\newline
Many years later a group theoretical criterium has been found~\cite{Bu} to
give a necessary condition for a PP and PV channels to be ''open door'',
according to $SU(6)_{CS}$ symmetry. Since a pseudoscalar and vector meson
transform as the singlet $1_{CS}$ or the adjoint $35_{CS}$ representation
of $SU(6)_{CS}$, respectively, only states, which transform as $1_{CS}$ (or
$35_{CS}$) of $SU(6)_{CS}$ may have ''open door'' amplitudes into PP (or VP)
final states.\newline
The contributions of the chromomagnetic interaction in the flavour 
symmetry limit are proportional to a combination of quadratic Casimir 
operators~\cite{HS} and depend mostly on the $SU(6)_{CS}$ Casimir operator. 
Therefore the eigenstates of the mass spectrum belong to almost irreducible 
representations of $SU(6)_{CS}$. This property is weakly affected by the
breaking of $SU(3)_F$. In particular, the lighter tetraquark meson scalar 
(or axial) states, which transform approximately as a singlet 
(or $35_{CS}$), have large ''open door'' amplitudes into PP (or VP)
channels~\cite{BHRS}.

These considerations can be extended to pentaquarks, as a consequence of the 
$SU(6)_{CS}$ transformation properties of the baryon $1/2^{+}$ octet and of
the $3/2^{+}$ decuplet, respectively in the $70_{CS}$ and the $20_{CS}$
representations. Since the pseudoscalars are colour-spin singlets only
pentaquarks with the same $SU(6)_{CS}$ transformation properties have ''open
door'' amplitudes into a channel consisting of one of these baryons and a
pseudoscalar meson~\cite{Bu}. This selection rule often coincide with the
one proposed in~\cite{BS} in analogy with the $SU(6)_{FS}$ selection rule
found in~\cite{SR}, but is more restrictive. \\
As seen in Eq.(9), for the negative parity pentaquarks there is
a relation among $SU(3)_{F}$ and $SU(6)_{CS}$ transformation properties of
the $4q$. This relation implies, according to Eq.(\ref{eq1}), larger masses
for higher dimension $SU(3)_{F}$ representations, since they correspond to
smaller $SU(6)_{CS}$ representations (more precisely with smaller quadratic
Casimir) for the $4q$, as a consequence of the sign of
the chromo-magnetic contribution proportional to $K^S$.
This implies that the lightest $J=S=1/2$ or $3/2$
states will be those transforming as the $70_{CS},J=S=1/2$ or the $20_{CS},J=S=3/2$
representations. Therefore there is a correlation between smaller mass and
large couplings to the final channels consisting of a baryon of the $56$ of
$SU(6)_{FS}$ and a pseudoscalar meson. For these negative parity pentaquarks
we expect the ''open door'' channels above threshold to be difficult to
detect for their broad width, as the long controversy about the $f_{0}$ has
shown.\newline
Instead we expect the more likely detectable positive parity pentaquarks to
be those with large couplings to the final states. In conclusion we expect P
and D-wave resonances to have been already found.\newline
As long as for the positive parity pentaquarks with the $\bar{q}$ in P-wave
with respect to the $4q$, there are is no ''open door'' channel, since the
$\bar{q}$ has no quark in S-wave to build a meson~\cite{HS}. So we expect
these states to be difficult to detect and for this reason we will not
discuss them here.

\subsection{Positive parity pentaquarks}

Let us consider the pentaquark with positive parity with $t$ in $P$-wave
and $\bar{q}$ in S-wave with respect to $t$.\\
In this case, the mass of the $Y=2$ pentaquark state, that we indicate with
$m^{P}$, can be calculated considering the system as composed of a pair
of diquarks with total orbital momentum $L=1$ and the
$\overline{s}$~\cite{JW}, whose chromo-magnetic interaction (with flavour
independent coupling costant $\overline{K}^{(P)}$) with the quarks can be
expressed in terms of Casimirs~\cite{BFT}.
The spin orbit term, proportional to $\sum_{i=1}^{4}\,1/m_{i}\ \vec{L}\vec{.S_{i}}$,
in the limit of equal masses depends only on the spin of the tetraquark
$\vec{S_{t}}$ and on the colour of the two diquarks.
Besides the nude masses and kinetic energy ($E_{kin} $) contributions, 
we must add the mass defects for the diquark clusters
$\Delta m_{qq}^{(12)}$ and $\Delta m_{qq}^{(34)}$~\cite{BFT}, so obtaining :
\begin{eqnarray} \label{mp1}
m^{(P)} &=&4m_{q}+m_{s}+\Delta m_{qq}^{(12)}+\Delta m_{qq}^{(34)}\,
-\,\frac{a}{4}\,\lambda_{b}^{(12)}\,\lambda_{b}^{(34)}\,
\vec{L}\vec{.S_{t}} + E_{kin} \\
&&+\,\frac{\overline{K}^{(P)}}{m_{q}\,m_{s}}\,
\left[C_{6}(p)\,-\,C_{6}(t)\,-\,\frac{1}{3}C_{2}(p)\,
+\,\frac{1}{3}C_{2}(t)\,-\,\frac{4}{3}\right] \notag
\end{eqnarray}
where the upper indices (12) and (34) refer to the two diquarks.
The mass defects $\Delta m_{qq}^{(12)}$ and $\Delta m_{qq}^{(34)}$ can be
equally calculated in terms of Casimirs according to the relation below: 
\begin{equation}
\frac{1}{4}\,\sum_{b=1}^{8}\,\sum_{k=1}^{3}\,
(\lambda _{b}\,\otimes \,\sigma_{k})^{(1)}\,
(\lambda _{b}\,\otimes \,\sigma _{k})^{(2)}\Rightarrow
\,\left(
C_{6}(q_{1}q_{2})\,-\,\frac{1}{2}C_{3}(q_{1}q_{2})\,-\,\frac{1}{3}
C_{2}(q_{1}q_{2})\,-\,4\right) .
\end{equation}
This contribution depends on the colour and spin of the pair of quarks
$q_{1}q_{2}$ and it is reported in Table~\ref{table2}.
\begin{table}[h!]
\begin{center}
\begin{tabular}{|c|c|}
\hline
&  \\[-9pt] 
$SU(3)_C {\times} SU(2)_S$ & $\frac{2 \Delta m_{qq}}{C_{qq}}$ \\[3pt] 
&  \\[-9pt] \hline
&  \\[-9pt] 
$(\bar{3},1)$ & $-2$ \\[3pt] 
$(6,3)$ & $-\frac{1}{3}$ \\[3pt] 
$(\bar{3},3)$ & $+\frac{2}{3}$ \\[3pt] 
$(6,1)$ & $+1$ \\[3pt] \hline
\end{tabular}
\end{center}
\caption{Chromomagnetic splittings for $2q$ states}
\label{table2}
\end{table}
It is assumed that the chromo-magnetic interaction concerns the quarks in
S-wave in the same pair~\cite{JW} and the $\bar{q}$ with both pairs. The
interaction among components not in $S$-wave is neglected; this is why the
interaction among the two diquark pairs does not contribute to 
Eq.(24). 
We shall take for the qq interaction in the pair the same coupling as in 
S-wave, i.e. $K^{S}$ and at difference from~\cite{BFT}, but as in~\cite{BS}, 
for the quark antiquark coupling half of that of S-wave. Indeed the factor
1/2 is a consequence of the total antisymmetrization of the tetraquark wave
function, which implies that the $\overline{q}$ has probability 1/2 to be in
S-wave with either pair: 
\begin{equation}
\overline{K}^{P}=\frac{1}{2}\,K^{S}
\end{equation}
In our treatment we shall consider the tetraquark state as two diquark
clusters , namely of quarks $(q_{1}q_{2})$ and $(q_{3}q_{4})$ of masses
$m_{12}$ and m$_{34}$ respectively, orbiting about each other with $L=1$, and
interacting chromo-magnetically only with the antiquark, denoted with index
5. The spin-orbit term arises, as in electrodynamics, from the interaction 
of the quarks with the coloured current. It is proportional to the
giro-chromomagnetic factor of the quarks in P wave as well to the product
of their colour matrices : more precisely, if the representation $3_{C}$ of
the $4q$ state is originated by the $\bar{3}_{C}\otimes \bar{3}_{C}$, or the 
$6_{C}\otimes \bar{3}_{C}$ representation of the two diquark pairs, the
coefficients will be in the ratio $2:5$. Since the colour and spatial 
degrees of freedom are independent, the interaction should be typically 
proportional to $\vec{L}\vec{.S_{ij}^{(\pm )}}$,
being
$\overrightarrow{S}_{ij}^{(\pm )}$ combinations of quarks 
spins,
$\overrightarrow{S}_{ij}^{(\pm )}=\overrightarrow{S_{i}}\pm \overrightarrow{S_{j}}$.
We include also the short range chomo-magnetic 
interaction between quarks in the same cluster and neglect the mass defects 
in the kinetic energy and the spin orbit Hamiltonian as a higher order 
effect. So $m^{(P)}=H_{0}+H_{CM}+H_{SO}$,
with
\bea
H_{0} &=& \sum_{i=1}^{4}m_{i}+m_{5}+(\frac{1}{m_{12}}\,+\,
\frac{1}{m_{34}})\,\frac{p^{2}}{2} \\
&\widetilde{-}&
\sum_{i=1}^{4}m_{i}+m_{5}+\left(\frac{1}{m_{1}+m_{2}}
+\frac{1}{m_{3}+m_{4}} \right)\,\frac{p^{2}}{2} \\
H_{CM} &=& K^{S} P \label{ondaP} \\
H_{SO} &=&
  a_{12} \, \vec{L}\vec{.S_{12}^{(+)}}
+ b_{12} \, \vec{L}\vec{.S_{12}^{(-)}} 
+ a_{34} \, \vec{L}\vec{.S_{34}^{(+)}}
+ b_{34} \, \vec{L}\vec{.S_{34}^{(-)}}
\eea
where $P$ is a $30 \times 30$ matrix, which splits into $15 \times 15$, $12 
\times 12$ and $3 \times 3$ matrices corresponding to spin $1/2$, $3/2$ 
and $5/2$, respectively. The matrix elements of $P$ between the states
in Appendix A may be computed through the use of Eq.'s(\ref{qq},\ref{qa}) 
and their values are reported in Appendix B.
As long for the spin orbit interaction,
$a_{12},~a_{34},~b_{12},~b_{34}$ are the appropriate kinematic factors
\bea
a_{ij} &=& -\frac{a}{4}\,\lambda_{b}^{(12)}\,\lambda_{b}^{(34)}\,
\frac{m_{q}^2}{2}\,\left(\frac{1}{m_{1}+m_{2}}+\frac{1}{m_{3}+m_{4}} \right)\,
\left(\frac{1}{m_{i}}+\frac{1}{m_j} \right) \\
b_{ij} &=& -\frac{a}{4}\,\lambda_{b}^{(12)}\,\lambda_{b}^{(34)}\,
\frac{m_{q}^2}{2}\, \left(\frac{1}{m_{1}+m_{2}}
+\frac{1}{m_{3}+m_{4}} \right)\, \left(\frac{1}{m_{i}}-\frac{1}{m_j} \right)
\eea
The total antisymmetry with respect to the 
exchange of the quarks, which are in the two S-wave pairs, and of the two 
pairs (which are in P-wave), fixes the $SU(3)_{F}$ quantum numbers of the 
pentaquarks. The $SU(3)_{F}$ breaking in the chromomagnetic interaction and in the 
spin-orbit term implies the mixing between different representations of 
$SU(3)_{F}$.
\newline
As it was the case of the negative parity states, the qualitative form of
the spectrum are shown in the symmetry limit: the lightest states will be 
the ones, where both the two diquarks transform as a $21_{CS}$ and the
pentaquark as the smallest possible representation of $SU(6)_{CS}$. From
Eq.'s(10,21,22,24) and the tensor products:
\bea
210_{CS}\otimes \,\bar{6}_{CS}\, &=&
\,1134_{CS}\,\oplus 70_{CS}\,\oplus 56_{CS} \\
\overline{15\,}_{CS}\otimes \,\bar{6}_{CS}\, &=&
\overline{70}_{CS}\,\oplus \,20_{CS}.
\eea
we deduce that the lightest $Y=2$ state has $J^P=1/2^+$ and $I=0$ and may be
identified with the $\Theta^+$. The corresponding state with a light $\bar{q}
$ can be identified with the $1/2^+$ $Y=1$ $I=1/2$ seen by BES~\cite{BES} at 
$1358$ MeV. At higher mass there are three $J^P=1/2^+$ $Y=2$ $I=1$ states,
one of which may be identified with the $P_{11}$ resonance seen in~\cite{AR1}
at $1720$ MeV; the $P_{13}$ $(1780)$ with the same internal quantum numbers,
seen in the same experiment, may be identified with the corresponding 
$3/2^+$ state. Finally the $\Xi^{--}$ state seen at CERN~\cite{NA} can be 
identified with his partner in the $27_F$. In the next section we shall 
fix the parameters to reproduce the values of the masses of the five 
states just quoted consistently with the ranges found for the $Y=2$ 
$D_{03}$ and $D_{15}$ previously mentioned states. Besides the positive 
parity states chosen to fix the parameters, we shall plot only the other 
"open door" states, which, according to Eq.'s.(10-14,21,22,33,34) will be the
flavour $J$ multiplets with positive parity
\bea
&&[\bar{10}+8,~8+1,~twice~(27+10+8+8+1),~35+10+\bar{10}+8~and~ 
27+10+8,~1/2 +3/2] \notag \\
&&[27+10+8+8+1,~35+10+\bar{10}+8~and~27+10+8,~1/2 +3/2 + 5/2] \notag
\eea
with "open door" decay into a pseudoscalar meson and a baryon of the octet
or the decuplet, respectively.

\section{Comparison with data}

To reproduce the masses of the four states $1/2^+$
mentioned at the end of the previous section, of the $(P_{13},Y=2,1780)$
and of the $D_{03}$ and $D_{15}$ resonances, we find the following values
for the parameters:
\begin{eqnarray}
\frac{K^S}{m_u^2} &=&74.5\,\mathrm{MeV} \\
a &=&42\,\mathrm{MeV} \\
<p^{2}>&=&(276\,\mathrm{MeV/c^2}) \\
m_{u} &=&346.8\,\mathrm{MeV} \\
m_{s} &=&480\,\mathrm{MeV}
\end{eqnarray}
With these values, we get for the states mentioned at the end of the previous
section the masses as a function of $J^P,Y,I$: 
\begin{eqnarray}
m(1/2^+,1,1/2)= 1356 MeV \\
m(1/2^+,2,0)= 1545 MeV \\
m(1/2^+,2,1)= 1732 MeV \\
m(3/2^+,2,1)= 1789 MeV \\
m(1/2^+,-1,3/2)= 1851 MeV \\
m(3/2^-,2,0)= 1858 MeV \\
m(5/2-,2,1)= 2088 MeV
\end{eqnarray}
With the same values of the parameters one may identify the resonances
seen in the photoproduction of $\Sigma K$ and
$\Lambda K$ resonances~\cite{Nic} with hidden strangeness 
partners of the ($1/2^+,Y=2,I=1,1734$) and the ($3/2^+,Y=2,I=1,1789$)
with masses $1862$ and $1908$, respectively.
The masses of all the negative parity and of the "open door" 
positive parity states corresponding to the paraneters just written are 
reported in Appendix~C, where we write a lower index $s$ for the 
states with hidden strangeness and the isospin for the values 
impossible for $qqq$ states. We put a * for the negative parity 
multiplets with " open door" decays and for the positive parity
states with "open door" decays into a $M B^*$ channel. \newline
Instead of grouping the multiplets according to their $SU(3)_{F}$
transformation properties, we group the different $I,Y$ multiplets in a
hybrid way following the same principle used for the vector
($\omega,\phi $) states and related to the fact that states with
or without hidden strangeness, which are components of the same 
$SU(3)_{F}$ multiplet, differ in mass by about $270$Mev: we combine $4q$ 
in $SU(3)_{F}$ multiplets either with $\bar{s}$ or with $\bar{u}$ and 
$\bar{d}$. \newline
For each $SU(3)_F$ reducible representation we report in Figures (1-8) 
the spectrum of at least one $J^P$ multiplet. More precisely:

1) For $1/2^+$ : 8 + 1, $\overline{10}$ + 8, 27 + 10 + 8 and 27 + 10 + 8 + 8 +1

2) For $3/2^+$ : 27 + 10 + 8 + 8 +1

3) For $3/2^-$ : $\overline{10}$ + 8, and 35 + 10

4) For $5/2^-$ : 27 + 10 + 8 
\\
\newline
There is evidence of two partners for the $Z_1~(1/2^+, 1734)$
(see Fig.4 in Appendix~D) and for one 
partner of the $Z_0~(1/2^+, 1545)$ (see Fig.2) and of the
$Z_1~(3/2^+, 1789)$ (see Fig.6).
The interpretation of the Roper resonance as a pentaquark was proposed
in~\cite{JW}. \\
As long as for $\Delta K$ states, it is not easy to find them in $KN$ 
reactions, since they have no common ''open door'' channel. The best way 
to find them should be in deep-inelastic reactions on strange partons, 
where the remaining $\bar{s}$ with the three valence quarks and another 
light quark may form a $Y=2$ state. \\
We conclude that the actual knowledge about the spectrum of the pentaquarks
is well consistent with the hypothesis that the chromo-magnetic interaction
plays the main role in describing their mass splittings. \\

\section{Conclusion}

The experimental situation is up to now controversial, as shown from the
disparition from $PDG$ of the $KN$ resonances in~\cite{AR1},\cite{AR2}
and~\cite{AR3}, the oscillating evidence for $\Theta ^{+}$ and the $\Xi ^{--}$ 
found at CERN.\newline
Indeed a recent report~\cite{Dan} is rather negative on the existence of 
the $\Theta^+$ and of the $\Xi^{--}$, as well as on the $C=-1$ pentaquark 
claimed in~\cite{Ak}. We show, anyway, that the spectrum of these states can be
described in the framework of QCD, as it happened for ordinary
hadrons~\cite{DGG}. \newline
Also the recent discoveries of the $(1/2^{+},Y=1,I=1/2)(1356)$ at
BES~\cite{BES} and of the $\Lambda K$ and $\Sigma K$ $P_{11}$ and $P_{13}$
resonances in photoproduction on proton~\cite{Nic} support the existence of 
pentaquarks with the spectrum well described in a constituent model, where 
the chromo-magnetic interaction and the spin orbit term, both expected within
QCD, play the main role.\newline
There is also an excess of $I=1/2, Y=1$ $N\pi $ positive parity states 
beyond the $56,L=2$ in the partial wave analysis performed at BES 
\cite{BES2} in $J/\psi \rightarrow p\bar{p}\pi ^{0}$, which may interpreted 
as pentaquarks, in particular the $P_{11}(1710)$ and the $P_{13}(1900)$. 
There are many states up to now escaped to ebservation, but the evidence 
shown here encourages further experimental research, for which this work 
can be a useful source of suggestions where to look for pentaquarks.

\section*{Acknowledgments}
One of us (MA) acknowledges G. D'Ambrosio and P. Santorelli for
very usefull discussions.

\newpage
\appendix

\section{States}

\begin{table}[h!]
{\tiny
\begin{tabular}{|l|c|l|}
\hline
State & $tetraq$ $Flavour$ \& $Spin$ & Wave function \\ \hline
\multicolumn{3}{|c|}{tates for $S=\frac{1}{2}$} \\ \hline
\, &  &  \\ 
$|$$1$, $S=\frac{1}{2}$ $\rangle$ & $3_F$ $S_t=0$ & $+\epsilon^{ABCD}~
\epsilon^{\alpha\beta\gamma}\,\epsilon^{ij}\,\epsilon^{hk}\,\cchi[\delta
,2]\, (+\ppsi[A,\alpha,h,u]\,\ppsi[B,\beta,i,u]\,\ppsi[C,\gamma,j,d]\,
\ppsi[D,\delta,k,s] $ \\ 
&  & $+\ppsi[A,\alpha,h,u]\,\ppsi[B,\beta,i,d]\,\ppsi[C,\gamma,j,s]\,
\ppsi[D,\delta,k,u] +\ppsi[A,\alpha,h,u]\,\ppsi[B,\beta,i,s]\,\ppsi[C,\gamma
,j,u]\,\ppsi[D,\delta,k,d])/96 $ \\ 
\, &  &  \\ \hline
\, &  &  \\ 
$|$$2$, $S=\frac{1}{2}$ $\rangle$ & $15_A$ $S_t=0$ & $+\epsilon^{ABCD}~
\epsilon^{\alpha\beta\gamma}\,\epsilon^{ij}\,\epsilon^{hk}\,\cchi[\delta,2]\,
$
\\ 
&  & $(+\ppsi[A,\alpha,h,s]\,\ppsi[B,\beta,i,u]\,\ppsi[C,\gamma,j,d]\,
\ppsi[D,\delta,k,u] +\ppsi[A,\alpha,h,u]\,\ppsi[B,\beta,i,u]\,\ppsi[C,\gamma
,j,d]\,\ppsi[D,\delta,k,s] $ \\ 
&  & $-\ppsi[A,\alpha,h,d]\,\ppsi[B,\beta,i,u]\,\ppsi[C,\gamma,j,s]\,
\ppsi[D,\delta,k,u] -\ppsi[A,\alpha,h,u]\,\ppsi[B,\beta,i,u]\,\ppsi[C,\gamma
,j,s]\,\ppsi[D,\delta,k,d] $ \\ 
&  & $+\ppsi[A,\alpha,h,u]\,\ppsi[B,\beta,i,s]\,\ppsi[C,\gamma,j,d]\,
\ppsi[D,\delta,k,u] -\ppsi[A,\alpha,h,u]\,\ppsi[B,\beta,i,d]\,\ppsi[C,\gamma
,j,s]\,\ppsi[D,\delta,k,u])/96/\sqrt{3} $ \\ 
\, &  &  \\ \hline
\, &  &  \\ 
$|$$3$, $S=\frac{1}{2}$ $\rangle$ & $15_S$ $S_t=0$ & $+\epsilon^{ABCD}~
\epsilon^{\alpha\beta\gamma}\,\epsilon^{ij}\,\epsilon^{hk}\,\cchi[\delta
,2]$\,\\
 &  & $(+\ppsi[A,\alpha,h,s]\,\ppsi[B,\beta,i,u]\,\ppsi[C,\gamma,j,d]\,
\ppsi[D,\delta,k,u] +\ppsi[A,\alpha,h,u]\,\ppsi[B,\beta,i,u]\,\ppsi[C,\gamma
,j,d]\,\ppsi[D,\delta,k,s] $ \\ 
&  & $+\ppsi[A,\alpha,h,d]\,\ppsi[B,\beta,i,u]\,\ppsi[C,\gamma,j,s]\,
\ppsi[D,\delta,k,u] +\ppsi[A,\alpha,h,u]\,\ppsi[B,\beta,i,u]\,\ppsi[C,\gamma
,j,s]\,\ppsi[D,\delta,k,d])/48/\sqrt{6} $ \\ 
\, &  &  \\ \hline
\, &  &  \\ 
$|$$4$, $S=\frac{1}{2}$ $\rangle$ & $3_F$ $S_t=1$ & $+\epsilon^{ABCD}~
\epsilon^{\alpha\beta\gamma}\,\epsilon^{ij}\,\cchi[\delta,h]\,$ \\ 
&  & $(+\ppsi[A,\alpha,h,u]\,\ppsi[B,\beta,i,u]\,\ppsi[C,\gamma,j,d]\,
\ppsi[D,\delta,1,s] +\ppsi[A,\alpha,h,u]\,\ppsi[B,\beta,i,d]\,\ppsi[C,\gamma
,j,s]\,\ppsi[D,\delta,1,u] $ \\ 
&  & $+\ppsi[A,\alpha,h,u]\,\ppsi[B,\beta,i,s]\,\ppsi[C,\gamma,j,u]\,
\ppsi[D,\delta,1,d] +\ppsi[A,\alpha,1,u]\,\ppsi[B,\beta,i,u]\,\ppsi[C,\gamma
,j,d]\,\ppsi[D,\delta,h,s] $ \\ 
&  & $+\ppsi[A,\alpha,1,u]\,\ppsi[B,\beta,i,d]\,\ppsi[C,\gamma,j,s]\,
\ppsi[D,\delta,h,u] +\ppsi[A,\alpha,1,u]\,\ppsi[B,\beta,i,s]\,\ppsi[C,\gamma
,j,u]\,\ppsi[D,\delta,h,d])/96/\sqrt{3} $ \\ 
\, &  &  \\ \hline
\, &  &  \\ 
$|$$5$, $S=\frac{1}{2}$ $\rangle$ & $15_A$ $S_t=1$ & $+\epsilon^{ABCD}~
\epsilon^{\alpha\beta\gamma}\,\epsilon^{ij}\,\cchi[\delta,h]\,$ \\ 
&  & ($+\ppsi[A,\alpha,h,s]\,\ppsi[B,\beta,i,u]\,\ppsi[C,\gamma,j,d]\,
\ppsi[D,\delta,1,u] -\ppsi[A,\alpha,h,d]\,\ppsi[B,\beta,i,u]\,\ppsi[C,\gamma
,j,s]\,\ppsi[D,\delta,1,u] $ \\ 
&  & $+\ppsi[A,\alpha,h,u]\,\ppsi[B,\beta,i,s]\,\ppsi[C,\gamma,j,d]\,
\ppsi[D,\delta,1,u] -\ppsi[A,\alpha,h,u]\,\ppsi[B,\beta,i,d]\,\ppsi[C,\gamma
,j,s]\,\ppsi[D,\delta,1,u] $ \\ 
&  & $+\ppsi[A,\alpha,h,u]\,\ppsi[B,\beta,i,u]\,\ppsi[C,\gamma,j,d]\,
\ppsi[D,\delta,1,s] -\ppsi[A,\alpha,h,u]\,\ppsi[B,\beta,i,u]\,\ppsi[C,\gamma
,j,s]\,\ppsi[D,\delta,1,d] $ \\ 
&  & $+\ppsi[A,\alpha,1,s]\,\ppsi[B,\beta,i,u]\,\ppsi[C,\gamma,j,d]\,
\ppsi[D,\delta,h,u] -\ppsi[A,\alpha,1,d]\,\ppsi[B,\beta,i,u]\,\ppsi[C,\gamma
,j,s]\,\ppsi[D,\delta,h,u] $ \\ 
&  & $+\ppsi[A,\alpha,1,u]\,\ppsi[B,\beta,i,s]\,\ppsi[C,\gamma,j,d]\,
\ppsi[D,\delta,h,u] -\ppsi[A,\alpha,1,u]\,\ppsi[B,\beta,i,d]\,\ppsi[C,\gamma
,j,s]\,\ppsi[D,\delta,h,u] $ \\ 
&  & $+\ppsi[A,\alpha,1,u]\,\ppsi[B,\beta,i,u]\,\ppsi[C,\gamma,j,d]\,
\ppsi[D,\delta,h,s] -\ppsi[A,\alpha,1,u]\,\ppsi[B,\beta,i,u]\,\ppsi[C,\gamma
,j,s]\,\ppsi[D,\delta,h,d])/96/\sqrt{3} $ \\ 
\, &  &  \\ \hline
\, &  &  \\ 
$|$$6$, $S=\frac{1}{2}$ $\rangle$ & $\bar{6}_F$ $S_t=1$ & $
+\epsilon^{ABCD}~\epsilon^{\alpha\beta\gamma}\,\epsilon^{ij}\,\cchi[\delta
,h]\,$ \\ 
&  & $(+\ppsi[A,\alpha,h,u]\,\ppsi[B,\beta,i,u]\,\ppsi[C,\gamma,j,d]\,
\ppsi[D,\delta,1,s] +\ppsi[A,\alpha,h,u]\,\ppsi[B,\beta,i,u]\,\ppsi[C,\gamma
,j,s]\,\ppsi[D,\delta,1,d] $ \\ 
&  & $-\ppsi[A,\alpha,h,d]\,\ppsi[B,\beta,i,u]\,\ppsi[C,\gamma,j,s]\,
\ppsi[D,\delta,1,u] -\ppsi[A,\alpha,h,s]\,\ppsi[B,\beta,i,u]\,\ppsi[C,\gamma
,j,d]\,\ppsi[D,\delta,1,u] $ \\ 
&  & $+\ppsi[A,\alpha,1,u]\,\ppsi[B,\beta,i,u]\,\ppsi[C,\gamma,j,d]\,
\ppsi[D,\delta,h,s] +\ppsi[A,\alpha,1,u]\,\ppsi[B,\beta,i,u]\,\ppsi[C,\gamma
,j,s]\,\ppsi[D,\delta,h,d] $ \\ 
&  & $-\ppsi[A,\alpha,1,d]\,\ppsi[B,\beta,i,u]\,\ppsi[C,\gamma,j,s]\,
\ppsi[D,\delta,h,u] -\ppsi[A,\alpha,1,s]\,\ppsi[B,\beta,i,u]\,\ppsi[C,\gamma
,j,d]\,\ppsi[D,\delta,h,u])/144/\sqrt{2} $ \\ 
\, &  &  \\ \hline
\, &  &  \\ 
$|$$7$, $S=\frac{1}{2}$ $\rangle$ & $15_S$ $S_t=1$ & $+\epsilon^{ABCD}~
\epsilon^{\alpha\beta\gamma}\,\epsilon^{ij}\,\cchi[\delta,h]\,$ \\ 
&  & $(+\ppsi[A,\alpha,h,s]\,\ppsi[B,\beta,i,u]\,\ppsi[C,\gamma,j,d]\,
\ppsi[D,\delta,1,u] +\ppsi[A,\alpha,h,d]\,\ppsi[B,\beta,i,u]\,\ppsi[C,\gamma
,j,s]\,\ppsi[D,\delta,1,u] $ \\ 
&  & $+\ppsi[A,\alpha,h,u]\,\ppsi[B,\beta,i,s]\,\ppsi[C,\gamma,j,d]\,
\ppsi[D,\delta,1,u] +\ppsi[A,\alpha,h,u]\,\ppsi[B,\beta,i,d]\,\ppsi[C,\gamma
,j,s]\,\ppsi[D,\delta,1,u] $ \\ 
&  & $+\ppsi[A,\alpha,h,u]\,\ppsi[B,\beta,i,u]\,\ppsi[C,\gamma,j,d]\,
\ppsi[D,\delta,1,s] +\ppsi[A,\alpha,h,u]\,\ppsi[B,\beta,i,u]\,\ppsi[C,\gamma
,j,s]\,\ppsi[D,\delta,1,d] $ \\ 
&  & $+\ppsi[A,\alpha,1,s]\,\ppsi[B,\beta,i,u]\,\ppsi[C,\gamma,j,d]\,
\ppsi[D,\delta,h,u] +\ppsi[A,\alpha,1,d]\,\ppsi[B,\beta,i,u]\,\ppsi[C,\gamma
,j,s]\,\ppsi[D,\delta,h,u] $ \\ 
&  & $+\ppsi[A,\alpha,1,u]\,\ppsi[B,\beta,i,s]\,\ppsi[C,\gamma,j,d]\,
\ppsi[D,\delta,h,u] +\ppsi[A,\alpha,1,u]\,\ppsi[B,\beta,i,d]\,\ppsi[C,\gamma
,j,s]\,\ppsi[D,\delta,h,u] $ \\ 
&  & $+\ppsi[A,\alpha,1,u]\,\ppsi[B,\beta,i,u]\,\ppsi[C,\gamma,j,d]\,
\ppsi[D,\delta,h,s] +\ppsi[A,\alpha,1,u]\,\ppsi[B,\beta,i,u]\,\ppsi[C,\gamma
,j,s]\,\ppsi[D,\delta,h,d])/48/\sqrt{6} $ \\ 
\, &  &  \\ \hline
\, &  &  \\ 
$|$$8$, $S=\frac{1}{2}$ $\rangle$ & $15^{\prime}_F$ $S_t=1$ & $
+\epsilon^{ABCD}~\epsilon^{\alpha\beta\gamma}\,\epsilon^{ij}\,\cchi[\delta
,h]\,$ \\ 
&  & $(+\ppsi[A,\alpha,h,d]\,\ppsi[B,\beta,1,s]\,\ppsi[C,\gamma,i,u]\,
\ppsi[D,\delta,j,u] +\ppsi[A,\alpha,h,s]\,\ppsi[B,\beta,1,d]\,\ppsi[C,\gamma
,i,u]\,\ppsi[D,\delta,j,u] $ \\ 
&  & $+\ppsi[A,\alpha,h,d]\,\ppsi[B,\beta,1,u]\,\ppsi[C,\gamma,i,s]\,
\ppsi[D,\delta,j,u] +\ppsi[A,\alpha,h,s]\,\ppsi[B,\beta,1,u]\,\ppsi[C,\gamma
,i,d]\,\ppsi[D,\delta,j,u] $ \\ 
&  & $+\ppsi[A,\alpha,h,d]\,\ppsi[B,\beta,1,u]\,\ppsi[C,\gamma,i,u]\,
\ppsi[D,\delta,j,s] +\ppsi[A,\alpha,h,s]\,\ppsi[B,\beta,1,u]\,\ppsi[C,\gamma
,i,u]\,\ppsi[D,\delta,j,d] $ \\ 
&  & $+\ppsi[A,\alpha,h,u]\,\ppsi[B,\beta,1,d]\,\ppsi[C,\gamma,i,s]\,
\ppsi[D,\delta,j,u] +\ppsi[A,\alpha,h,u]\,\ppsi[B,\beta,1,s]\,\ppsi[C,\gamma
,i,d]\,\ppsi[D,\delta,j,u] $ \\ 
&  & $+\ppsi[A,\alpha,h,u]\,\ppsi[B,\beta,1,d]\,\ppsi[C,\gamma,i,u]\,
\ppsi[D,\delta,j,s] +\ppsi[A,\alpha,h,u]\,\ppsi[B,\beta,1,s]\,\ppsi[C,\gamma
,i,u]\,\ppsi[D,\delta,j,d] $ \\ 
&  & $+\ppsi[A,\alpha,h,u]\,\ppsi[B,\beta,1,u]\,\ppsi[C,\gamma,i,d]\,
\ppsi[D,\delta,j,s] +\ppsi[A,\alpha,h,u]\,\ppsi[B,\beta,1,u]\,\ppsi[C,\gamma
,i,s]\,\ppsi[D,\delta,j,d] $ \\ 
&  & $+\ppsi[A,\alpha,1,d]\,\ppsi[B,\beta,h,s]\,\ppsi[C,\gamma,i,u]\,
\ppsi[D,\delta,j,u] +\ppsi[A,\alpha,1,s]\,\ppsi[B,\beta,h,d]\,\ppsi[C,\gamma
,i,u]\,\ppsi[D,\delta,j,u] $ \\ 
&  & $+\ppsi[A,\alpha,1,d]\,\ppsi[B,\beta,h,u]\,\ppsi[C,\gamma,i,s]\,
\ppsi[D,\delta,j,u] +\ppsi[A,\alpha,1,s]\,\ppsi[B,\beta,h,u]\,\ppsi[C,\gamma
,i,d]\,\ppsi[D,\delta,j,u] $ \\ 
&  & $+\ppsi[A,\alpha,1,d]\,\ppsi[B,\beta,h,u]\,\ppsi[C,\gamma,i,u]\,
\ppsi[D,\delta,j,s] +\ppsi[A,\alpha,1,s]\,\ppsi[B,\beta,h,u]\,\ppsi[C,\gamma
,i,u]\,\ppsi[D,\delta,j,d] $ \\ 
&  & $+\ppsi[A,\alpha,1,u]\,\ppsi[B,\beta,h,d]\,\ppsi[C,\gamma,i,s]\,
\ppsi[D,\delta,j,u] +\ppsi[A,\alpha,1,u]\,\ppsi[B,\beta,h,s]\,\ppsi[C,\gamma
,i,d]\,\ppsi[D,\delta,j,u] $ \\ 
&  & $+\ppsi[A,\alpha,1,u]\,\ppsi[B,\beta,h,d]\,\ppsi[C,\gamma,i,u]\,
\ppsi[D,\delta,j,s] +\ppsi[A,\alpha,1,u]\,\ppsi[B,\beta,h,s]\,\ppsi[C,\gamma
,i,u]\,\ppsi[D,\delta,j,d] $ \\ 
&  & $+\ppsi[A,\alpha,1,u]\,\ppsi[B,\beta,h,u]\,\ppsi[C,\gamma,i,d]\,
\ppsi[D,\delta,j,s] +\ppsi[A,\alpha,1,u]\,\ppsi[B,\beta,h,u]\,\ppsi[C,\gamma
,i,s]\,\ppsi[D,\delta,j,d])/576 $ \\ 
\, &  &  \\ \hline
\end{tabular}
}
\caption{Pentaquark states with $J^P=\frac{1}{2}^-$}
\label{t1S}
\end{table}
\begin{table}[h!]
{\scriptsize 
\begin{tabular}{|l|l|l|}
\hline
State & $t$ $F$ \& $S$ & Wave function \\ \hline
\multicolumn{3}{|c|}{tates for $S=\frac{3}{2}$} \\ \hline
\, &  &  \\ 
$|$$1$, $S=\frac{3}{2}$ $\rangle$ & $3_F$ $S_t=1$ & $+\epsilon^{ABCD}~
\epsilon^{\alpha\beta\gamma}\,\epsilon^{ij}\,\cchi[\delta,2]\, (+
\ppsi[A,\alpha,1,u]\,\ppsi[B,\beta,i,u]\,\ppsi[C,\gamma,j,d]\,\ppsi[D,\delta
,1,s] $ \\ 
&  & $+\ppsi[A,\alpha,1,u]\,\ppsi[B,\beta,i,d]\,\ppsi[C,\gamma,j,s]\,
\ppsi[D,\delta,1,u] +\ppsi[A,\alpha,1,u]\,\ppsi[B,\beta,i,s]\,\ppsi[C,\gamma
,j,u]\,\ppsi[D,\delta,1,d])/48/\sqrt{2} $ \\ 
\, &  &  \\ \hline
\, &  &  \\ 
$|$$2$, $S=\frac{3}{2}$ $\rangle$ & $15_A$ $S_t=1$ & $+\epsilon^{ABCD}~
\epsilon^{\alpha\beta\gamma}\,\epsilon^{ij}\,\cchi[\delta,2] $\\ 
 &  & $( +\ppsi[A,\alpha,1,s]\,\ppsi[B,\beta,i,u]\,\ppsi[C,\gamma,j,d]\,
\ppsi[D,\delta,1,u] -\ppsi[A,\alpha,1,d]\,\ppsi[B,\beta,i,u]\,\ppsi[C,\gamma
,j,s]\,\ppsi[D,\delta,1,u] $ \\ 
&  & $+\ppsi[A,\alpha,1,u]\,\ppsi[B,\beta,i,s]\,\ppsi[C,\gamma,j,d]\,
\ppsi[D,\delta,1,u] -\ppsi[A,\alpha,1,u]\,\ppsi[B,\beta,i,d]\,\ppsi[C,\gamma
,j,s]\,\ppsi[D,\delta,1,u] $ \\ 
&  & $+\ppsi[A,\alpha,1,u]\,\ppsi[B,\beta,i,u]\,\ppsi[C,\gamma,j,d]\,
\ppsi[D,\delta,1,s] -\ppsi[A,\alpha,1,u]\,\ppsi[B,\beta,i,u]\,\ppsi[C,\gamma
,j,s]\,\ppsi[D,\delta,1,d])/48/\sqrt{2} $ \\ 
\, &  &  \\ \hline
\, &  &  \\ 
$|$$3$, $S=\frac{3}{2}$ $\rangle$ & $15_A$ $S_t=2$ & $+\epsilon^{ABCD}~
\epsilon^{\alpha\beta\gamma}\,\cchi[\delta,h]\,$ \\ 
&  & $(+\ppsi[A,\alpha,h,u]\,\ppsi[B,\beta,1,u]\,\ppsi[C,\gamma,1,d]\,
\ppsi[D,\delta,1,s] +\ppsi[A,\alpha,1,u]\,\ppsi[B,\beta,h,u]\,\ppsi[C,\gamma
,1,d]\,\ppsi[D,\delta,1,s] $ \\ 
&  & $+\ppsi[A,\alpha,1,u]\,\ppsi[B,\beta,1,u]\,\ppsi[C,\gamma,h,d]\,
\ppsi[D,\delta,1,s] +\ppsi[A,\alpha,1,u]\,\ppsi[B,\beta,1,u]\,\ppsi[C,\gamma
,1,d]\,\ppsi[D,\delta,h,s] $ \\ 
&  & $-\ppsi[A,\alpha,h,u]\,\ppsi[B,\beta,1,u]\,\ppsi[C,\gamma,1,s]\,
\ppsi[D,\delta,1,d] -\ppsi[A,\alpha,1,u]\,\ppsi[B,\beta,h,u]\,\ppsi[C,\gamma
,1,s]\,\ppsi[D,\delta,1,d] $ \\ 
&  & $-\ppsi[A,\alpha,1,u]\,\ppsi[B,\beta,1,u]\,\ppsi[C,\gamma,h,s]\,
\ppsi[D,\delta,1,d] -\ppsi[A,\alpha,1,u]\,\ppsi[B,\beta,1,u]\,\ppsi[C,\gamma
,1,s]\,\ppsi[D,\delta,h,d])/96/\sqrt{5} $ \\ 
\, &  &  \\ \hline
\, &  &  \\ 
$|$$4$, $S=\frac{3}{2}$ $\rangle$ & $\bar{6}_F$ $S_t=1$ & $
+\epsilon^{ABCD}~\epsilon^{\alpha\beta\gamma}\,\epsilon^{ij}\,\cchi[\delta
,2]$ \\ 
&  & $(+\ppsi[A,\alpha,1,u]\,\ppsi[B,\beta,i,u]\,\ppsi[C,\gamma,j,d]\,
\ppsi[D,\delta,1,s] +\ppsi[A,\alpha,1,u]\,\ppsi[B,\beta,i,u]\,\ppsi[C,\gamma
,j,s]\,\ppsi[D,\delta,1,d] $ \\ 
&  & $-\ppsi[A,\alpha,1,d]\,\ppsi[B,\beta,i,u]\,\ppsi[C,\gamma,j,s]\,
\ppsi[D,\delta,1,u] -\ppsi[A,\alpha,1,s]\,\ppsi[B,\beta,i,u]\,\ppsi[C,\gamma
,j,d]\,\ppsi[D,\delta,1,u])/48/\sqrt{3} $ \\ 
\, &  &  \\ \hline
\, &  &  \\ 
$|$$5$, $S=\frac{3}{2}$ $\rangle$ & $15_S$ $S_t=1$ & $+\epsilon^{ABCD}~
\epsilon^{\alpha\beta\gamma}\,\epsilon^{ij}\,\cchi[\delta,2]$\\ 
&  & $(+\ppsi[A,\alpha,1,s]\,\ppsi[B,\beta,i,u]\,\ppsi[C,\gamma,j,d]\,
\ppsi[D,\delta,1,u] +\ppsi[A,\alpha,1,d]\,\ppsi[B,\beta,i,u]\,\ppsi[C,\gamma
,j,s]\,\ppsi[D,\delta,1,u] $ \\ 
&  & $+\ppsi[A,\alpha,1,u]\,\ppsi[B,\beta,i,s]\,\ppsi[C,\gamma,j,d]\,
\ppsi[D,\delta,1,u] +\ppsi[A,\alpha,1,u]\,\ppsi[B,\beta,i,d]\,\ppsi[C,\gamma
,j,s]\,\ppsi[D,\delta,1,u] $ \\ 
&  & $+\ppsi[A,\alpha,1,u]\,\ppsi[B,\beta,i,u]\,\ppsi[C,\gamma,j,d]\,
\ppsi[D,\delta,1,s] +\ppsi[A,\alpha,1,u]\,\ppsi[B,\beta,i,u]\,\ppsi[C,\gamma
,j,s]\,\ppsi[D,\delta,1,d])/48 $ \\ 
\, &  &  \\ \hline
\, &  &  \\ 
$|$$6$, $S=\frac{3}{2}$ $\rangle$ & $15_S$ $S_t=2$ & $+\epsilon^{ABCD}~
\epsilon^{\alpha\beta\gamma}\,\cchi[\delta,h]$\\ 
&  & $(+\ppsi[A,\alpha,h,u]\,\ppsi[B,\beta,1,u]\,\ppsi[C,\gamma,1,d]\,
\ppsi[D,\delta,1,s] +\ppsi[A,\alpha,1,u]\,\ppsi[B,\beta,h,u]\,\ppsi[C,\gamma
,1,d]\,\ppsi[D,\delta,1,s] $ \\ 
&  & $+\ppsi[A,\alpha,1,u]\,\ppsi[B,\beta,1,u]\,\ppsi[C,\gamma,h,d]\,
\ppsi[D,\delta,1,s] +\ppsi[A,\alpha,1,u]\,\ppsi[B,\beta,1,u]\,\ppsi[C,\gamma
,1,d]\,\ppsi[D,\delta,h,s] $ \\ 
&  & $+\ppsi[A,\alpha,h,u]\,\ppsi[B,\beta,1,u]\,\ppsi[C,\gamma,1,s]\,
\ppsi[D,\delta,1,d] +\ppsi[A,\alpha,1,u]\,\ppsi[B,\beta,h,u]\,\ppsi[C,\gamma
,1,s]\,\ppsi[D,\delta,1,d] $ \\ 
&  & $+\ppsi[A,\alpha,1,u]\,\ppsi[B,\beta,1,u]\,\ppsi[C,\gamma,h,s]\,
\ppsi[D,\delta,1,d] +\ppsi[A,\alpha,1,u]\,\ppsi[B,\beta,1,u]\,\ppsi[C,\gamma
,1,s]\,\ppsi[D,\delta,h,d])/48/\sqrt{10} $ \\ 
\, &  &  \\ \hline
\, &  &  \\ 
$|$$7$, $S=\frac{3}{2}$ $\rangle$ & $15^{\prime}_F$ $S_t=1$ & $
+\epsilon^{ABCD}~\epsilon^{\alpha\beta\gamma}\,\epsilon^{ij}\,\cchi[\delta
,2]$ \\ 
&  & $(+\ppsi[A,\alpha,1,d]\,\ppsi[B,\beta,1,s]\,\ppsi[C,\gamma,i,u]\,
\ppsi[D,\delta,j,u] +\ppsi[A,\alpha,1,s]\,\ppsi[B,\beta,1,d]\,\ppsi[C,\gamma
,i,u]\,\ppsi[D,\delta,j,u] $ \\ 
&  & $+\ppsi[A,\alpha,1,d]\,\ppsi[B,\beta,1,u]\,\ppsi[C,\gamma,i,s]\,
\ppsi[D,\delta,j,u] +\ppsi[A,\alpha,1,s]\,\ppsi[B,\beta,1,u]\,\ppsi[C,\gamma
,i,d]\,\ppsi[D,\delta,j,u] $ \\ 
&  & $+\ppsi[A,\alpha,1,d]\,\ppsi[B,\beta,1,u]\,\ppsi[C,\gamma,i,u]\,
\ppsi[D,\delta,j,s] +\ppsi[A,\alpha,1,s]\,\ppsi[B,\beta,1,u]\,\ppsi[C,\gamma
,i,u]\,\ppsi[D,\delta,j,d] $ \\ 
&  & $+\ppsi[A,\alpha,1,u]\,\ppsi[B,\beta,1,d]\,\ppsi[C,\gamma,i,s]\,
\ppsi[D,\delta,j,u] +\ppsi[A,\alpha,1,u]\,\ppsi[B,\beta,1,s]\,\ppsi[C,\gamma
,i,d]\,\ppsi[D,\delta,j,u] $ \\ 
&  & $+\ppsi[A,\alpha,1,u]\,\ppsi[B,\beta,1,d]\,\ppsi[C,\gamma,i,u]\,
\ppsi[D,\delta,j,s] +\ppsi[A,\alpha,1,u]\,\ppsi[B,\beta,1,s]\,\ppsi[C,\gamma
,i,u]\,\ppsi[D,\delta,j,d] $ \\ 
&  & $+\ppsi[A,\alpha,1,u]\,\ppsi[B,\beta,1,u]\,\ppsi[C,\gamma,i,d]\,
\ppsi[D,\delta,j,s] +\ppsi[A,\alpha,1,u]\,\ppsi[B,\beta,1,u]\,\ppsi[C,\gamma
,i,s]\,\ppsi[D,\delta,j,d])/96/\sqrt{6} $ \\ 
\, &  &  \\ \hline
\end{tabular}
}
\caption{Pentaquark states with $J^P=\frac{3}{2}^-$}
\label{t2S}
\end{table}
\begin{table}[h]
{\scriptsize 
\begin{tabular}{|l|l|l|}
\hline
State & $t$ $F$ \& $S$ & Wave function \\ \hline
\multicolumn{3}{|c|}{tates for $S=\frac{5}{2}$} \\ \hline
\, &  &  \\ 
$|$$1$, $S=\frac{5}{2}$ $\rangle$ & $15_S$ $S_t=2$ & $+\epsilon^{ABCD}~
\epsilon^{\alpha\beta\gamma}\,\cchi[\delta,2] $\\ 
&  & $(+\ppsi[A,\alpha,1,u]\,\ppsi[B,\beta,1,u]\,\ppsi[C,\gamma,1,d]\,
\ppsi[D,\delta,1,s] +\ppsi[A,\alpha,1,u]\,\ppsi[B,\beta,1,u]\,\ppsi[C,\gamma
,1,s]\,\ppsi[D,\delta,1,d])/24/\sqrt{2} $ \\ 
\, &  &  \\ \hline
\, &  &  \\ 
$|$$2$, $S=\frac{5}{2}$ $\rangle$ & $15_A$ $S_t=2$ & $+\epsilon^{ABCD}~
\epsilon^{\alpha\beta\gamma}\,\cchi[\delta,2] $\\ 
&  & $(+\ppsi[A,\alpha,1,u]\,\ppsi[B,\beta,1,u]\,\ppsi[C,\gamma,1,d]\,
\ppsi[D,\delta,1,s] -\ppsi[A,\alpha,1,u]\,\ppsi[B,\beta,1,u]\,\ppsi[C,\gamma
,1,s]\,\ppsi[D,\delta,1,d])/48 $ \\ 
\, &  &  \\ \hline
\end{tabular}
}
\caption{Pentaquark states with $J^P=\frac{5}{2}^-$}
\label{t3S}
\end{table}
\begin{center}
\begin{table}
{\scriptsize 
\begin{tabular}{|c|l|}
\hline
$\left[(d_C, d_S)_{qq} \; (d_C, d_S)_{qq} \right]_S $ & States \\ 
\hline
$\left[(6,3)\; (\bar{3},1) \right]_{S_q =1} $ & $|\, 1> =
\varepsilon^{\alpha\beta\gamma}\, \varepsilon^{ij} (\psi_{\alpha 1}
\psi_{\eta p} + \psi_{\eta 1} \psi_{\alpha p} + \psi_{\alpha p} \psi_{\eta
1} + \psi_{\eta p} \psi_{\alpha 1} ) \, \phi_{\beta i}\, \phi_{\gamma j}\,
\chi^{\eta p}$ \\ 
$\left[(\bar{3},1) \; (6,3) \right]_{S_q =1} $ & $|\, 2> =
\varepsilon^{\alpha\beta\gamma}\, \varepsilon^{ij} 
\psi_{\beta i}\, \psi_{\gamma j}\,(\phi_{\alpha 1}
\phi_{\eta p} + \phi_{\eta 1} \phi_{\alpha p} + \phi_{\alpha p} \phi_{\eta
1} + \phi_{\eta p} \phi_{\alpha 1} ) \, 
\chi^{\eta p} $ \\ 
$\left[ (\bar{3},1) \; (\bar{3},1) \right]_{S_q =0}$ & $|\, 3> =
\varepsilon^{\alpha\beta\gamma}\, \varepsilon^{\delta\epsilon\rho}\,
\varepsilon^{ij}\, \varepsilon^{hk}\,\psi_{\beta i}\, \psi_{\gamma j} \,
\phi_{\epsilon h}\, \phi_{\rho k}\, \varepsilon_{\alpha \delta \eta} \,
\chi^{\eta 2} $ \\ 
$\left[ (6,1)\; (\bar{3},1)\right]_{S_q =0} $ & $|\,4> =
\varepsilon^{\alpha\beta\gamma}\, \varepsilon^{ij} \, \varepsilon^{hk}
\,(\psi_{\alpha h} \psi_{\eta k} + \psi_{\eta k} \psi_{\alpha h}) \,
\phi_{\beta i}\, \phi_{\gamma j}\, \chi^{\eta 2} $ \\ 
$\left[(\bar{3},3) \; (\bar{3},1) \right]_{S_q =1}$ & $|\, 5> =
\varepsilon^{\alpha\beta\gamma}\, \varepsilon^{ij} (\psi_{\alpha 1}
\psi_{\eta p} - \psi_{\eta 1} \psi_{\alpha p} + \psi_{\alpha p} \psi_{\eta
1} - \psi_{\eta p} \psi_{\alpha 1} ) \, \phi_{\beta i}\, \phi_{\gamma j} \,
\chi^{\eta p}$ \\ 
$\left[ (\bar{3},3)\; (6,3)\right]_{S_q =1} $ & $|\, 6> =
\varepsilon^{\alpha\beta\gamma}\, \varepsilon^{ij} \left[ (\psi_{\beta 1}
\psi_{\gamma i} + \psi_{\beta i} \psi_{\gamma 1} )\, (\phi_{\alpha p}
\phi_{\eta j} + \phi_{\eta p} \phi_{\alpha j} + \phi_{\alpha j} \phi_{\eta
p} + \phi_{\eta j} \phi_{\alpha p} )+ \right. $ \\ 
& $\left. \qquad + (\psi_{\beta p} \psi_{\gamma i} + \psi_{\beta i}
\psi_{\gamma p} )\, (\phi_{\alpha 1} \phi_{\eta j} + \phi_{\eta 1}
\phi_{\alpha j} + \phi_{\alpha j} \phi_{\eta 1} + \phi_{\eta j} \phi_{\alpha
1} ) \right] \, \chi^{\eta p} $ \\ 
$\left[ (\bar{3},3)\; (6,3)\right]_{S_q =0} $ & $|\, 7> =
\varepsilon^{\alpha\beta\gamma}\, \varepsilon^{hk}\, \varepsilon^{ij} 
\left[ (\psi_{\beta h}\psi_{\gamma i} + \psi_{\beta i} \psi_{\gamma h} )
\, (\phi_{\alpha k}
\phi_{\eta j} + \phi_{\eta k} \phi_{\alpha j} + \phi_{\alpha j} \phi_{\eta
k} + \phi_{\eta j} \phi_{\alpha k} ) \right] \chi^{\eta~2} $ \\ 
$\left[ (6,3)\; (\bar{3},3)\right]_{S_q =1} $ & $|\, 8> =
\varepsilon^{\alpha\beta\gamma}\, \varepsilon^{ij} \, \left[ (\psi_{\alpha
p} \psi_{\eta j} + \psi_{\eta p} \psi_{\alpha j} + \psi_{\alpha j}
\psi_{\eta p} + \psi_{\eta j} \psi_{\alpha p} ) \, (\phi_{\beta 1}
\phi_{\gamma i} + \phi_{\beta i} \phi_{\gamma 1} )+ \right.$ \\ 
& $\left. \qquad + (\psi_{\alpha 1} \psi_{\eta j} + \psi_{\eta 1}
\psi_{\alpha j} + \psi_{\alpha j} \psi_{\eta 1} + \psi_{\eta j} \psi_{\alpha
1} ) \, (\phi_{\beta p} \phi_{\gamma i} + \phi_{\beta i} \phi_{\gamma p} ) 
\right] \, \chi^{\eta p} $ \\ 
$\left[ (6,3)\; (\bar{3},3)\right]_{S_q =0} $ & $|\, 9> =
\varepsilon^{\alpha\beta\gamma}\, \varepsilon^{hk} \, \varepsilon^{ij} \, 
\left[ (\psi_{\alpha h} \psi_{\eta i} + \psi_{\eta h} \psi_{\alpha i} + 
\psi_{\alpha i} \psi_{\eta h} + \psi_{\eta i} \psi_{\alpha h} ) \, 
(\phi_{\beta k} \phi_{\gamma j} + \phi_{\beta j} \phi_{\gamma k} )
\right]  \chi^{\eta p} $ \\ 
$\left[ (\bar{3},1)\; (6,1) \right]_{S_q =0} $ & $|\,10> =
\varepsilon^{\alpha\beta\gamma}\, \varepsilon^{ij} \,
\varepsilon^{hk}\,\psi_{\beta i}\, \psi_{\gamma j}\,(\phi_{\alpha h}
\phi_{\eta k} + \phi_{\eta h} \phi_{\alpha k}) \, \chi^{\eta 2}$ \\ 
$\left[ (\bar{3},1) \; (\bar{3},3)\right]_{S_q =1}$ & $|\, 11> =
\varepsilon^{\alpha\beta\gamma}\, \varepsilon^{ij} \, \psi_{\beta i}\,
\psi_{\gamma j}\,(\phi_{\alpha 1} \phi_{\eta p} - \phi_{\eta 1} \phi_{\alpha
p} + \phi_{\alpha p} \phi_{\eta 1} - \phi_{\eta p} \phi_{\alpha 1} ) \,
\chi^{\eta p} $ \\ 
$\left[ (6,1) \; (\bar{3},3)\right]_{S_q =1}$ & $|\, 12> =
\varepsilon^{\alpha\beta\gamma}\, \varepsilon^{ij} \, (\psi_{\alpha i}\,
\psi_{\eta j} + \psi_{\eta i} \psi_{\alpha j} ) \, (\phi_{\beta 1}\,
\phi_{\gamma p} + \phi_{\beta p} \phi_{\gamma 1} ) \, \chi^{\eta p} $ \\ 
$\left[ (\bar{3},3) \; (\bar{3},3)\right]_{S_q =0}$ & $|\, 13>
=\varepsilon^{\alpha\beta\gamma}\, \varepsilon^{hk} \, \varepsilon^{ij} \, 
\left[ \,(\psi_{\beta h} \psi_{\gamma i} + \psi_{\beta i} \psi_{\gamma h} ) 
\, (\phi_{\eta k} \, \phi_{\alpha j} - \phi_{\alpha k}\,\phi_{\eta j} +
\phi_{\eta j} \, \phi_{\alpha k} - \phi_{\alpha j} \,\phi_{\eta k}) +
\right]  \chi^{\eta 2} $ \\ 
$\left[ (\bar{3},3) \; (\bar{3},3)\right]_{S_q =1}$ & $|\, 14>
=\varepsilon^{\alpha\beta\gamma}\, \varepsilon^{ij} \, \left[ \,
(\psi_{\beta 1} \psi_{\gamma i} + \psi_{\beta i} \psi_{\gamma 1} ) \, (
\phi_{\eta p} \, \phi_{\alpha j} - \phi_{\alpha p}\,\phi_{\eta j} +
\phi_{\eta j} \, \phi_{\alpha p} - \phi_{\alpha j} \,\phi_{\eta p})+ \right.$
\\ 
& $\left. \qquad + (\psi_{\beta p} \psi_{\gamma i} + \psi_{\beta i}
\psi_{\gamma p} ) \, ( \phi_{\eta 1} \, \phi_{\alpha j} - \phi_{\alpha
1}\,\phi_{\eta j} + \phi_{\eta j} \, \phi_{\alpha 1} - \phi_{\alpha j}
\,\phi_{\eta 1}) \right] \, \chi^{\eta p} $ \\ 
$\left[ (\bar{3},3) \; (6,1) \right]_{S_q =1}$ & $|\, 15> =
\varepsilon^{\alpha\beta\gamma}\, 
\varepsilon^{ij} 
\, (\psi_{\beta 1}\,
\psi_{\gamma p} + \psi_{\beta p} \psi_{\gamma 1} )\, (\phi_{\eta i}
\phi_{\alpha j}+\phi_{\alpha i}\, \phi_{\eta j} )\, \chi^{\eta p} $ \\ 
$\left[ (\bar{3},3) \; (\bar{3},3)\right]_{S_q =2}$ & $|\, 16>
=\varepsilon^{\alpha\beta\gamma}\, [ \psi_{\beta 1} \psi_{\gamma 1} \, 
\left
( \phi_{\eta 1} \, \phi_{\alpha p} - \phi_{\alpha 1}\,\phi_{\eta p} +
\phi_{\eta p} \, \phi_{\alpha 1} - \phi_{\alpha p} \,\phi_{\eta 1})+ \right. 
$ \\ 
& $\left.\qquad + (\psi_{\beta 1} \psi_{\gamma p} + \psi_{\beta p}
\psi_{\gamma 1} ) \, ( \phi_{\eta 1} \, \phi_{\alpha 1} - \phi_{\alpha
1}\,\phi_{\eta 1} ) \right] \, \chi^{\eta p} $ \\ 
$\left[ (\bar{3},3) \; (\bar{3},3)\right]_{S_q =1}$ & $|\, 17>
=\varepsilon^{\alpha\beta\gamma}\, \varepsilon^{ij} 
( \psi_{\beta 1} \psi_{\gamma i} + \, \psi_{\beta i} \psi_{\gamma 1} )
\, \left( \phi_{\eta 1} \, \phi_{\alpha j} - \phi_{\alpha j}\,\phi_{\eta p} +
\phi_{\eta p} \, \phi_{\alpha j} - \phi_{\alpha p} \,\phi_{\eta j}) 
\right) \, \chi^{\eta 2} $ \\ 
$\left[ (\bar{3},3) \; (6,1) \right]_{S_q =1}$ & $|\, 18> =
\varepsilon^{\alpha\beta\gamma}\, \varepsilon^{ij} \,\psi_{\beta 1}\,
\psi_{\gamma 1} 
\, (\phi_{\alpha i}\,
\phi_{\eta j} + \phi_{\eta i} \phi_{\alpha j} ) \, \chi^{\eta 2} $ \\ 
$\left[ (6,1) \; (\bar{3},3) \right]_{S_q =1}$ & $|\, 19> =
\varepsilon^{\alpha\beta\gamma}\, \varepsilon^{ij} \, (\psi_{\eta i}
\psi_{\alpha j}+\psi_{\alpha i}\, \psi_{\eta j} )\,\phi_{\beta 1}\,
\phi_{\gamma 1} \, \chi^{\eta 2} $ \\ 
$\left[(\bar{3},1) \; (\bar{3},3) \right]_{S_q =1}$ & $|\,20> =
\varepsilon^{\alpha\beta\gamma}\, 
\varepsilon^{ij} 
\, \psi_{\beta i}\,
\psi_{\gamma j}\, (\phi_{\alpha 1} \phi_{\eta 1} - \phi_{\eta 1}
\phi_{\alpha 1} ) \, \chi^{\eta 2} $ \\ 
$\left[ (6,3)\; (\bar{3},3)\right]_{S_q =2} $ & $|\, 21> =
\varepsilon^{\alpha\beta\gamma}\, \, \left[ (\psi_{\alpha 1} \psi_{\eta 1} +
\psi_{\eta 1} \psi_{\alpha 1} ) \, (\phi_{\beta 1} \phi_{\gamma p} +
\phi_{\beta p} \phi_{\gamma 1} )+ \right. $ \\ 
& $\left. \qquad + (\psi_{\alpha 1} \psi_{\eta p} + \psi_{\eta 1}
\psi_{\alpha p} + \psi_{\alpha p} \psi_{\eta 1} + \psi_{\eta p} \psi_{\alpha
1} ) \, \phi_{\beta 1}\, \phi_{\gamma 1} \right] \, \chi^{\eta p} $ \\ 
$\left[ (6,3)\; (\bar{3},3)\right]_{S_q =1} $ & $|\, 22> =
\varepsilon^{\alpha\beta\gamma} 
\varepsilon^{ij} $ \\
& $\left. \qquad  (\psi_{\alpha 1} \psi_{\eta i} + \psi_{\eta 1}
\psi_{\alpha i} + \psi_{\alpha i} \psi_{\eta 1} + \psi_{\eta i} 
\psi_{\alpha1} ) \, 
( \phi_{\beta 1}\, \phi_{\gamma j} + \phi_{\beta j}\, \phi_{\gamma 1} )
\right. \, \chi^{\eta j} $ \\ 
$\left[ (\bar{3},3)\; (\bar{3},1)\right]_{S_q =1} $ & $|\, 23> =
\varepsilon^{\alpha\beta\gamma}\, \varepsilon^{ij} 
\psi_{\beta 1} \psi_{\gamma 1}\, (\phi_{\alpha i} \phi_{\eta j} - 
\phi_{\eta i} \phi_{\alpha j} )  \, \chi^{\eta 2} $ \\ 
$\left[ (\bar{3},3)\; (6,3)\right]_{S_q =2} $ & $|\, 24> =
\varepsilon^{\alpha\beta\gamma}\, \left[ \psi_{\beta 1} \psi_{\gamma 1}\,
(\phi_{\eta 1} \phi_{\alpha p} + \phi_{\alpha 1} \phi_{\eta p} + \phi_{\eta
p} \phi_{\alpha 1} + \phi_{\alpha p} \phi_{\eta 1} )+ \right.$ \\ 
& $\left. \qquad + (\psi_{\beta 1} \psi_{\gamma p} + \psi_{\beta p}
\psi_{\gamma 1} )\, (\phi_{\eta 1} \phi_{\alpha 1} + \phi_{\alpha 1}
\phi_{\eta 1} ) \right] \, \chi^{\eta p} $ \\ 
$\left[ (\bar{3},3)\; (6,3)\right]_{S_q =1} $ & $|\, 25> =
\varepsilon^{\alpha\beta\gamma}\varepsilon^{ij} 
\, \left[ ( \psi_{\beta 1} \psi_{\gamma i}\, +  \psi_{\beta i} 
\psi_{\gamma 1} ) \,(\phi_{\eta 1} \phi_{\alpha j} + 
\phi_{\alpha 1} \phi_{\eta j} +  \phi_{\eta j} \phi_{\alpha 1} 
+ \phi_{\alpha j} \phi_{\eta 1} )  \right. \chi^{\eta p} $ \\ 
$\left[ (6,3) \; (\bar{3},1)\right]_{S_q =1}$ & $|\, 26> =
\varepsilon^{\alpha\beta\gamma}\, \varepsilon^{ij} \, (\psi_{\alpha 1}\,
\psi_{\eta 1} + \psi_{\eta 1} \psi_{\alpha 1} ) \, \phi_{\beta i}\,
\phi_{\gamma j} \, \chi^{\eta p} $ \\ 
$\left[ (\bar{3},1) \; (6,3) \right]_{S_q =1}$ & $|\, 27> =
\varepsilon^{\alpha\beta\gamma}\, \varepsilon^{ij} \, \psi_{\beta i}\,
\psi_{\gamma j}  \, (\phi_{\eta 1}\,
\phi_{\alpha 1} + \phi_{\alpha 1} \phi_{\eta 1} ) \, \chi^{\eta 2} $ \\ 
$\left[ (\bar{3},3) \; (\bar{3},3)\right]_{S_q =2}$ & $|\, 28>
=\varepsilon^{\alpha\beta\gamma}\, \psi_{\beta 1} \psi_{\gamma 1} \, (
\phi_{\eta 1} \, \phi_{\alpha 1}- \phi_{\alpha 1} \phi_{\eta 1} ) \,
\chi^{\eta 2} $ \\ 
$\left[ (\bar{3},3) \; (6,3) \right]_{S_q =2}$ & $|\, 29>
=\varepsilon^{\alpha\beta\gamma}\, \psi_{\beta 1} \psi_{\gamma 1} \, (
\phi_{\eta 1} \, \phi_{\alpha 1}+ \phi_{\alpha 1} \phi_{\eta 1} ) \,
\chi^{\eta 2} $ \\ 
$\left[ (6,3) \; (\bar{3},3) \right]_{S_q =2}$ & $|\, 30>
=\varepsilon^{\alpha\beta\gamma}\, ( \psi_{\alpha 1} \, \psi_{\eta 1}+
\psi_{\eta 1} \psi_{\alpha 1} ) \phi_{\beta 1} \phi_{\gamma 1} \, \,
\chi^{\eta 2} $ \\ \hline
\end{tabular}
}
\caption{Pentaquark states. The first two quarks are
indicated with $\protect\psi$, the other two quarks with $\protect\phi$, the
antiquark with $\protect\chi$; the Greek indexes refer to colour and the
other ones to spin. In the first column, the group representations of $
SU(3)_C$ and $SU(2)_S$, named after their dimension $d$, are represented,
for the first and the second pair of quarks, respectively. The total spin $S_q$
of the $4$ quark state is also indicated. In the second column the
pentaquark states are listed. The $kets$ $|i>$ have total spin $S=S_z=1/2$ for
$i \leq 15$, $S=S_z=3/2$ for $16 \leq i \leq 27$, $S=S_z=5/2$ for $i \geq 28$}
\label{table1}
\end{table}
\end{center}

\section{Matrix elements}

The non vanishing matrix elements of the chromo-magnetic interaction
for negative parity pentaquarks with $J=1/2,3/2$ and $5/2$ where
$C_{q_i q_j} = \frac{1}{m_{q_i} m_{q_j}}$ and 
$C_{q_i} = \frac{1}{m_{q_i} m_{\bar{q}}}$.

\subsubsection*{4 quarks in S-wave\\
J=1/2}

\noindent\(S_{1/2}\text{ [}1,1\text{] = }-\frac{3 \text{Cds}}{4}-2 \text{Cud}-2 \text{Cus}+\frac{3 \text{Cuu}}{4}\)

\noindent\(S_{1/2}\text{ [}1,2\text{] = }\frac{5 \text{Cds}}{4 \sqrt{3}}-\frac{\text{Cud}}{2 \sqrt{3}}-\frac{\text{Cus}}{2 \sqrt{3}}-\frac{\text{Cuu}}{4
\sqrt{3}}\)

\noindent\(S_{1/2}\text{ [}1,3\text{] = }\frac{1}{2} \sqrt{\frac{3}{2}} \text{Cus}-\frac{1}{2} \sqrt{\frac{3}{2}} \text{Cud}\)

\noindent\(S_{1/2}\text{ [}1,4\text{] = }-\frac{2 \text{Cd}}{\sqrt{3}}-\frac{2 \text{Cs}}{\sqrt{3}}-\frac{2 \text{Cu}}{\sqrt{3}}\)

\noindent\(S_{1/2}\text{ [}1,5\text{] = }-\frac{\text{Cd}}{2 \sqrt{3}}-\frac{\text{Cs}}{2 \sqrt{3}}+\frac{\text{Cu}}{\sqrt{3}}\)

\noindent\(S_{1/2}\text{ [}1,6\text{] = }\frac{3 \text{Cd}}{2 \sqrt{2}}-\frac{3 \text{Cs}}{2 \sqrt{2}}\)

\noindent\(S_{1/2}\text{ [}1,7\text{] = }\frac{\text{Cd}}{\sqrt{6}}-\frac{\text{Cs}}{\sqrt{6}}\)

%
\noindent\(S_{1/2}\text{ [}2,2\text{] = }-\frac{19 \text{Cds}}{12}+\frac{\text{Cud}}{3}+\frac{\text{Cus}}{3}+\frac{11 \text{Cuu}}{12}\)

\noindent\(S_{1/2}\text{ [}2,3\text{] = }\frac{5 \text{Cus}}{2 \sqrt{2}}-\frac{5 \text{Cud}}{2 \sqrt{2}}\)

\noindent\(S_{1/2}\text{ [}2,4\text{] = }-\frac{2 \text{Cd}}{3}-\frac{2 \text{Cs}}{3}+\frac{4 \text{Cu}}{3}\)

\noindent\(S_{1/2}\text{ [}2,5\text{] = }-\frac{\text{Cd}}{6}-\frac{\text{Cs}}{6}-\frac{5 \text{Cu}}{3}\)

\noindent\(S_{1/2}\text{ [}2,6\text{] = }\frac{7 \text{Cd}}{6 \sqrt{6}}-\frac{7 \text{Cs}}{6 \sqrt{6}}\)

\noindent\(S_{1/2}\text{ [}2,7\text{] = }\frac{\sqrt{2} \text{Cd}}{3}-\frac{\sqrt{2} \text{Cs}}{3}\)

\noindent\(S_{1/2}\text{ [}2,8\text{] = }\frac{\text{Cd}}{3 \sqrt{3}}-\frac{\text{Cs}}{3 \sqrt{3}}\)

\noindent\(S_{1/2}\text{ [}3,3\text{] = }\frac{2 \text{Cds}}{3}-\frac{2 \text{Cud}}{3}-\frac{2 \text{Cus}}{3}+\frac{2 \text{Cuu}}{3}\)

\noindent\(S_{1/2}\text{ [}3,4\text{] = }\frac{2 \sqrt{2} \text{Cd}}{3}-\frac{2 \sqrt{2} \text{Cs}}{3}\)

\noindent\(S_{1/2}\text{ [}3,5\text{] = }\frac{\sqrt{2} \text{Cd}}{3}-\frac{\sqrt{2} \text{Cs}}{3}\)

\noindent\(S_{1/2}\text{ [}3,6\text{] = }-\frac{7 \text{Cd}}{6 \sqrt{3}}-\frac{7 \text{Cs}}{6 \sqrt{3}}+\frac{7 \text{Cu}}{3 \sqrt{3}}\)

\noindent\(S_{1/2}\text{ [}3,7\text{] = }-\frac{\text{Cd}}{2}-\frac{\text{Cs}}{2}-\text{Cu}\)

\noindent\(S_{1/2}\text{ [}3,8\text{] = }\frac{\text{Cd}}{3 \sqrt{6}}+\frac{\text{Cs}}{3 \sqrt{6}}-\frac{1}{3} \sqrt{\frac{2}{3}} \text{Cu}\)

\noindent\(S_{1/2}\text{ [}4,4\text{] = }-\frac{17 \text{Cd}}{12}-\frac{7 \text{Cds}}{6}-\frac{17 \text{Cs}}{12}-\frac{\text{Cu}}{2}-\frac{17 \text{Cud}}{12}-\frac{17
\text{Cus}}{12}+\frac{2 \text{Cuu}}{3}\)

\noindent\(S_{1/2}\text{ [}4,5\text{] = }-\frac{5 \text{Cd}}{12}+\frac{5 \text{Cds}}{6}-\frac{5 \text{Cs}}{12}+\frac{5 \text{Cu}}{6}-\frac{5 \text{Cud}}{12}-\frac{5
\text{Cus}}{12}\)

\noindent\(S_{1/2}\text{ [}4,6\text{] = }\frac{5 \text{Cd}}{2 \sqrt{6}}-\frac{5 \text{Cs}}{2 \sqrt{6}}-\frac{5 \text{Cud}}{2 \sqrt{6}}+\frac{5 \text{Cus}}{2
\sqrt{6}}\)

\noindent\(S_{1/2}\text{ [}4,7\text{] = }\frac{5 \text{Cd}}{6 \sqrt{2}}-\frac{5 \text{Cs}}{6 \sqrt{2}}-\frac{5 \text{Cud}}{6 \sqrt{2}}+\frac{5 \text{Cus}}{6
\sqrt{2}}\)
%

\noindent\(S_{1/2}\text{ [}5,5\text{] = }\frac{7 \text{Cd}}{12}-\frac{7 \text{Cds}}{6}+\frac{7 \text{Cs}}{12}-\frac{\text{Cu}}{2}+\frac{7 \text{Cud}}{12}+\frac{7
\text{Cus}}{12}+\frac{2 \text{Cuu}}{3}\)

\noindent\(S_{1/2}\text{ [}5,6\text{] = }\frac{7 \text{Cd}}{6 \sqrt{6}}-\frac{7 \text{Cs}}{6 \sqrt{6}}-\frac{\text{Cud}}{6 \sqrt{6}}+\frac{\text{Cus}}{6
\sqrt{6}}\)

\noindent\(S_{1/2}\text{ [}5,7\text{] = }\frac{5 \text{Cd}}{6 \sqrt{2}}-\frac{5 \text{Cs}}{6 \sqrt{2}}-\frac{11 \text{Cud}}{6 \sqrt{2}}+\frac{11
\text{Cus}}{6 \sqrt{2}}\)

\noindent\(S_{1/2}\text{ [}5,8\text{] = }-\frac{2 \text{Cd}}{3 \sqrt{3}}+\frac{2 \text{Cs}}{3 \sqrt{3}}-\frac{\text{Cud}}{3 \sqrt{3}}+\frac{\text{Cus}}{3
\sqrt{3}}\)

\noindent\(S_{1/2}\text{ [}6,6\text{] = }-\frac{5 \text{Cd}}{6}+\frac{13 \text{Cds}}{18}-\frac{5 \text{Cs}}{6}-\frac{5 \text{Cu}}{3}-\frac{25 \text{Cud}}{18}-\frac{25
\text{Cus}}{18}+\frac{13 \text{Cuu}}{18}\)

\noindent\(S_{1/2}\text{ [}6,7\text{] = }-\frac{7 \text{Cd}}{6 \sqrt{3}}+\frac{\text{Cds}}{6 \sqrt{3}}-\frac{7 \text{Cs}}{6 \sqrt{3}}+\frac{7 \text{Cu}}{3
\sqrt{3}}-\frac{\text{Cuu}}{6 \sqrt{3}}\)

\noindent\(S_{1/2}\text{ [}6,8\text{] = }-\frac{\text{Cds}}{9 \sqrt{2}}+\frac{\text{Cud}}{9 \sqrt{2}}+\frac{\text{Cus}}{9 \sqrt{2}}-\frac{\text{Cuu}}{9
\sqrt{2}}\)

\noindent\(S_{1/2}\text{ [}7,7\text{] = }\frac{\text{Cd}}{6}+\frac{5 \text{Cds}}{6}+\frac{\text{Cs}}{6}+\frac{\text{Cu}}{3}-\frac{\text{Cud}}{2}-\frac{\text{Cus}}{2}+\frac{5
\text{Cuu}}{6}\)

\noindent\(S_{1/2}\text{ [}7,8\text{] = }-\frac{1}{3} \sqrt{\frac{2}{3}} \text{Cd}-\frac{\text{Cds}}{3 \sqrt{6}}-\frac{1}{3} \sqrt{\frac{2}{3}} \text{Cs}+\frac{2}{3}
\sqrt{\frac{2}{3}} \text{Cu}+\frac{\text{Cuu}}{3 \sqrt{6}}\)

\noindent\(S_{1/2}\text{ [}8,8\text{] = }\frac{2 \text{Cd}}{3}+\frac{7 \text{Cds}}{9}+\frac{2 \text{Cs}}{3}+\frac{4 \text{Cu}}{3}+\frac{14 \text{Cud}}{9}+\frac{14
\text{Cus}}{9}+\frac{7 \text{Cuu}}{9}\)

~
\subsubsection*{4 quarks in S-wave\\
J=3/2}

\noindent\(S_{3/2}\text{ [}1,1\text{] = }\frac{17 \text{Cd}}{24}-\frac{7 \text{Cds}}{6}+\frac{17 \text{Cs}}{24}+\frac{\text{Cu}}{4}-\frac{17 \text{Cud}}{12}-\frac{17
\text{Cus}}{12}+\frac{2 \text{Cuu}}{3}\)

\noindent\(S_{3/2}\text{ [}1,2\text{] = }\frac{5 \text{Cd}}{24}+\frac{5 \text{Cds}}{6}+\frac{5 \text{Cs}}{24}-\frac{5 \text{Cu}}{12}-\frac{5 \text{Cud}}{12}-\frac{5
\text{Cus}}{12}\)

\noindent\(S_{3/2}\text{ [}1,3\text{] = }\frac{1}{12} \sqrt{\frac{5}{2}} \text{Cd}+\frac{1}{12} \sqrt{\frac{5}{2}} \text{Cs}-\frac{1}{6} \sqrt{\frac{5}{2}}
\text{Cu}\)

\noindent\(S_{3/2}\text{ [}1,4\text{] = }-\frac{5 \text{Cd}}{4 \sqrt{6}}+\frac{5 \text{Cs}}{4 \sqrt{6}}-\frac{5 \text{Cud}}{2 \sqrt{6}}+\frac{5 \text{Cus}}{2
\sqrt{6}}\)

\noindent\(S_{3/2}\text{ [}1,5\text{] = }-\frac{5 \text{Cd}}{12 \sqrt{2}}+\frac{5 \text{Cs}}{12 \sqrt{2}}-\frac{5 \text{Cud}}{6 \sqrt{2}}+\frac{5
\text{Cus}}{6 \sqrt{2}}\)

\noindent\(S_{3/2}\text{ [}1,6\text{] = }\frac{\sqrt{5} \text{Cd}}{12}-\frac{\sqrt{5} \text{Cs}}{12}\)
%

\noindent\(S_{3/2}\text{ [}2,2\text{] = }-\frac{7 \text{Cd}}{24}-\frac{7 \text{Cds}}{6}-\frac{7 \text{Cs}}{24}+\frac{\text{Cu}}{4}+\frac{7 \text{Cud}}{12}+\frac{7
\text{Cus}}{12}+\frac{2 \text{Cuu}}{3}\)

\noindent\(S_{3/2}\text{ [}2,3\text{] = }-\frac{11}{12} \sqrt{\frac{5}{2}} \text{Cd}-\frac{11}{12} \sqrt{\frac{5}{2}} \text{Cs}-\frac{1}{6} \sqrt{\frac{5}{2}}
\text{Cu}\)

\noindent\(S_{3/2}\text{ [}2,4\text{] = }-\frac{7 \text{Cd}}{12 \sqrt{6}}+\frac{7 \text{Cs}}{12 \sqrt{6}}-\frac{\text{Cud}}{6 \sqrt{6}}+\frac{\text{Cus}}{6
\sqrt{6}}\)

\noindent\(S_{3/2}\text{ [}2,5\text{] = }-\frac{5 \text{Cd}}{12 \sqrt{2}}+\frac{5 \text{Cs}}{12 \sqrt{2}}-\frac{11 \text{Cud}}{6 \sqrt{2}}+\frac{11
\text{Cus}}{6 \sqrt{2}}\)

\noindent\(S_{3/2}\text{ [}2,6\text{] = }\frac{5 \sqrt{5} \text{Cd}}{12}-\frac{5 \sqrt{5} \text{Cs}}{12}\)

\noindent\(S_{3/2}\text{ [}2,7\text{] = }\frac{\text{Cd}}{3 \sqrt{3}}-\frac{\text{Cs}}{3 \sqrt{3}}-\frac{\text{Cud}}{3 \sqrt{3}}+\frac{\text{Cus}}{3
\sqrt{3}}\)

\noindent\(S_{3/2}\text{ [}3,3\text{] = }-\frac{5 \text{Cd}}{4}-\frac{\text{Cds}}{3}-\frac{5 \text{Cs}}{4}+\frac{\text{Cu}}{2}+\frac{5 \text{Cud}}{6}+\frac{5
\text{Cus}}{6}+\frac{2 \text{Cuu}}{3}\)

\noindent\(S_{3/2}\text{ [}3,4\text{] = }\frac{1}{12} \sqrt{\frac{5}{3}} \text{Cd}-\frac{1}{12} \sqrt{\frac{5}{3}} \text{Cs}\)

\noindent\(S_{3/2}\text{ [}3,5\text{] = }\frac{5 \sqrt{5} \text{Cs}}{12}-\frac{5 \sqrt{5} \text{Cd}}{12}\)

\noindent\(S_{3/2}\text{ [}3,6\text{] = }\frac{3 \text{Cd}}{2 \sqrt{2}}-\frac{3 \text{Cs}}{2 \sqrt{2}}+\frac{\text{Cud}}{\sqrt{2}}-\frac{\text{Cus}}{\sqrt{2}}\)

\noindent\(S_{3/2}\text{ [}3,7\text{] = }\frac{2}{3} \sqrt{\frac{10}{3}} \text{Cd}-\frac{2}{3} \sqrt{\frac{10}{3}} \text{Cs}\)

\noindent\(S_{3/2}\text{ [}4,4\text{] = }\frac{5 \text{Cd}}{12}+\frac{13 \text{Cds}}{18}+\frac{5 \text{Cs}}{12}+\frac{5 \text{Cu}}{6}-\frac{25 \text{Cud}}{18}-\frac{25
\text{Cus}}{18}+\frac{13 \text{Cuu}}{18}\)

\noindent\(S_{3/2}\text{ [}4,5\text{] = }\frac{7 \text{Cd}}{12 \sqrt{3}}+\frac{\text{Cds}}{6 \sqrt{3}}+\frac{7 \text{Cs}}{12 \sqrt{3}}-\frac{7 \text{Cu}}{6
\sqrt{3}}-\frac{\text{Cuu}}{6 \sqrt{3}}\)

\noindent\(S_{3/2}\text{ [}4,6\text{] = }\frac{1}{6} \sqrt{\frac{5}{6}} \text{Cd}+\frac{1}{6} \sqrt{\frac{5}{6}} \text{Cs}-\frac{1}{3} \sqrt{\frac{5}{6}}
\text{Cu}\)

\noindent\(S_{3/2}\text{ [}4,7\text{] = }-\frac{\text{Cds}}{9 \sqrt{2}}+\frac{\text{Cud}}{9 \sqrt{2}}+\frac{\text{Cus}}{9 \sqrt{2}}-\frac{\text{Cuu}}{9
\sqrt{2}}\)

\noindent\(S_{3/2}\text{ [}5,5\text{] = }-\frac{\text{Cd}}{12}+\frac{5 \text{Cds}}{6}-\frac{\text{Cs}}{12}-\frac{\text{Cu}}{6}-\frac{\text{Cud}}{2}-\frac{\text{Cus}}{2}+\frac{5
\text{Cuu}}{6}\)

\noindent\(S_{3/2}\text{ [}5,6\text{] = }\frac{1}{2} \sqrt{\frac{5}{2}} \text{Cd}+\frac{1}{2} \sqrt{\frac{5}{2}} \text{Cs}+\sqrt{\frac{5}{2}} \text{Cu}\)

\noindent\(S_{3/2}\text{ [}5,7\text{] = }\frac{\text{Cd}}{3 \sqrt{6}}-\frac{\text{Cds}}{3 \sqrt{6}}+\frac{\text{Cs}}{3 \sqrt{6}}-\frac{1}{3} \sqrt{\frac{2}{3}}
\text{Cu}+\frac{\text{Cuu}}{3 \sqrt{6}}\)

\noindent\(S_{3/2}\text{ [}6,6\text{] = }-\frac{\text{Cd}}{2}+\frac{2 \text{Cds}}{3}-\frac{\text{Cs}}{2}-\text{Cu}+\frac{\text{Cud}}{3}+\frac{\text{Cus}}{3}+\frac{2
\text{Cuu}}{3}\)

\noindent\(S_{3/2}\text{ [}6,7\text{] = }-\frac{2}{3} \sqrt{\frac{5}{3}} \text{Cd}-\frac{2}{3} \sqrt{\frac{5}{3}} \text{Cs}+\frac{4}{3} \sqrt{\frac{5}{3}}
\text{Cu}\)

\noindent\(S_{3/2}\text{ [}7,7\text{] = }-\frac{\text{Cd}}{3}+\frac{7 \text{Cds}}{9}-\frac{\text{Cs}}{3}-\frac{2 \text{Cu}}{3}+\frac{14 \text{Cud}}{9}+\frac{14
\text{Cus}}{9}+\frac{7 \text{Cuu}}{9}\)

~
\subsubsection*{4 quarks in S-wave\\
J=5/2}

\noindent\(S_{5/2}\text{ [}1,1\text{] = }\frac{\text{Cd}}{3}+\frac{2 \text{Cds}}{3}+\frac{\text{Cs}}{3}+\frac{2 \text{Cu}}{3}+\frac{\text{Cud}}{3}+\frac{\text{Cus}}{3}+\frac{2
\text{Cuu}}{3}\)

\noindent\(S_{5/2}\text{ [}1,2\text{] = }-\frac{\text{Cd}}{\sqrt{2}}+\frac{\text{Cs}}{\sqrt{2}}+\frac{\text{Cud}}{\sqrt{2}}-\frac{\text{Cus}}{\sqrt{2}}\)

\noindent\(S_{5/2}\text{ [}2,2\text{] = }\frac{5 \text{Cd}}{6}-\frac{\text{Cds}}{3}+\frac{5 \text{Cs}}{6}-\frac{\text{Cu}}{3}+\frac{5 \text{Cud}}{6}+\frac{5
\text{Cus}}{6}+\frac{2 \text{Cuu}}{3}\)

~\\
\noindent
In the following we report the non vanishing matrix elements of the
chromo-magnetic interaction for positive parity pentaquarks with
Spin $1/2,3/2$ and $5/2$ where
$c12 = \frac{1}{m_{1} m_{2}}$, $c34 = \frac{1}{m_{3} m_{4}}$ and 
$ci5 = \frac{1}{m_{i} m_{\bar{q}}}$.

\subsubsection*{4 quarks in P-wave\\
Block from [1,1] to [15,15] with $S=1/2$}

\noindent\(\text{P[}1,1\text{] = }-\frac{\text{c12}}{3}-\frac{5 \text{c15}}{3}-\frac{5 \text{c25}}{3}-2 \text{c34}\)
%

\noindent\(\text{P[}1,3\text{] = }-\sqrt{\frac{3}{2}} \text{c15}-\sqrt{\frac{3}{2}} \text{c25}\)

\noindent\(\text{P[}1,4\text{] = }\frac{5 \text{c15}}{2 \sqrt{3}}-\frac{5 \text{c25}}{2 \sqrt{3}}\)

\noindent\(\text{P[}1,5\text{] = }\sqrt{2} \text{c15}-\sqrt{2} \text{c25}\)
%
%

\noindent\(\text{P[}1,8\text{] = }\frac{\text{c35}}{3 \sqrt{2}}-\frac{\text{c45}}{3 \sqrt{2}}\)

\noindent\(\text{P[}1,9\text{] = }\frac{\text{c35}}{6}-\frac{\text{c45}}{6}\)
%
%
%
%
%
%

\noindent\(\text{P[}2,2\text{] = }-2 \text{c12}-\frac{\text{c34}}{3}-\frac{5 \text{c35}}{3}-\frac{5 \text{c45}}{3}\)

\noindent\(\text{P[}2,3\text{] = }\sqrt{\frac{3}{2}} \text{c35}+\sqrt{\frac{3}{2}} \text{c45}\)
%
%

\noindent\(\text{P[}2,6\text{] = }\frac{\text{c25}}{3 \sqrt{2}}-\frac{\text{c15}}{3 \sqrt{2}}\)

\noindent\(\text{P[}2,7\text{] = }\frac{\text{c15}}{6}-\frac{\text{c25}}{6}\)
%
%

\noindent\(\text{P[}2,10\text{] = }\frac{5 \text{c35}}{2 \sqrt{3}}-\frac{5 \text{c45}}{2 \sqrt{3}}\)

\noindent\(\text{P[}2,11\text{] = }\sqrt{2} \text{c35}-\sqrt{2} \text{c45}\)
%
%
%
%

\noindent\(\text{P[}3,3\text{] = }-2 \text{c12}-2 \text{c34}\)
%

\noindent\(\text{P[}3,5\text{] = }\frac{\text{c15}}{\sqrt{3}}-\frac{\text{c25}}{\sqrt{3}}\)
%
%
%
%
%

\noindent\(\text{P[}3,11\text{] = }\frac{\text{c45}}{\sqrt{3}}-\frac{\text{c35}}{\sqrt{3}}\)
%
%
%
%

\noindent\(\text{P[}4,4\text{] = }\text{c12}-2 \text{c34}\)

\noindent\(\text{P[}4,5\text{] = }-\sqrt{\frac{3}{2}} \text{c15}-\sqrt{\frac{3}{2}} \text{c25}\)
%
%
%
%
%
%

\noindent\(\text{P[}4,12\text{] = }\frac{\text{c45}}{2 \sqrt{3}}-\frac{\text{c35}}{2 \sqrt{3}}\)
%
%
%

\noindent\(\text{P[}5,5\text{] = }\frac{2 \text{c12}}{3}-\frac{2 \text{c15}}{3}-\frac{2 \text{c25}}{3}-2 \text{c34}\)

\noindent\(\text{P[}5,6\text{] = }-\text{c35}-\text{c45}\)

\noindent\(\text{P[}5,7\text{] = }-\frac{\text{c35}}{\sqrt{2}}-\frac{\text{c45}}{\sqrt{2}}\)
%
%
%
%
%

\noindent\(\text{P[}5,13\text{] = }\frac{\text{c45}}{3}-\frac{\text{c35}}{3}\)

\noindent\(\text{P[}5,14\text{] = }\frac{\sqrt{2} \text{c45}}{3}-\frac{\sqrt{2} \text{c35}}{3}\)
%

\noindent\(\text{P[}6,6\text{] = }\frac{2 \text{c12}}{3}+\frac{\text{c15}}{6}+\frac{\text{c25}}{6}-\frac{\text{c34}}{3}-\frac{5 \text{c35}}{6}-\frac{5
\text{c45}}{6}\)

\noindent\(\text{P[}6,7\text{] = }-\frac{\text{c15}}{3 \sqrt{2}}-\frac{\text{c25}}{3 \sqrt{2}}-\frac{5 \text{c35}}{3 \sqrt{2}}-\frac{5 \text{c45}}{3
\sqrt{2}}\)
%
%
%
%
%

\noindent\(\text{P[}6,13\text{] = }\text{c45}-\text{c35}\)

\noindent\(\text{P[}6,14\text{] = }\frac{\text{c45}}{\sqrt{2}}-\frac{\text{c35}}{\sqrt{2}}\)

\noindent\(\text{P[}6,15\text{] = }\frac{5 \text{c45}}{3 \sqrt{2}}-\frac{5 \text{c35}}{3 \sqrt{2}}\)

\noindent\(\text{P[}7,7\text{] = }\frac{2 \text{c12}}{3}-\frac{\text{c34}}{3}\)
%
%
%
%
%
%

\noindent\(\text{P[}7,14\text{] = }\text{c45}-\text{c35}\)

\noindent\(\text{P[}7,15\text{] = }\frac{5 \text{c45}}{6}-\frac{5 \text{c35}}{6}\)

\noindent\(\text{P[}8,8\text{] = }-\frac{\text{c12}}{3}-\frac{5 \text{c15}}{6}-\frac{5 \text{c25}}{6}+\frac{2 \text{c34}}{3}+\frac{\text{c35}}{6}+\frac{\text{c45}}{6}\)

\noindent\(\text{P[}8,9\text{] = }\frac{5 \text{c15}}{3 \sqrt{2}}+\frac{5 \text{c25}}{3 \sqrt{2}}+\frac{\text{c35}}{3 \sqrt{2}}+\frac{\text{c45}}{3
\sqrt{2}}\)
%

\noindent\(\text{P[}8,11\text{] = }\text{c15}+\text{c25}\)

\noindent\(\text{P[}8,12\text{] = }\frac{5 \text{c15}}{3 \sqrt{2}}-\frac{5 \text{c25}}{3 \sqrt{2}}\)

\noindent\(\text{P[}8,13\text{] = }\text{c25}-\text{c15}\)

\noindent\(\text{P[}8,14\text{] = }\frac{\text{c15}}{\sqrt{2}}-\frac{\text{c25}}{\sqrt{2}}\)
%

\noindent\(\text{P[}9,9\text{] = }\frac{2 \text{c34}}{3}-\frac{\text{c12}}{3}\)
%

\noindent\(\text{P[}9,11\text{] = }-\frac{\text{c15}}{\sqrt{2}}-\frac{\text{c25}}{\sqrt{2}}\)

\noindent\(\text{P[}9,12\text{] = }\frac{5 \text{c25}}{6}-\frac{5 \text{c15}}{6}\)
%

\noindent\(\text{P[}9,14\text{] = }\text{c25}-\text{c15}\)
%

\noindent\(\text{P[}10,10\text{] = }\text{c34}-2 \text{c12}\)

\noindent\(\text{P[}10,11\text{] = }-\sqrt{\frac{3}{2}} \text{c35}-\sqrt{\frac{3}{2}} \text{c45}\)
%
%
%

\noindent\(\text{P[}10,15\text{] = }\frac{\text{c25}}{2 \sqrt{3}}-\frac{\text{c15}}{2 \sqrt{3}}\)

\noindent\(\text{P[}11,11\text{] = }-2 \text{c12}+\frac{2 \text{c34}}{3}-\frac{2 \text{c35}}{3}-\frac{2 \text{c45}}{3}\)
%

\noindent\(\text{P[}11,13\text{] = }\frac{\text{c15}}{3}-\frac{\text{c25}}{3}\)

\noindent\(\text{P[}11,14\text{] = }\frac{\sqrt{2} \text{c25}}{3}-\frac{\sqrt{2} \text{c15}}{3}\)
%

\noindent\(\text{P[}12,12\text{] = }\text{c12}+\frac{2 \text{c34}}{3}+\frac{\text{c35}}{3}+\frac{\text{c45}}{3}\)

\noindent\(\text{P[}12,13\text{] = }\frac{\text{c15}}{\sqrt{2}}+\frac{\text{c25}}{\sqrt{2}}\)

\noindent\(\text{P[}12,14\text{] = }-\text{c15}-\text{c25}\)
%

\noindent\(\text{P[}13,13\text{] = }\frac{2 \text{c12}}{3}+\frac{2 \text{c34}}{3}\)

\noindent\(\text{P[}13,14\text{] = }\frac{\sqrt{2} \text{c15}}{3}+\frac{\sqrt{2} \text{c25}}{3}-\frac{\sqrt{2} \text{c35}}{3}-\frac{\sqrt{2} \text{c45}}{3}\)

\noindent\(\text{P[}13,15\text{] = }-\frac{\text{c35}}{\sqrt{2}}-\frac{\text{c45}}{\sqrt{2}}\)

\noindent\(\text{P[}14,14\text{] = }\frac{2 \text{c12}}{3}-\frac{\text{c15}}{3}-\frac{\text{c25}}{3}+\frac{2 \text{c34}}{3}-\frac{\text{c35}}{3}-\frac{\text{c45}}{3}\)

\noindent\(\text{P[}14,15\text{] = }-\text{c35}-\text{c45}\)

\noindent\(\text{P[}15,15\text{] = }\frac{2 \text{c12}}{3}+\frac{\text{c15}}{3}+\frac{\text{c25}}{3}+\text{c34}\)

\subsubsection*{4 quarks in P-wave\\
Block from [16,16] to [27,27] with $S=3/2$}

\noindent\(\text{P[}16,16\text{] = }\frac{2 \text{c12}}{3}-\frac{\text{c15}}{2}-\frac{\text{c25}}{2}+\frac{2 \text{c34}}{3}-\frac{\text{c35}}{2}-\frac{\text{c45}}{2}\)

\noindent\(\text{P[}16,17\text{] = }\frac{\sqrt{5} \text{c15}}{6}+\frac{\sqrt{5} \text{c25}}{6}-\frac{\sqrt{5} \text{c35}}{6}-\frac{\sqrt{5} \text{c45}}{6}\)

\noindent\(\text{P[}16,18\text{] = }\frac{\sqrt{5} \text{c35}}{2}+\frac{\sqrt{5} \text{c45}}{2}\)

\noindent\(\text{P[}16,19\text{] = }-\frac{\sqrt{5} \text{c15}}{2}-\frac{\sqrt{5} \text{c25}}{2}\)

\noindent\(\text{P[}16,20\text{] = }\frac{1}{3} \sqrt{\frac{5}{2}} \text{c15}-\frac{1}{3} \sqrt{\frac{5}{2}} \text{c25}\)

\noindent\(\text{P[}16,21\text{] = }\frac{3 \text{c15}}{2 \sqrt{2}}-\frac{3 \text{c25}}{2 \sqrt{2}}\)

\noindent\(\text{P[}16,22\text{] = }\frac{1}{2} \sqrt{\frac{5}{2}} \text{c25}-\frac{1}{2} \sqrt{\frac{5}{2}} \text{c15}\)

\noindent\(\text{P[}16,23\text{] = }\frac{1}{3} \sqrt{\frac{5}{2}} \text{c45}-\frac{1}{3} \sqrt{\frac{5}{2}} \text{c35}\)

\noindent\(\text{P[}16,24\text{] = }\frac{3 \text{c45}}{2 \sqrt{2}}-\frac{3 \text{c35}}{2 \sqrt{2}}\)

\noindent\(\text{P[}16,25\text{] = }\frac{1}{2} \sqrt{\frac{5}{2}} \text{c45}-\frac{1}{2} \sqrt{\frac{5}{2}} \text{c35}\)
%
%

\noindent\(\text{P[}17,17\text{] = }\frac{2 \text{c12}}{3}+\frac{\text{c15}}{6}+\frac{\text{c25}}{6}+\frac{2 \text{c34}}{3}+\frac{\text{c35}}{6}+\frac{\text{c45}}{6}\)

\noindent\(\text{P[}17,18\text{] = }\frac{\text{c35}}{2}+\frac{\text{c45}}{2}\)

\noindent\(\text{P[}17,19\text{] = }\frac{\text{c15}}{2}+\frac{\text{c25}}{2}\)

\noindent\(\text{P[}17,20\text{] = }\frac{\text{c25}}{3 \sqrt{2}}-\frac{\text{c15}}{3 \sqrt{2}}\)

\noindent\(\text{P[}17,21\text{] = }\frac{1}{2} \sqrt{\frac{5}{2}} \text{c25}-\frac{1}{2} \sqrt{\frac{5}{2}} \text{c15}\)

\noindent\(\text{P[}17,22\text{] = }\frac{\text{c25}}{2 \sqrt{2}}-\frac{\text{c15}}{2 \sqrt{2}}\)

\noindent\(\text{P[}17,23\text{] = }\frac{\text{c45}}{3 \sqrt{2}}-\frac{\text{c35}}{3 \sqrt{2}}\)

\noindent\(\text{P[}17,24\text{] = }\frac{1}{2} \sqrt{\frac{5}{2}} \text{c45}-\frac{1}{2} \sqrt{\frac{5}{2}} \text{c35}\)

\noindent\(\text{P[}17,25\text{] = }\frac{\text{c35}}{2 \sqrt{2}}-\frac{\text{c45}}{2 \sqrt{2}}\)
%
%

\noindent\(\text{P[}18,18\text{] = }\frac{2 \text{c12}}{3}-\frac{\text{c15}}{6}-\frac{\text{c25}}{6}+\text{c34}\)
%
%
%
%
%

\noindent\(\text{P[}18,24\text{] = }\frac{5}{6} \sqrt{\frac{5}{2}} \text{c35}-\frac{5}{6} \sqrt{\frac{5}{2}} \text{c45}\)

\noindent\(\text{P[}18,25\text{] = }\frac{5 \text{c35}}{6 \sqrt{2}}-\frac{5 \text{c45}}{6 \sqrt{2}}\)
%
%

\noindent\(\text{P[}19,19\text{] = }\text{c12}+\frac{2 \text{c34}}{3}-\frac{\text{c35}}{6}-\frac{\text{c45}}{6}\)
%

\noindent\(\text{P[}19,21\text{] = }\frac{5}{6} \sqrt{\frac{5}{2}} \text{c15}-\frac{5}{6} \sqrt{\frac{5}{2}} \text{c25}\)

\noindent\(\text{P[}19,22\text{] = }\frac{5 \text{c25}}{6 \sqrt{2}}-\frac{5 \text{c15}}{6 \sqrt{2}}\)
%
%
%
%
%

\noindent\(\text{P[}20,20\text{] = }-2 \text{c12}+\frac{2 \text{c34}}{3}+\frac{\text{c35}}{3}+\frac{\text{c45}}{3}\)

\noindent\(\text{P[}20,21\text{] = }-\frac{\sqrt{5} \text{c15}}{2}-\frac{\sqrt{5} \text{c25}}{2}\)

\noindent\(\text{P[}20,22\text{] = }\frac{\text{c15}}{2}+\frac{\text{c25}}{2}\)
%
%
%
%

\noindent\(\text{P[}20,27\text{] = }\frac{\text{c35}}{\sqrt{2}}-\frac{\text{c45}}{\sqrt{2}}\)

\noindent\(\text{P[}21,21\text{] = }-\frac{\text{c12}}{3}-\frac{5 \text{c15}}{4}-\frac{5 \text{c25}}{4}+\frac{2 \text{c34}}{3}+\frac{\text{c35}}{4}+\frac{\text{c45}}{4}\)

\noindent\(\text{P[}21,22\text{] = }\frac{5 \sqrt{5} \text{c15}}{12}+\frac{5 \sqrt{5} \text{c25}}{12}+\frac{\sqrt{5} \text{c35}}{12}+\frac{\sqrt{5}
\text{c45}}{12}\)
%
%
%

\noindent\(\text{P[}21,26\text{] = }\frac{1}{6} \sqrt{\frac{5}{2}} \text{c45}-\frac{1}{6} \sqrt{\frac{5}{2}} \text{c35}\)
%

\noindent\(\text{P[}22,22\text{] = }-\frac{\text{c12}}{3}+\frac{5 \text{c15}}{12}+\frac{5 \text{c25}}{12}+\frac{2 \text{c34}}{3}-\frac{\text{c35}}{12}-\frac{\text{c45}}{12}\)
%
%
%

\noindent\(\text{P[}22,26\text{] = }\frac{\text{c45}}{6 \sqrt{2}}-\frac{\text{c35}}{6 \sqrt{2}}\)
%

\noindent\(\text{P[}23,23\text{] = }\frac{2 \text{c12}}{3}+\frac{\text{c15}}{3}+\frac{\text{c25}}{3}-2 \text{c34}\)

\noindent\(\text{P[}23,24\text{] = }-\frac{\sqrt{5} \text{c35}}{2}-\frac{\sqrt{5} \text{c45}}{2}\)

\noindent\(\text{P[}23,25\text{] = }-\frac{\text{c35}}{2}-\frac{\text{c45}}{2}\)

\noindent\(\text{P[}23,26\text{] = }\frac{\text{c15}}{\sqrt{2}}-\frac{\text{c25}}{\sqrt{2}}\)
%

\noindent\(\text{P[}24,24\text{] = }\frac{2 \text{c12}}{3}+\frac{\text{c15}}{4}+\frac{\text{c25}}{4}-\frac{\text{c34}}{3}-\frac{5 \text{c35}}{4}-\frac{5
\text{c45}}{4}\)

\noindent\(\text{P[}24,25\text{] = }-\frac{\sqrt{5} \text{c15}}{12}-\frac{\sqrt{5} \text{c25}}{12}-\frac{5 \sqrt{5} \text{c35}}{12}-\frac{5 \sqrt{5}
\text{c45}}{12}\)
%

\noindent\(\text{P[}24,27\text{] = }\frac{1}{6} \sqrt{\frac{5}{2}} \text{c25}-\frac{1}{6} \sqrt{\frac{5}{2}} \text{c15}\)

\noindent\(\text{P[}25,25\text{] = }\frac{2 \text{c12}}{3}-\frac{\text{c15}}{12}-\frac{\text{c25}}{12}-\frac{\text{c34}}{3}+\frac{5 \text{c35}}{12}+\frac{5
\text{c45}}{12}\)
%

\noindent\(\text{P[}25,27\text{] = }\frac{\text{c15}}{6 \sqrt{2}}-\frac{\text{c25}}{6 \sqrt{2}}\)

\noindent\(\text{P[}26,26\text{] = }-\frac{\text{c12}}{3}+\frac{5 \text{c15}}{6}+\frac{5 \text{c25}}{6}-2 \text{c34}\)
%

\noindent\(\text{P[}27,27\text{] = }-2 \text{c12}-\frac{\text{c34}}{3}+\frac{5 \text{c35}}{6}+\frac{5 \text{c45}}{6}\)

~
\subsubsection*{4 quarks in P-wave\\
Block from [28,28] to [30,30] with $S=5/2$}

\noindent\(\text{P[}28,28\text{] = }\frac{2 \text{c12}}{3}+\frac{\text{c15}}{3}+\frac{\text{c25}}{3}+\frac{2 \text{c34}}{3}+\frac{\text{c35}}{3}+\frac{\text{c45}}{3}\)

\noindent\(\text{P[}28,29\text{] = }\frac{\text{c35}}{\sqrt{2}}-\frac{\text{c45}}{\sqrt{2}}\)

\noindent\(\text{P[}28,30\text{] = }\frac{\text{c25}}{\sqrt{2}}-\frac{\text{c15}}{\sqrt{2}}\)

\noindent\(\text{P[}29,29\text{] = }\frac{2 \text{c12}}{3}-\frac{\text{c15}}{6}-\frac{\text{c25}}{6}-\frac{\text{c34}}{3}+\frac{5 \text{c35}}{6}+\frac{5
\text{c45}}{6}\)
%

\noindent\(\text{P[}30,30\text{] = }-\frac{\text{c12}}{3}+\frac{5 \text{c15}}{6}+\frac{5 \text{c25}}{6}+\frac{2 \text{c34}}{3}-\frac{\text{c35}}{6}-\frac{\text{c45}}{6}\)

\section{Spectrum}

The spectrum of the $SU(3)_F$ reducible multiplets of
negative parity and ``open door'' positive parity.
We call $O$ the $s=-4$ states with quark content
$ssss\bar{q}$. A star has been put for the negative
parity states with ``open door'' decays and for the positive
parity states with ``open door'' decays into a pseudoscalar
meson and bayon of the decuplet.

\begin{table}[h!]
\begin{tabular}{l|c|c|c|c}
$3 \times\bar{3}_F$ & $N_s$ & $\Lambda_s$ & $\Lambda + \Sigma$ & $\Xi$ \\
$ 1/2^{-*}$  & 1525  & 1636  & 1289  & 1397 \\
$ 1/2^-$     & 1878  & 1995  & 1794  & 1918 \\
$ 3/2^-$     & 1884  & 2011  & 1775  & 1907 \\
$ 1/2^+$     & 1775  & 1944  & 1611  & 1780 \\
$ 3/2^+$     & 1870  & 2025  & 1707  & 1865
\end{tabular}
\end{table}

\begin{table}[h!]
\begin{tabular}{l|c|c|c|c|c|c}
$\bar{6} \times \bar{3}$ & $Z^0$ & $N_s$ & $\Sigma_s$ & $N$ &
$\Lambda + \Sigma$ & $\Xi + \Xi_{3/2}$ \\
$ 1/2^{-*}$  & 1588  & 1744  & 1907 & 1386  & 1547  & 1712  \\
$ 3/2^-$     & 1858  & 2005  & 2142 & 1759  & 1905  & 2038  \\
$ 1/2^+$     & 1545  & 1719  & 1893 & 1356  & 1534  & 1711  \\
$ 3/2^+$     & 1614  & 1779  & 1994 & 1434  & 1608  & 1761
\end{tabular}
\end{table}

\begin{table}[h!]
\begin{tabular}{l|c|c|c|c|c|c}
$15 \times \bar{3}$ & $Z^1$ & $N_s$ & $\Delta_s$ & $\Lambda_s$ &
$\Sigma_s$ & $\Xi_s$ \\
$1/2^{-*}$  & 1794  & 1920  & 1964  & 2045  & 2077  & 2184 \\
$1/2^-$     & 2026  & 2126  & 2189  & 2229  & 2269  & 2355 \\
$3/2^{-*}$  & 1733  & 1831  & 2175  & 1939  & 2008  & 2088 \\
$3/2^-$     & 2074  & 2175  & 2201  & 2278  & 2293  & 2391 \\
$5/2^-$     & 2088  & 2201  & 2193  & 2307  & 2301  & 2416 \\
$1/2^+$     & 1922  & 2052  & 2052  & 2180  & 2180  & 2306 \\
$3/2^+$     & 2030  & 2120  & 2120  & 2236  & 2236  & 2358 \\
$1/2^{+*}$  & 1959  & 2082  & 2082  & 2204  & 2204  & 2336 \\
$3/2^{+*}$  & 1999  & 2132  & 2132  & 2246  & 2246  & 2367 \\
$5/2^{+*}$  & 2101  & 2208  & 2208  & 2323  & 2323  & 2426
\end{tabular}
\end{table}

\newpage

\begin{table}[h!]
\begin{tabular}{l|c|c|c|c|c|c}
$15 \times \bar{3}$ & $N + \Delta$ & $\Lambda + \Sigma$ &
$\Sigma + \Sigma_2$ & $\Xi$ & $\Xi + \Xi_{3/2}$ & $\Omega + \Omega_1 $\\
$1/2^{-*}$  & 1627  & 1749  & 1793  & 1871  & 1916  & 2030 \\
$1/2^-$     & 1941  & 2032  & 2102  & 2137  & 2177  & 2259 \\
$3/2^{-*}$  & 1511  & 1612  & 1727  & 1719  & 1796  & 1869 \\
$3/2^-$     & 1943  & 2082  & 2123  & 2184  & 2219  & 2311 \\
$5/2^-$     & 1982  & 2088  & 2088  & 2197  & 2197  & 2311 \\
$1/2^+$     & 1761  & 1892  & 1892  & 2023  & 2021  & 2151 \\
$3/2^+$     & 1841  & 1960  & 1960  & 2086  & 2076  & 2203 \\
$1/2^{+*}$  & 1810  & 1932  & 1932  & 2059  & 2057  & 2187 \\
$3/2^{+*}$  & 1869  & 1984  & 1984  & 2104  & 2101  & 2225 \\
$5/2^{+*}$  & 1958  & 2068  & 2068  & 2173  & 2179  & 2285
\end{tabular}
\end{table}

\begin{table}[h!]
\begin{tabular}{l|c|c|c|c|c}
$15' \times \bar{3}$ & $Z_2$ & $\Delta_s$ & $\Sigma_s$ & $\Xi_s$ &
$\Omega_s$\\
$1/2^-$     & 2358  & 2434  & 2513  & 2597  & 2685 \\
$3/2^{-*}$  & 2143  & 2242  & 2338  & 2434  & 2529
\end{tabular}
\end{table}

\begin{table}[h!]
\begin{tabular}{l|c|c|c|c|c}
$15' \times \bar{3}$ & $\Delta + \Delta_{5/2}$ & $\Sigma + \Sigma_2$ &
$\Xi + \Xi_{3/2}$ & $\Omega + \Omega_1$ & $O $ \\
$1/2^-$     & 2280  & 2352  & 2428  & 2508  & 2591 \\
$3/2^{-*}$  & 1943  & 2066  & 2161  & 2263  & 2376
\end{tabular}
\end{table}

\begin{table}[h!]
\begin{tabular}{l|c|c|c|c|c|c}
$15 + 3 \times \bar{3}$ & $Z^1$ & $N_s$ & $N_s + \Delta_s$ & $\Lambda_s$ &
$ \Lambda_s +\Sigma_s$ & $\Xi_s $ \\
$1/2^+$      & 1732  & 1862  & 1901  & 1994  & 2030       & 2161 \\
$1/2^+$      & 1851  & 1979  & 2001  & 2106  & 2126       & 2256 \\
$3/2^+$      & 1789  & 1908  & 1957  & 2027  & 2077-2069  & 2194 \\
$3/2^+$      & 1908  & 2014  & 2063  & 2123  & 2172-2168  & 2281 \\
$1/2^{+*}$   & 1683  & 1862  & 1901  & 1958  & 1992-1996  & 2133 \\
$3/2^{+*}$   & 1767  & 1892  & 1934  & 2016  & 2061-2058  & 2185 \\
$5/2^{+*}$   & 1888  & 1989  & 2049  & 2098  & 2199-2201  & 2259
\end{tabular}
\end{table}


\begin{table}[h!]
\begin{tabular}{l|c|c|c|c|c|c}
$15 + 3 \times \bar{3}$ & $N + \Delta$ & $\Lambda + \Sigma$ &
$\Sigma + \Sigma_2$ & $\Xi$ & $\Xi + \Xi_{3/2}$ & $\Omega + \Omega_1$ \\
$1/2^+$     & 1547  & 1681  & 1719  & 1814-1852  & 1851  & 1984 \\
$1/2^+$     & 1705  & 1833  & 1851  & 1960-1980  & 1p81  & 2110 \\
$3/2^+$     & 1607  & 1736  & 1779  & 1853-1902  & 1897  & 2019 \\
$3/2^+$     & 1761  & 1870  & 1912  & 1980-2020  & 2026  & 2134 \\
$1/2^{+*}$  & 1506  & 1644  & 1688  & 1779-1826  & 1823  & 1960 \\
$3/2^{+*}$  & 1597  & 1722  & 1766  & 1839-1892  & 1889  & 2013 \\
$5/2^{+*}$  & 1729  & 1828  & 1888  & 1935-1991  & 1986  & 2096 \\
\end{tabular}
\end{table}

\begin{table}[h!]
\begin{tabular}{l|c|c|c|c|c|c}
$15' + \bar{6} \times \bar{3}$ & $Z_0 + Z_2$ & $N_s + \Delta_s$ &
$\Sigma_s$ & $\Xi_s $ & $\Omega_s$ \\
$1/2^+$     & 1983  & 2107  & 2229-2235  & 2354  & 2476 \\
$3/2^+$     & 2048  & 2165  & 2275-2280  & 2395  & 2508 \\
$1/2^{+*}$  & 1863  & 1994  & 2125-2130  & 2262  & 2390 \\
$3/2^{+*}$  & 1920  & 2045  & 2163-2173  & 2298  & 2426 \\
$5/2^{+*}$  & 2018  & 2129  & 2242-2243  & 2360  & 2546 \\
\end{tabular}
\end{table}

\begin{table}[h!]
\begin{tabular}{l|c|c|c|c|c|c}
$15' + \bar{6} \times \bar{3}$ & $N + \Delta + \Delta_{5/2}$ &
$\Lambda + \Sigma + \Sigma + \Sigma_2$ & $\Xi + \Xi_{3/2}$ &
$\Omega + \Omega_1$ & $ O $ \\
$1/2^+$     & 1838  & 1964  & 2086-2091  & 2210  & 2332 \\
$3/2^+$     & 1899  & 2017  & 2130-2149  & 2245  & 2366 \\
$1/2^{+*}$  & 1685  & 1821  & 1957-1959  & 2094  & 2230 \\
$3/2^{+*}$  & 1742  & 1872  & 2001-2001  & 2130  & 2260 \\
$5/2^{+*}$  & 1843  & 1958  & 2075-2075  & 2192  & 2313
\end{tabular}
\end{table}

\section{Diagrams}
\begin{wrapfigure}{c}{\columnwidth}
\begin{center}
\includegraphics[width=0.58\textwidth]{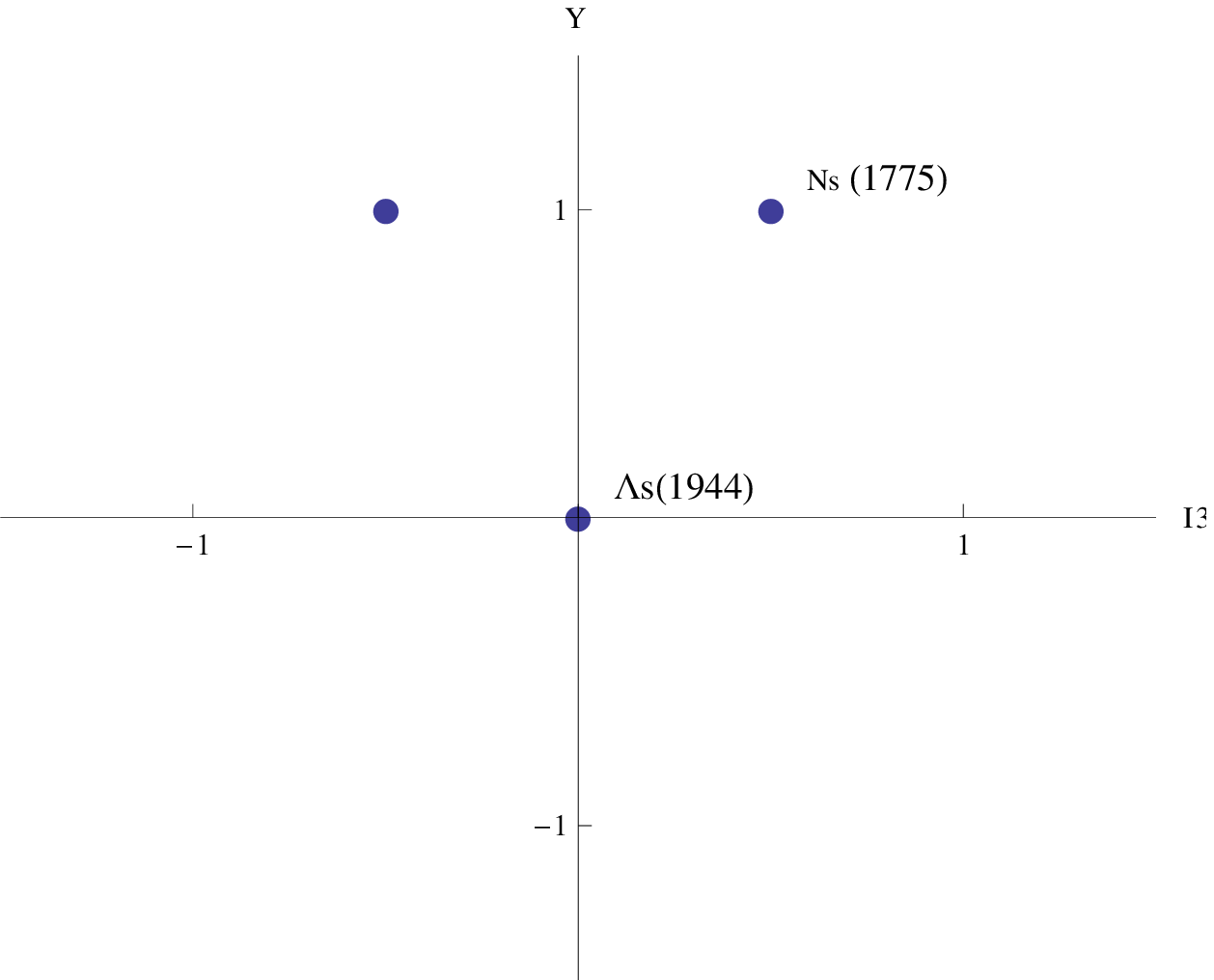}
\vspace{1.5cm}
\includegraphics[width=0.58\columnwidth]{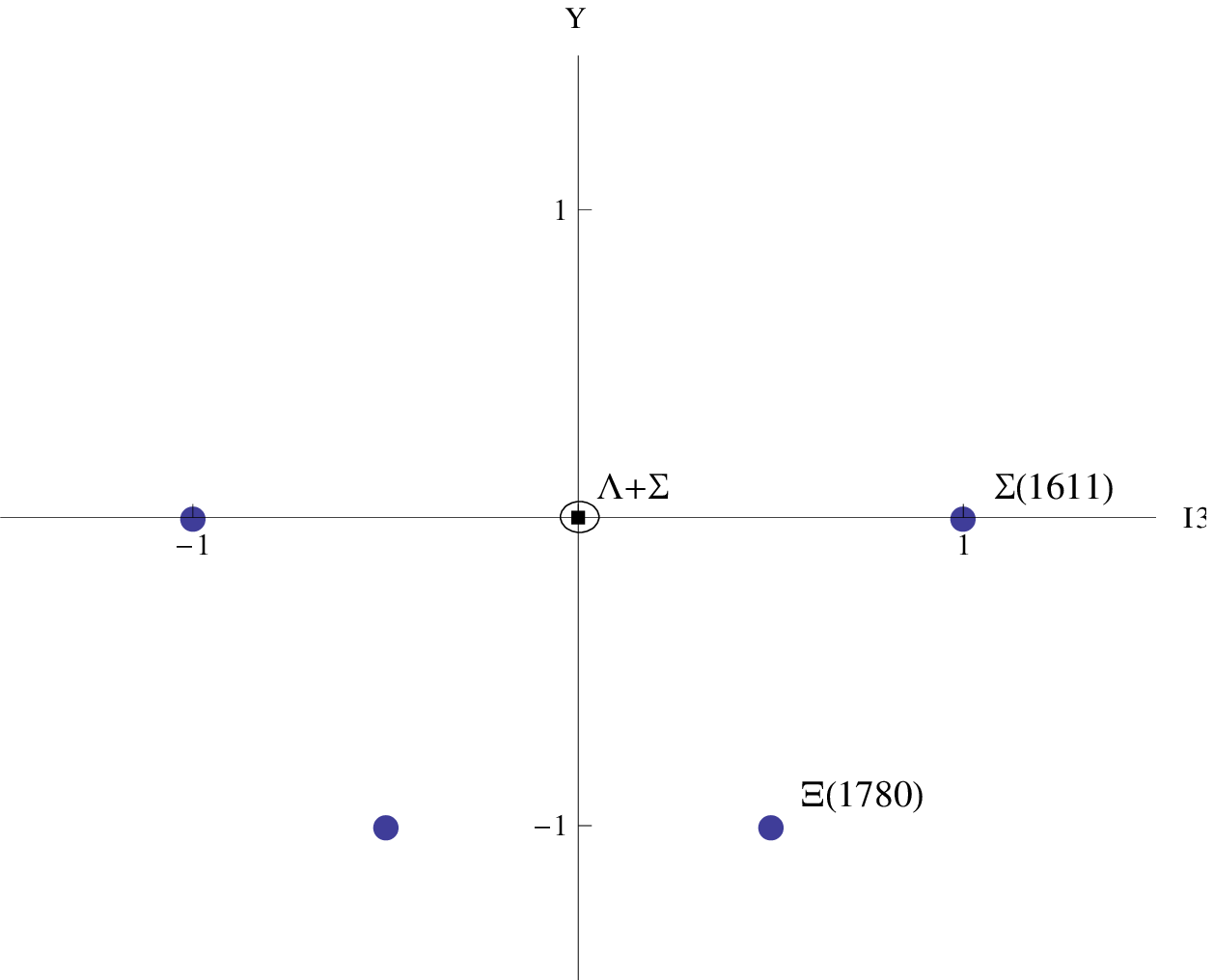}
\caption{
States with $J^P = 1/2^+$
obtained with the tetraquark in $3_F$ representation.
The upper diagram (1A) correspond to $\bar{s}$ and the
lower one (1B) to $\bar{u}$ and $\bar{d}$. Small circles denote
a weight degeneracy.}
\end{center}
\end{wrapfigure}

\begin{wrapfigure}{c}{\columnwidth}
\begin{center}
\includegraphics[width=0.6\textwidth]{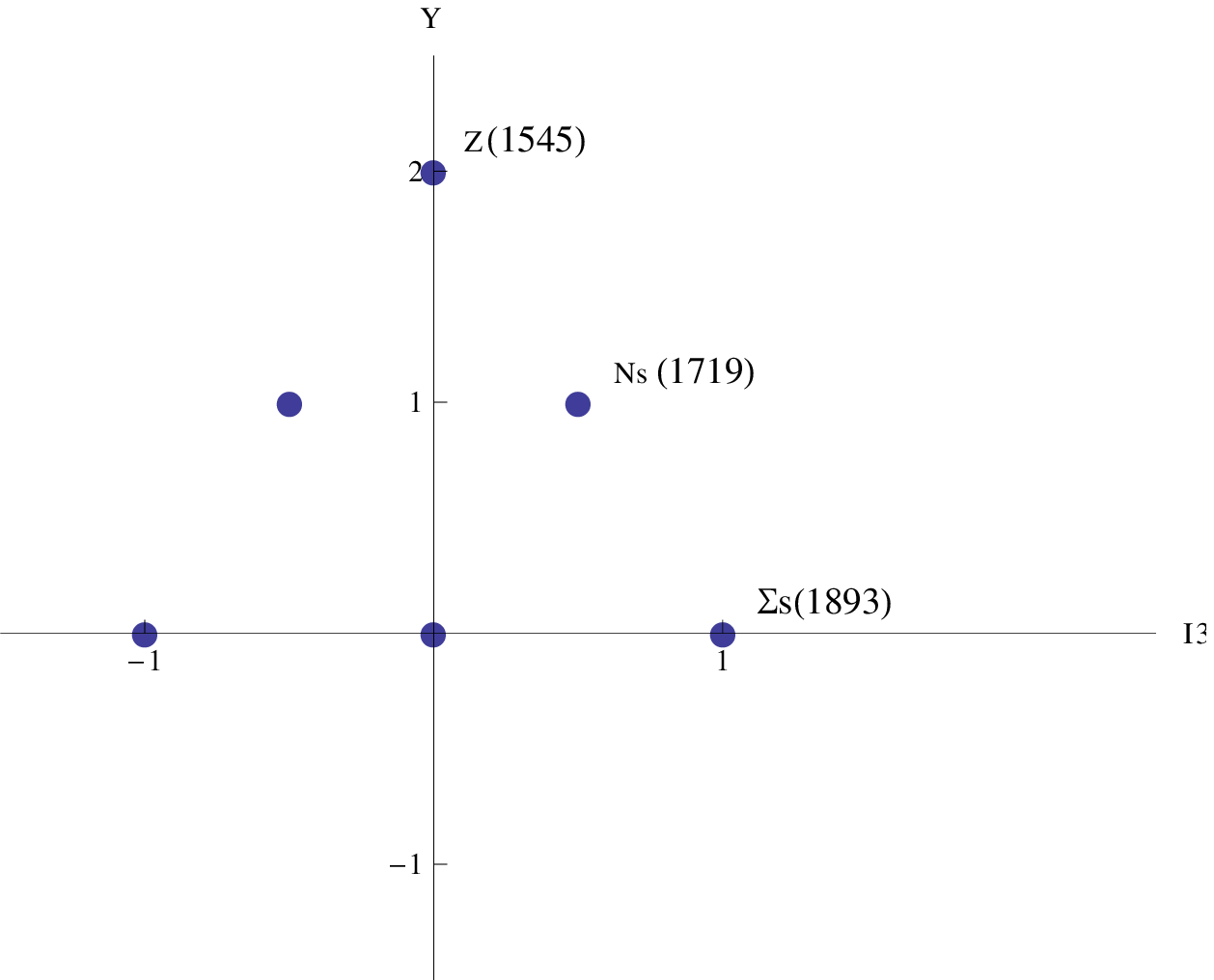}
\vspace{1.5cm}
\includegraphics[width=0.6\columnwidth]{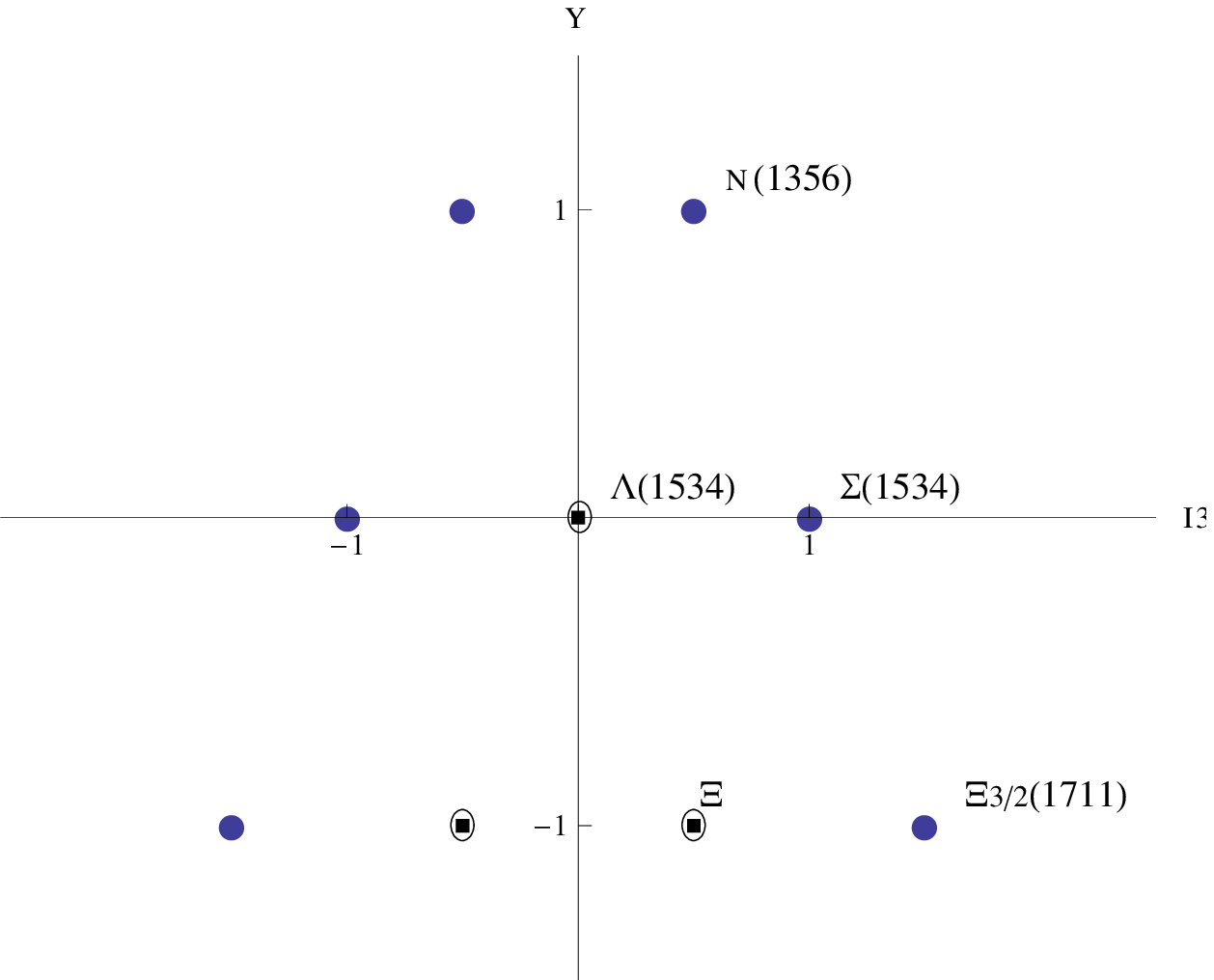}
\caption{
States with $J^P = 1/2^+$
obtained with the tetraquark in $\bar{6}_F$ representation.
The upper diagram (2A) correspond to $\bar{s}$ and the
lower one (2B) to $\bar{u}$ and $\bar{d}$. Small circles denote
a weight degeneracy.}
\end{center}
\end{wrapfigure}

\begin{wrapfigure}{c}{\columnwidth}
\begin{center}
\includegraphics[width=0.6\textwidth]{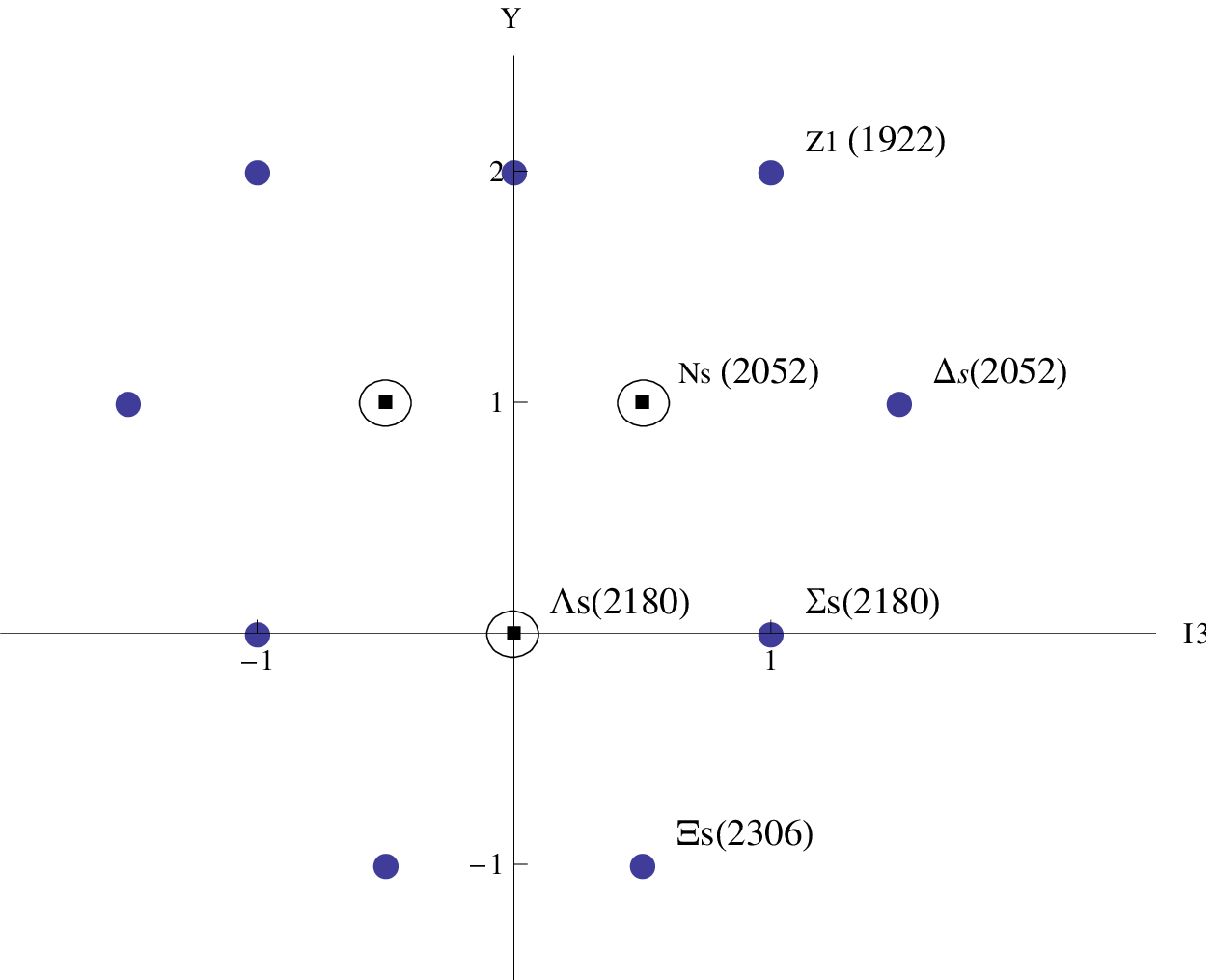}
\vspace{1.5cm}
\includegraphics[width=0.6\columnwidth]{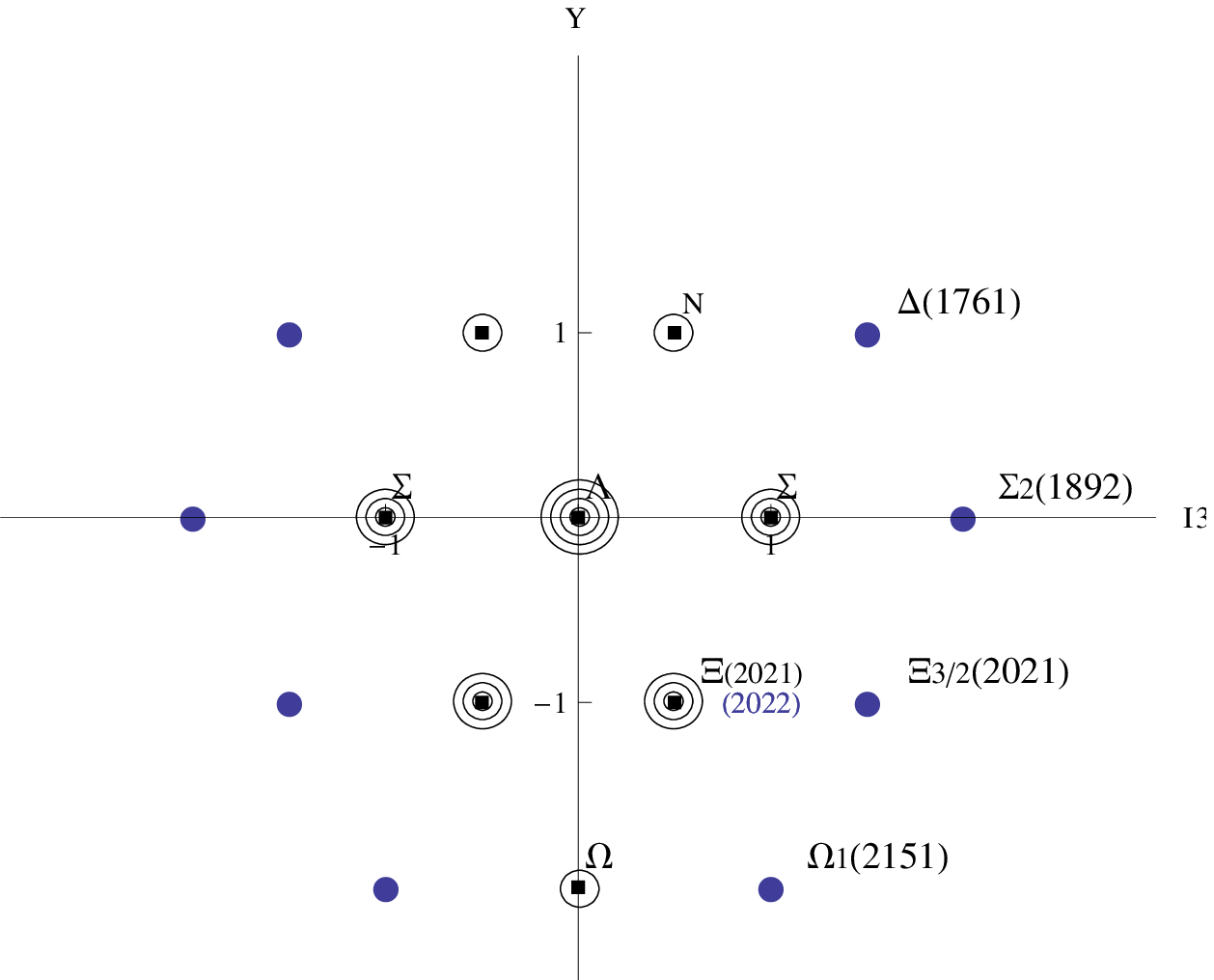}
\caption{
States with $J^P = 1/2^+$
obtained with the tetraquark in $15_F$ representation.
The upper diagram (3A) correspond to $\bar{s}$ and the
lower one (3B) to $\bar{u}$ and $\bar{d}$. Small circles denote
a weight degeneracy.}
\end{center}
\end{wrapfigure}

\begin{wrapfigure}{c}{\columnwidth}
\begin{center}
\includegraphics[width=0.6\textwidth]{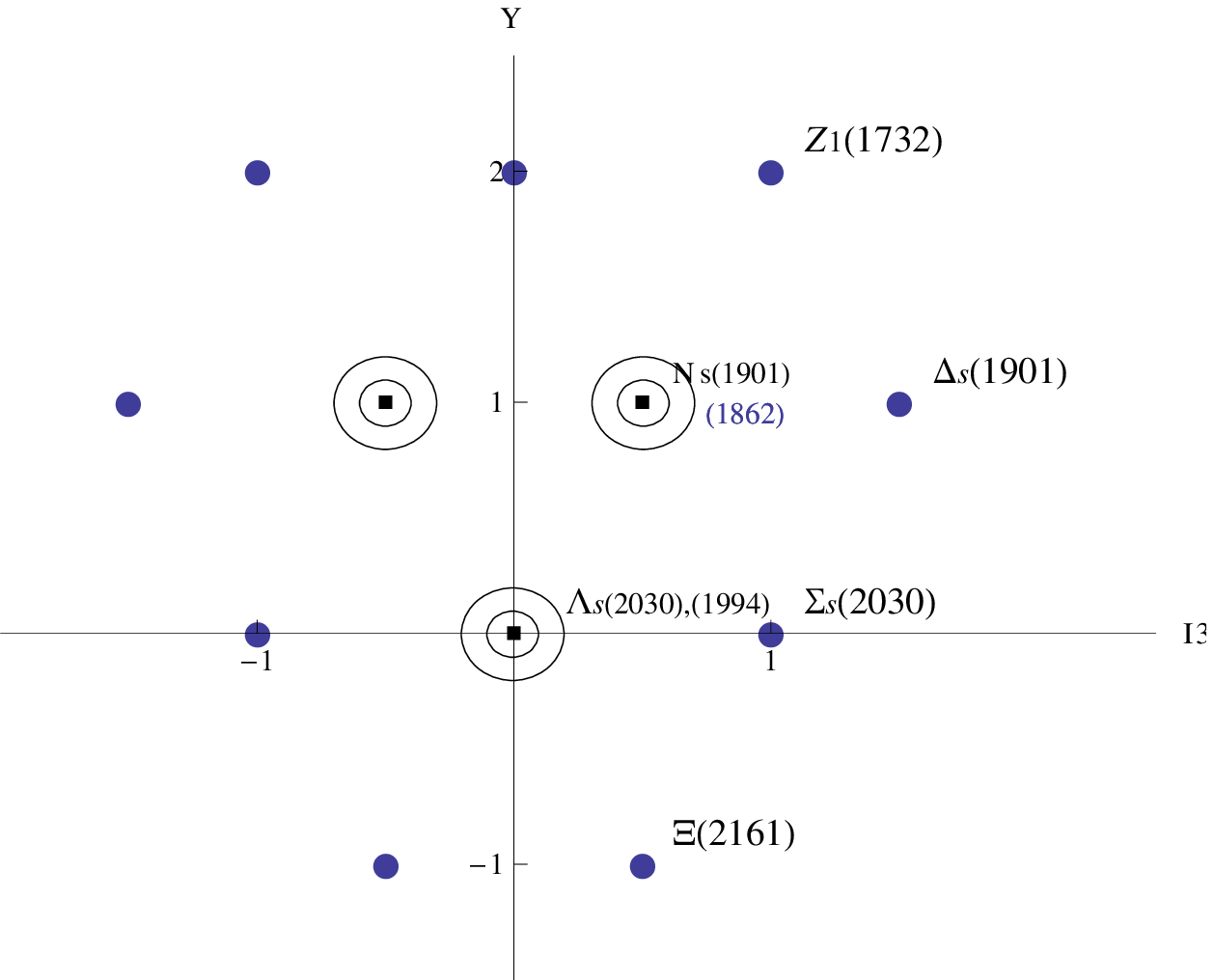}
\vspace{1.5cm}
\includegraphics[width=0.6\columnwidth]{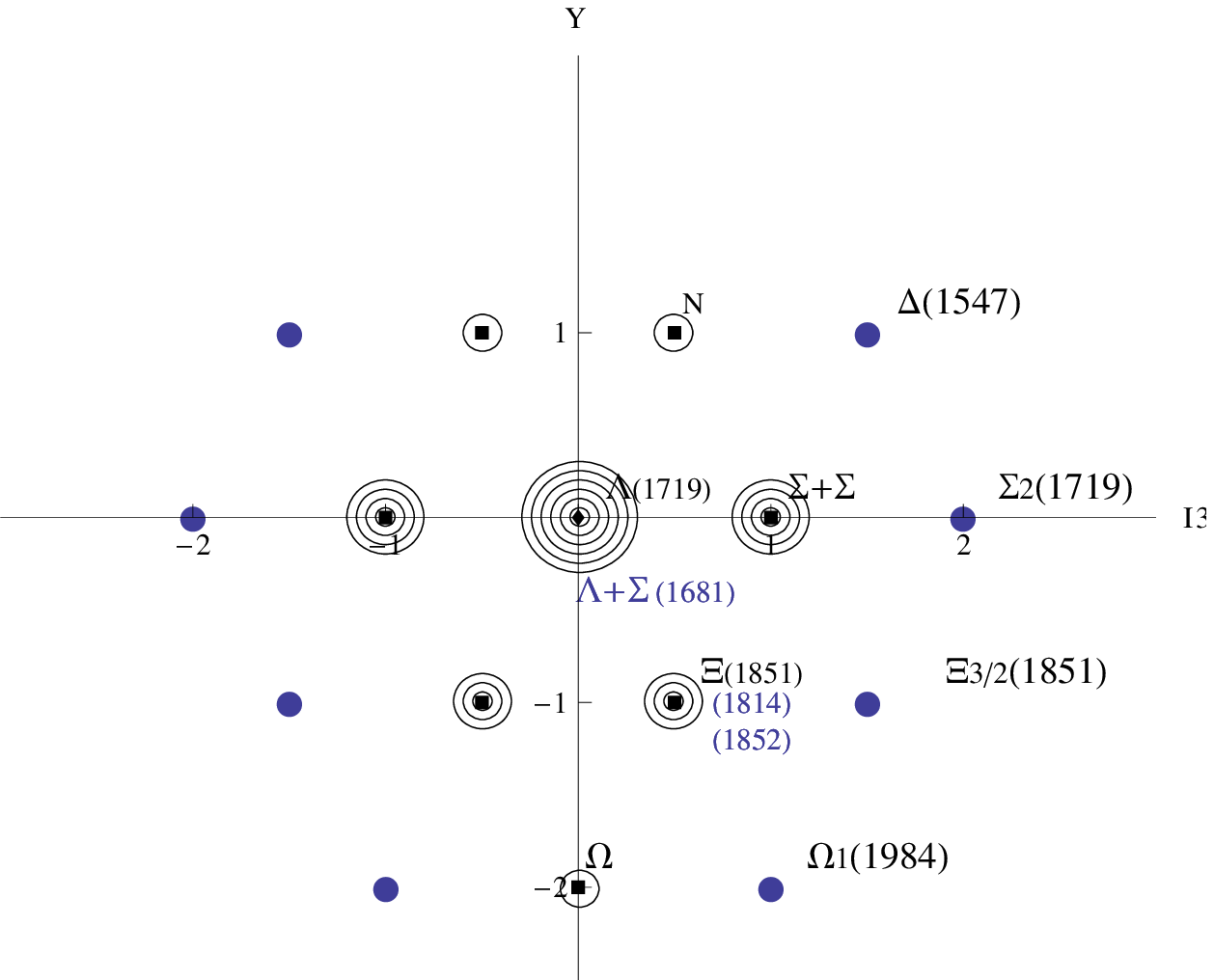}
\caption{
States with $J^P = 1/2^+$
obtained with the tetraquark in mixed representation $15_F + 3_F$.
The upper diagram (4A) correspond to $\bar{s}$ and the
lower one (4B) to $\bar{u}$ and $\bar{d}$. Small circles denote
a weight degeneracy.}
\end{center}
\end{wrapfigure}

\begin{wrapfigure}{c}{\columnwidth}
\begin{center}
\includegraphics[width=0.6\textwidth]{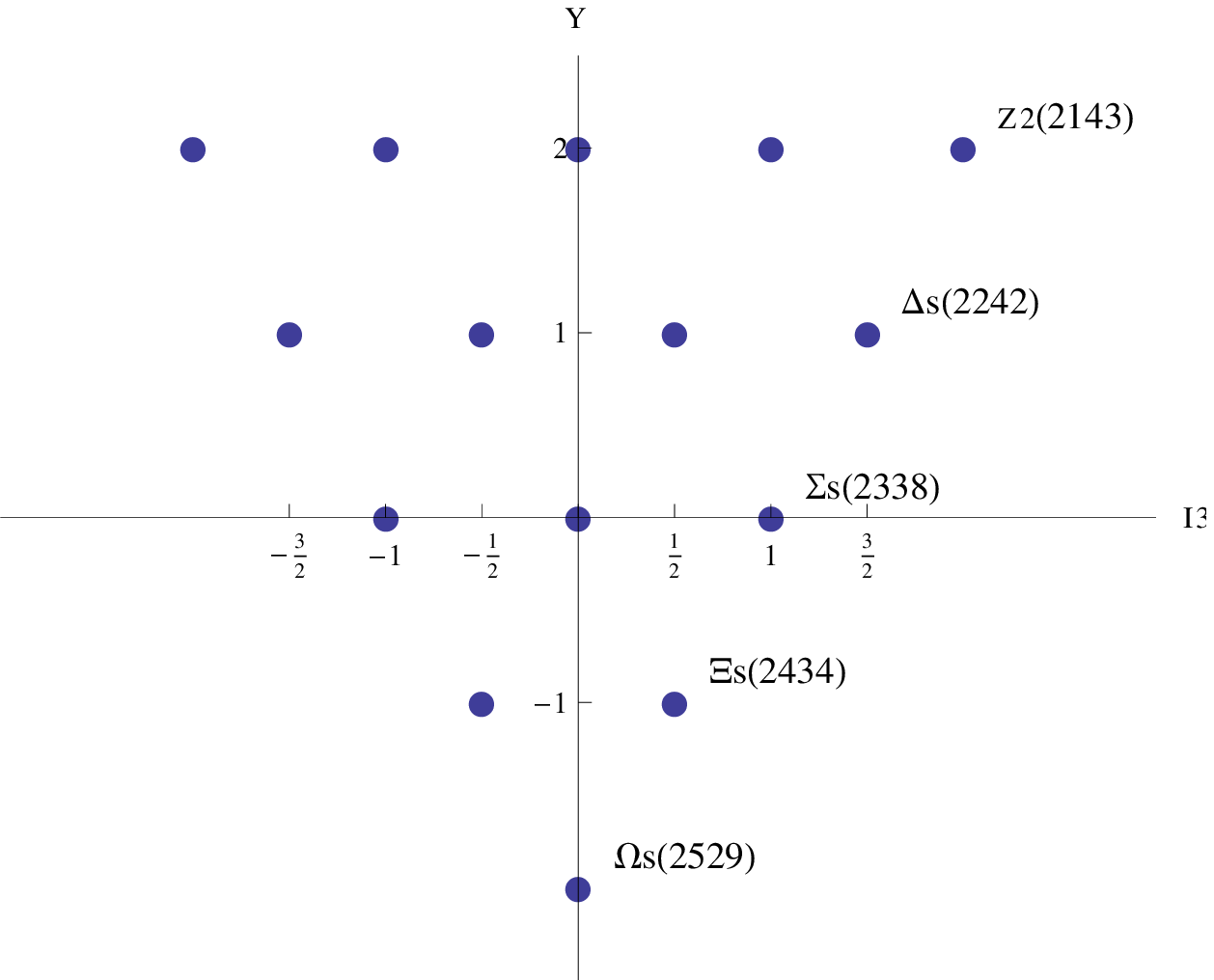}
\vspace{1.5cm}
\includegraphics[width=0.6\columnwidth]{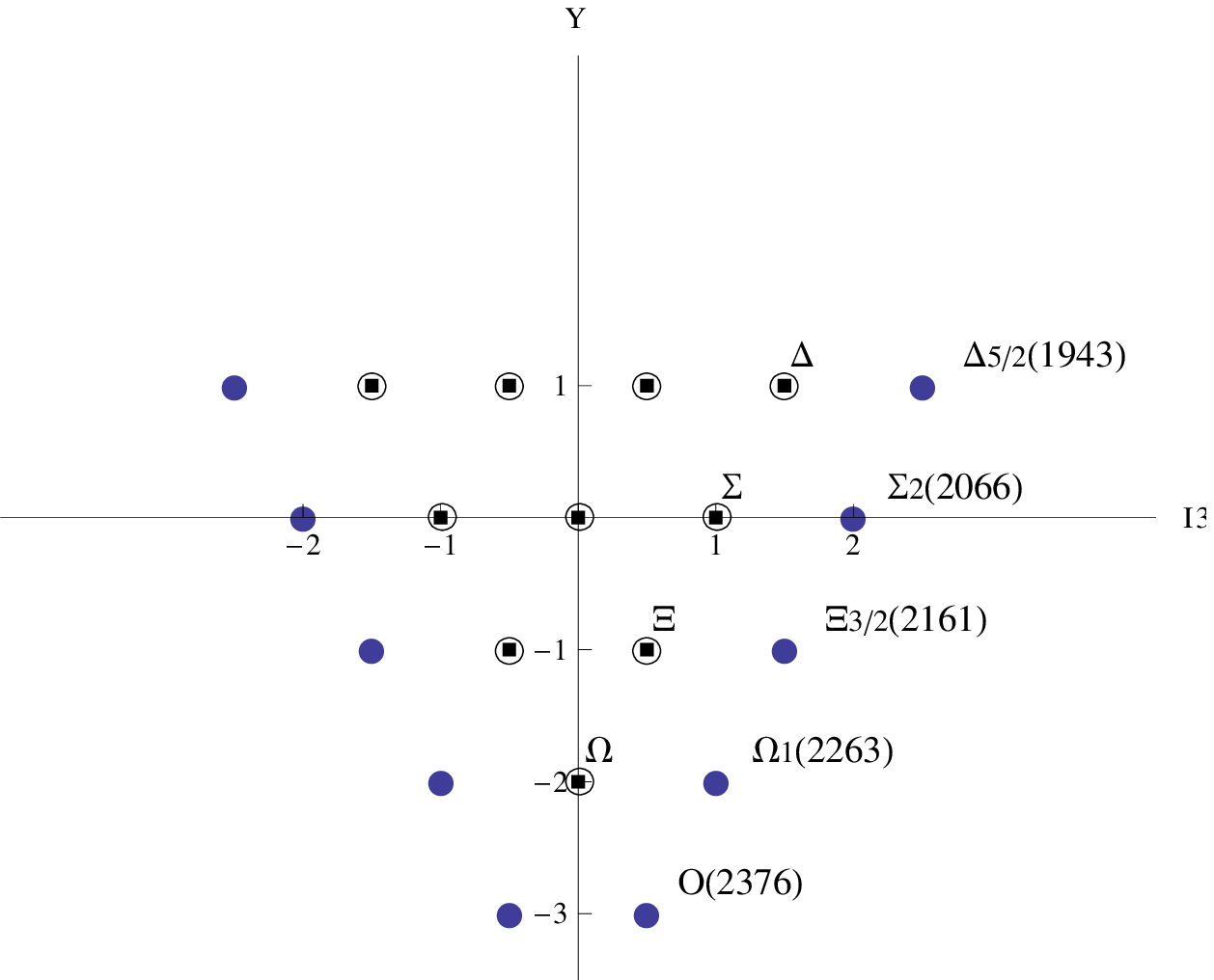}
\caption{
States with $J^P = 3/2^-$
obtained with the tetraquark in a $15'_F$ representation.
The upper diagram (5A) correspond to $\bar{s}$ and the
lower one (5B) to $\bar{u}$ and $\bar{d}$. Small circles denote
a weight degeneracy.}
\end{center}
\end{wrapfigure}

\begin{wrapfigure}{c}{\columnwidth}
\begin{center}
\includegraphics[width=0.6\textwidth]{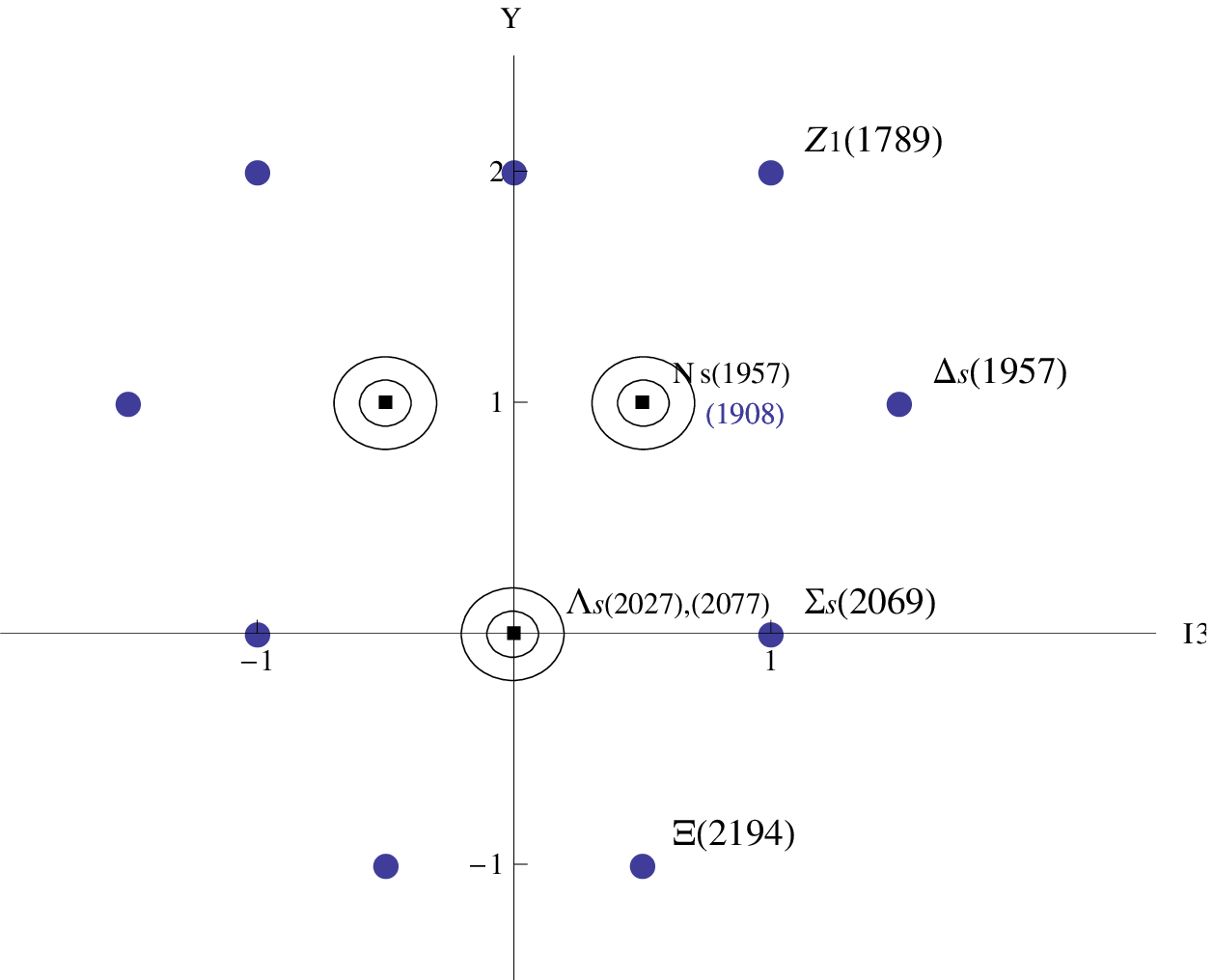}
\vspace{1.5cm}
\includegraphics[width=0.6\columnwidth]{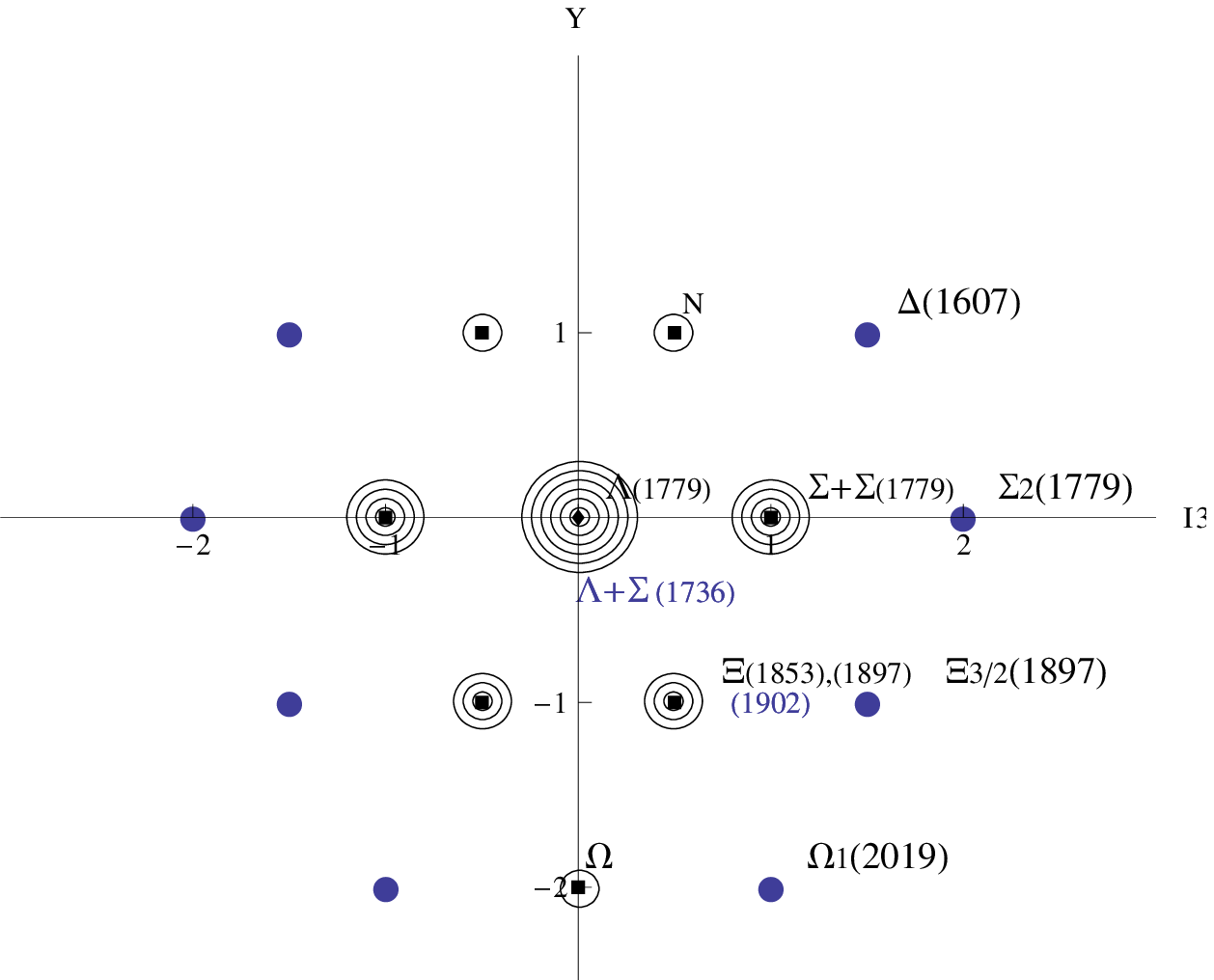}
\caption{
States with $J^P = 3/2^+$
obtained with the tetraquark in a mixed $15_F + \bar{3}_F$ representation.
The upper diagram (6A) correspond to $\bar{s}$ and the
lower one (6B) to $\bar{u}$ and $\bar{d}$. Small circles denote
a weight degeneracy.}
\end{center}
\end{wrapfigure}

\begin{wrapfigure}{c}{\columnwidth}
\begin{center}
\includegraphics[width=0.6\textwidth]{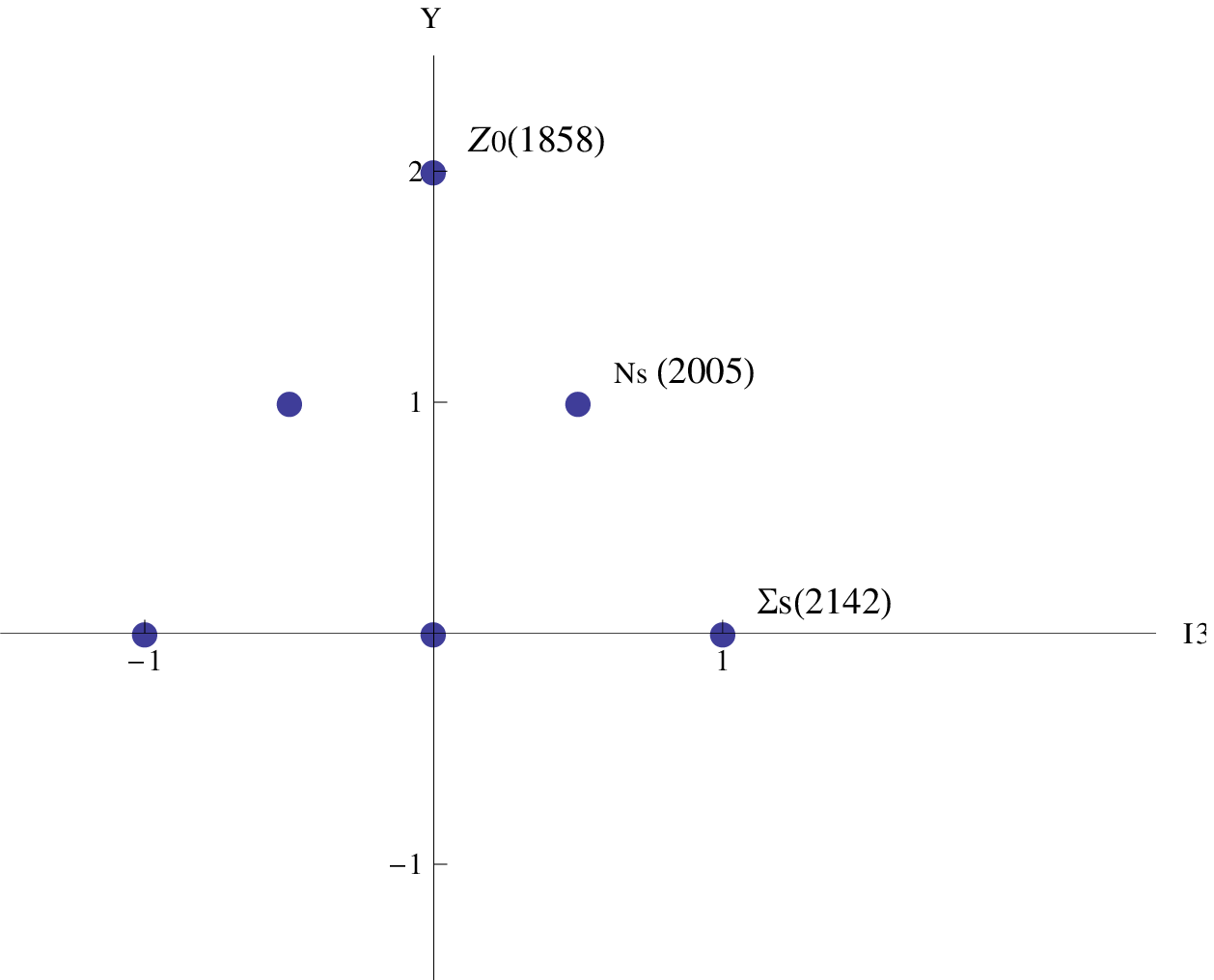}
\vspace{1.5cm}
\includegraphics[width=0.6\columnwidth]{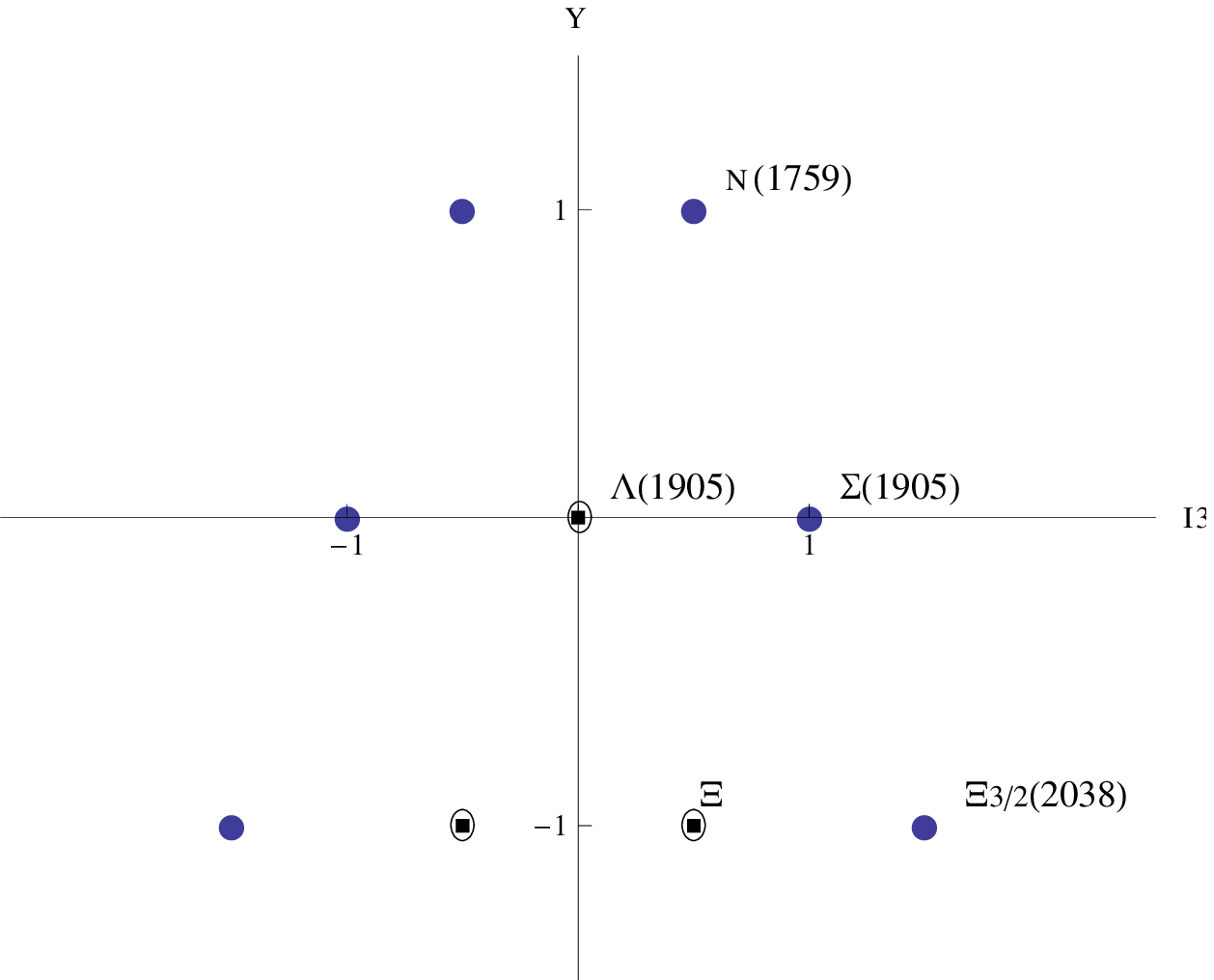}
\caption{
States with $J^P = 3/2^-$
obtained with the tetraquark in a $\bar{6}_F$ representation.
The upper diagram (7A) correspond to $\bar{s}$ and the
lower one (7B) to $\bar{u}$ and $\bar{d}$. Small circles denote
a weight degeneracy.}
\end{center}
\end{wrapfigure}

\begin{wrapfigure}{c}{\columnwidth}
\begin{center}
\includegraphics[width=0.6\textwidth]{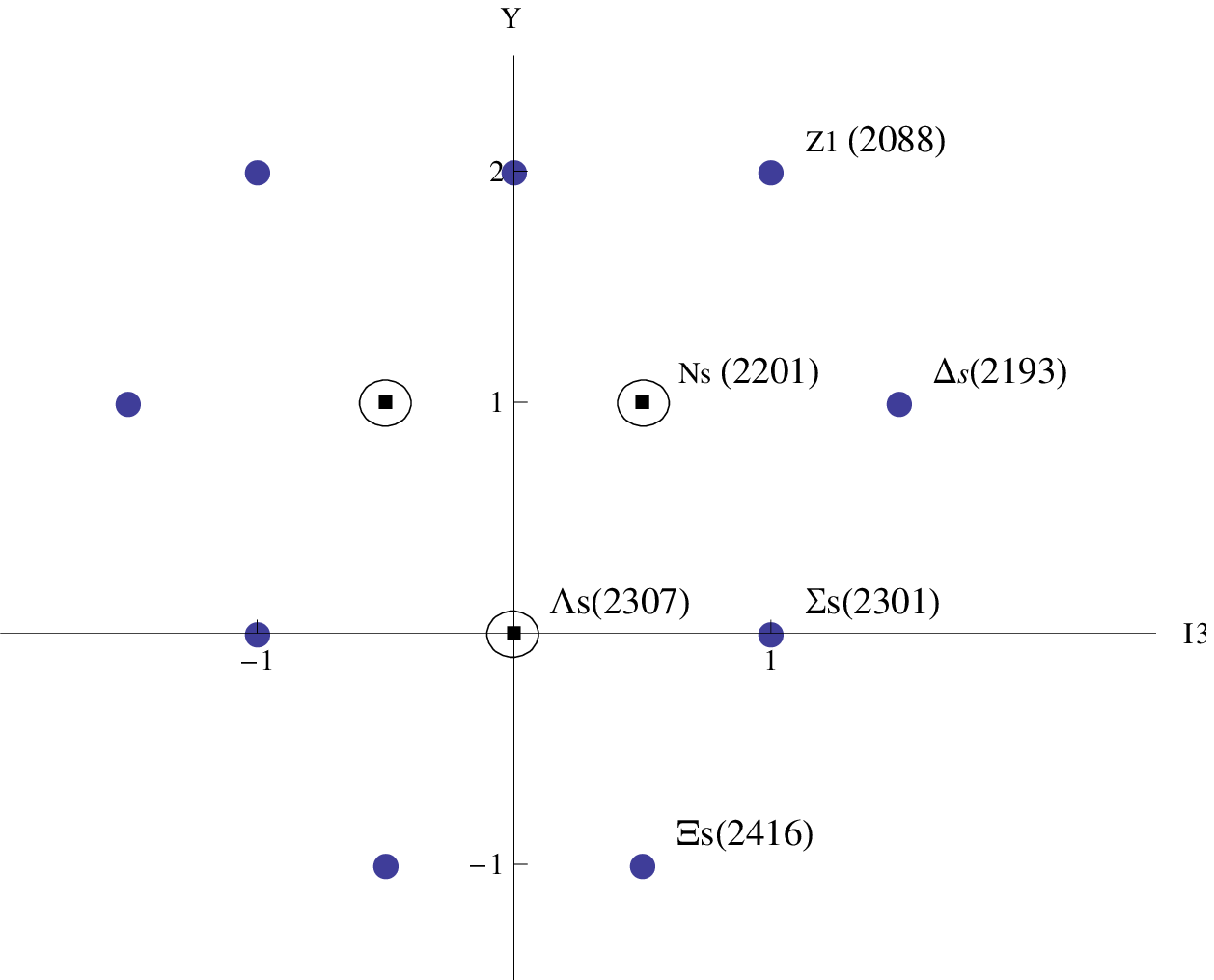}
\vspace{1.5cm}
\includegraphics[width=0.6\columnwidth]{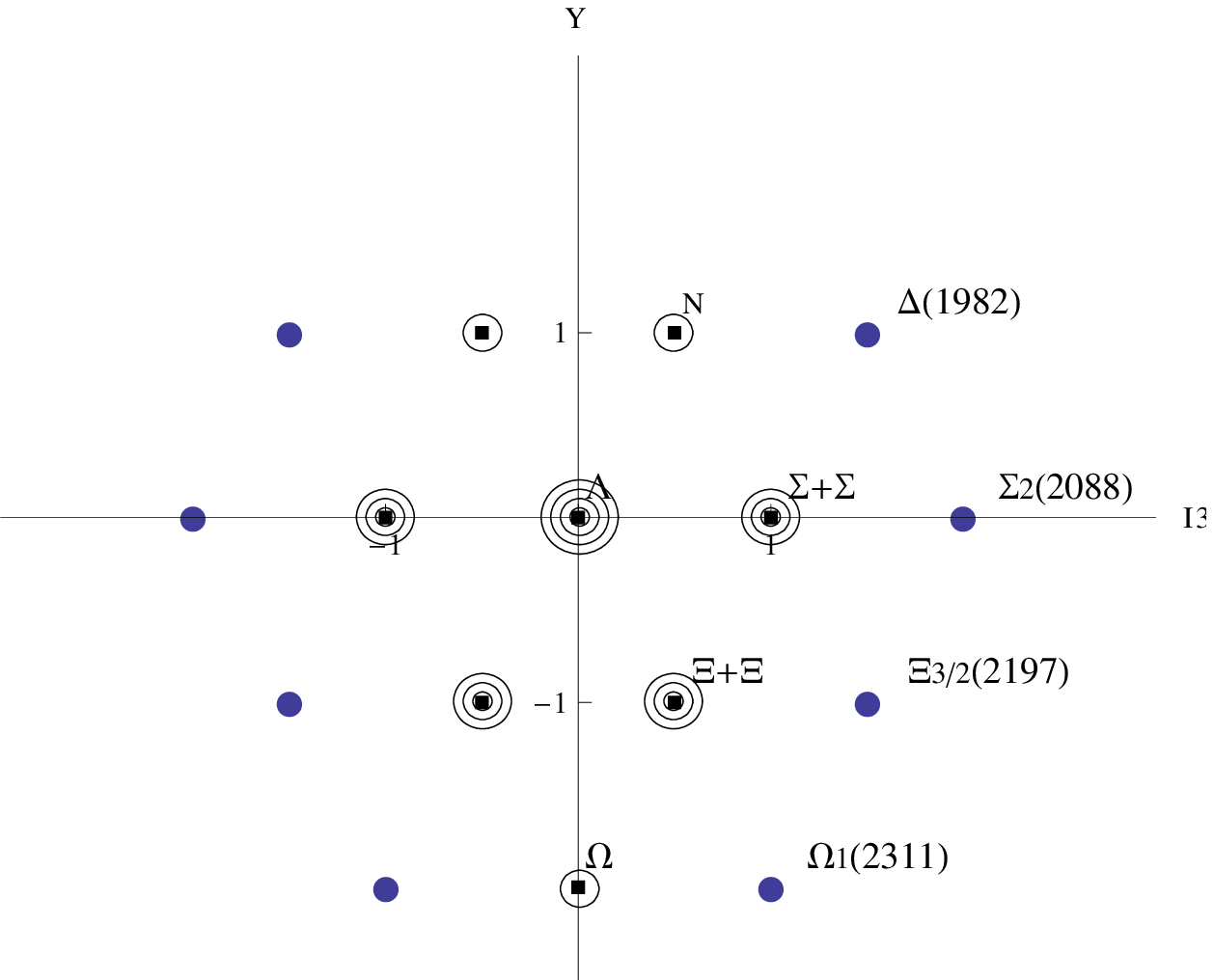}
\caption{
States with $J^P = 5/2^-$
obtained with the tetraquark in a $15_F$ representation.
The upper diagram (8A) correspond to $\bar{s}$ and the
lower one (8B) to $\bar{u}$ and $\bar{d}$. Small circles denote
a weight degeneracy.}
\end{center}
\end{wrapfigure}


\newpage

\begin{thebibliography}{99}

\bibitem{AR1}  The Particle Data Group \textit{Phys. Lett.} \textbf{B111}
(1982) 1.

\bibitem{AR2}  R. A. Arndt and L. D. Roper \textit{Phys.~Rev.}~\textbf{D31}
(1985) 2230.

\bibitem{AR3}  J. S. Hyslop, R. A. Arndt , L. D. Roper and R. L. Workman 
\textit{Phys.~Rev.} \textbf{D46} (1992) 961.

\bibitem{NA}  NA49 Collaboration, C. Alt et al., \textit{Phys.~Rev.~Lett.}~
\textbf{92} (2004) 042003

\bibitem{nakano}  LEPS Collaboration, T. Nakano et al., \textit{Phys. Rev.
Lett.} \textbf{91} (2003) 012002; \\
DIANA collaboration, V.V. Barmin et al.,
Phys Atom. Nucl. 66, 1715 (2003)

\bibitem{N}  CLAS Collaboration, S. Stepanyan et al., \textit{Phys. Rev.
Lett.} \textbf{91} (2003) 252001;\newline
SAPHIR Collaboration, J. Barth et al., \textit{Phys. Lett.} \textbf{B572}
(2003) 127;\newline
CLAS Collaboration, V. Kurakovsky and S. Stepanyan, 8th International
Conference on Interactions of Particle and Nuclear Physics, New York (2003),
AIP \textit{Conf. Proc.} \textbf{698} (2004) 543.

\bibitem{JW}  
R. L. Jaffe and F. Wilczek, \textit{Phys. Rev. Lett.} \textbf{91}
(2003) 232003;  0307341

\bibitem{SR}  F. Stancu and D. O. Riska, \textit{Phys.Lett.} \textbf{B575}
(2003) 242;

\bibitem{All} B. K. Jennings and K. Maltman, \textit{Phys. Rev.}\textbf{D69} (2004)
094020; \\
R. Bijker, M.M. Giannini and E. Santopinto, \textit{Eur. Phys. J.} \textbf{
A22}, (2004) 319; \\
C. E. Carlson, D. Carone, J. Kwee and V. Nazaryan, \textit{Phys. Lett.} 
\textbf{B573} (2003) 101 and \textbf{B579} (2004) 52; \\ 
F. Close and J.J. Dudek, \textit{Phys. Lett.} \textbf{B583} (2004) 278; \\
H. H\"{o}gaasen and P. Sorba, {Mod. Phys.Lett.} \textbf{A19} (2004) 2403; \\
I. M. Narodetskii, C. Semay, B. Silvestre-Brac and Yu. A. Simonov,
\textit{Nucl. Phys. Proc. Suppl.} \textbf{142} (2005) 383;

\bibitem{CLAS}  CLAS Collaboration, B. Mc Kinnon et al., Phys. Rev. Lett. 
\textbf{96} (2006) 2120031

\bibitem{news}  V.V. Barmin et al.,DIANA Coll., Phys.Atom.Nucl. 70 (2007)
35-43; A. Kubarovsky, SVD Coll., arXiv:hep-ex/0610050v1

\bibitem{BES}  BES Collaboration, 
M. Ablikim et al., 
\textit{Phys. Rev. Lett.} \textbf{97} (2006) 062001.

\bibitem{Nic}  V.~A.~ Nikonov,
{\it Proceedings to the 12th International Conference on hadron
spectroscopy (HADRON 07), Frascati, Italy, 8-13 Oct 2007}

\bibitem{HS}  H. H\"{o}gaasen and P. Sorba, \textit{Nucl. Phys.} \textbf{B145
} (1978) 119; \newline
M. De Crombrugghe, H. H\"{o}gaasen and P. Sorba, \textit{Nucl. Phys.} 
\textbf{B156} (1979) 347.

\bibitem{DGG}  A. De R\'{u}jula, H. Georgi and S.L. Glashow, \textit{Phys.
Rev.}\textbf{D12} (1975) 147.

\bibitem{GR}  F. Gursey and L.A. Radicati,\textit{Phys. Rev. Lett.} \textbf{
13} (1964) 173.

\bibitem{BFT}  F.~Buccella, D.~Falcone and F.~Tramontano, \textit{Central 
Eur. J. Phys.} \textbf{3} (2005) 525.

\bibitem{BHRS}  F. Buccella , H. H\"{o}gaasen, J. M. Richard and P. Sorba, 
\textit{Eur. Phys. J.} \textbf{C49} (2007) 743

\bibitem{J}  R. L. Jaffe, \textit{Phys. Rev.} \textbf{D17} (1978) 1444.

\bibitem{Bu}  F. Buccella, \textit{Mod. Phys. Lett.} \textbf{A21} (2006) 831

\bibitem{BS}  F. Buccella and P. Sorba, \textit{Mod. Phys. Lett.}
\textbf{A19} (2004) 1547

\bibitem{Dan}  M. Danilov and R. Mizuk hep-ex 0704.3531v1

\bibitem{Ak}  A. Aktas et al. \textit{Phys. Lett.} \textbf{B588} (2004) 17.

\bibitem{BES2}  BES Collaboration,
{\it Proceedings to the 12th International Conference on hadron
spectroscopy (HADRON 07), Frascati, Italy, 8-13 Oct 2007}


\end{thebibliography}
\end{document}